\documentclass[superscriptaddress,
 reprint,
superscriptaddress,
 amsmath,amssymb,
 aps,
prb,
floatfix,
]{revtex4-2}

\usepackage{silence}
\usepackage{graphicx}
\usepackage{dcolumn}
\usepackage{bm}
\usepackage{bbm} 
\usepackage{mathtools}
\usepackage{leftindex} 

\graphicspath{{./Figures/}}

\usepackage{float} 
\usepackage{silence}
\usepackage{fix-cm} 
\usepackage[dvipsnames]{xcolor}
\usepackage[normalem]{ulem}
\usepackage[english]{babel}
\usepackage{graphicx}
\usepackage{dcolumn}
\usepackage{verbatim}
\usepackage{mathrsfs}
\usepackage{cancel}
\usepackage{epstopdf}
\usepackage{color}
\usepackage{tensor}
\usepackage{relsize} 
\usepackage{ytableau}
\usepackage{slashed} 
\usepackage{caption}
\usepackage{subcaption} 
\usepackage{boldline} 

\usepackage{pifont}
\usepackage[dvipsnames]{xcolor}
\definecolor{Verde}{RGB}{0, 200, 0} 

\newcommand{\eq}[1]{(\ref{#1})}

\newcommand{\dd}{\mathrm{d}}
\newcommand{\ii}{\mathrm{i}}

\newcommand{\Tr}[1]{\text{Tr}\left(#1\right)}
\newcommand{\Abs}[1]{\left\vert #1\right\vert}

\newcommand{\Mu}{\text{M}}
\newcommand{\surf}{\parallel}
\newcommand{\Eqref}[1]{Eq. \eqref{#1}}
\newcommand{\dlim}{\displaystyle\lim}
\newcommand{\dint}{\displaystyle\int}
\newcommand{\dsum}{\displaystyle\sum}
\newcommand{\mean}[1]{\big\langle #1\big\rangle}
\newcommand{\abs}[1]{\vert #1\vert}

\newcommand{\Imag}[1]{\mathbb{I}\text{m}\left[ #1 \right]}
\newcommand{\Real}[1]{\mathbb{R}\text{e}\left[ #1 \right]}
\newcommand{\Eq}[1]{Eq.~\eqref{#1}}
\newcommand{\Eqs}[1]{Eqs.~\eqref{#1}}
\newcommand{\Fig}[1]{Fig.~\ref{#1}}
\newcommand{\Sect}[1]{Sect.~\ref{#1}}

\newcommand{\eV}{\text{ eV}}
\newcommand{\KK}{\text{ K}}
\newcommand{\nm}{\text{ nm}}

\usepackage[usenames,dvipsnames]{pstricks}
\usepackage{epsfig}
\usepackage{pst-grad} 
\usepackage{pst-plot} 

\usepackage{hyperref}

\usepackage{verbatim}

\begin{document}

\preprint{APS/123-QED}

\title{Casimir-Lifshitz force with graphene: theory versus experiment, role of spatial non-locality and of losses}

\author{Pablo Rodriguez-Lopez}
\email{pablo.ropez@urjc.es}
\affiliation{{\'A}rea de Electromagnetismo and Grupo Interdisciplinar de Sistemas Complejos (GISC), Universidad Rey Juan Carlos, 28933, M{\'o}stoles, Madrid, Spain}
\affiliation{Laboratoire Charles Coulomb (L2C), UMR 5221 CNRS-University of Montpellier, F-34095 Montpellier, France}

\author{Mauro Antezza}
\email{mauro.antezza@umontpellier.fr}
\affiliation{Laboratoire Charles Coulomb (L2C), UMR 5221 CNRS-University of Montpellier, F-34095 Montpellier, France}
\affiliation{Institut Universitaire de France, Ministère de  l’Enseignement Supérieur et de la Recherche, 1 rue Descartes, F-75231, Paris, France}

\date{\today}
\begin{abstract}
We analyze the impact of spatial non-locality and losses in the electromagnetic response of graphene on the Casimir-Lifshitz interaction. To this purpose, we calculate the Casimir-Lifshitz force (CLF) between a gold sphere and a graphene-coated SiO$_2$ plane and compare our finding with the recent experiment in PRL {\bf 126}, 206802 (2021) and PRB {\bf 104}, 085436 (2021). We calculated the CLF using three different models for the electromagnetic response of graphene: electric conductivity using a non-local and lossy Kubo model, electric conductivity using the local and lossy Kubo model, and the non-local and lossless polarization operator model. The relation between these three models has been recently explored in PRB {\bf 111}, 115428 (2025). We show that, for the parameters of the available experiments, the theoretical predictions for the Casimir-Lifshitz force using the three models are practically identical (having a relative differences smaller than $10^{-3}$). This implies that for those given experiments, both non-local and lossy effects in the graphene response are completely negligible. We also find that this experiment cannot distinguish between the Drude and Plasma prescriptions for the involved materials (gold and graphene). Our findings are relevant for present and future comparisons with experimental measurement of the Casimir-Lifshitz force involving graphene structures. Indeed, we show that an extremely simple local Kubo model for the electric conductivity, explicitly depending on Dirac mass, chemical potential, losses and temperature, is largely enough for a totally comprehensive comparison with typical experimental configurations. We also show how the Polarization tensor must be used and modified in general, for phenomena needing a more fine response function, i.e. requiring both the spatial non-locality and losses.
\end{abstract}

\maketitle


\section{Introduction}
\begin{figure}[H]
\centering
\includegraphics[width=0.9\linewidth]{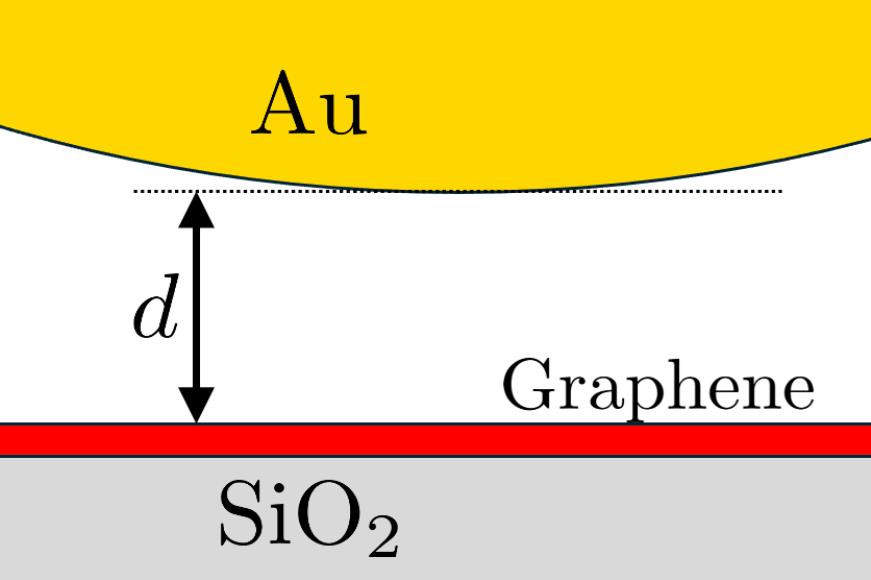}
\caption{ Scheme of the systems, as in the experiment \cite{PRL_Mohideen, PRB_Mohideen}. An SiO$_{2}$ plate covered with a single sheet of graphene placed at a distance $d$ of a gold covered sphere of radius $R$.}
\label{Fig_setup}
\end{figure}

The Casimir-Lifshitz force (CLF) is an ubiquitous dispersion interaction acting between polarizable objects \cite{DLP_1961,RevModPhys.88.045003}. Originated by the quantum and thermal fluctuations of the electromagnetic field, it strongly depends on both the geometry and the dielectric properties of the involved bodies. In recent years, with the arrival and extended investigation of 2D materials in different contexts, particular attention has been devoted to the CLF in graphene-based systems\cite{MAntezza17,Rodriguez-Lopez2017,GomezSantos2009, Bimonte2017, Bordag2009, Woods2010,Arxivwang2024photon,PabloRL2024,PhysRevA.108.062811,PhysRevB.108.115412,YoussefPRB}. Remarkably, it has been predicted that the CLF between two parallel graphene sheets has an extraordinary high thermal effect already at very short separation \cite{GomezSantos2009}. Recently, M. Liu \emph{et al.} \cite{PRL_Mohideen, PRB_Mohideen} measured the CLF gradient between a metallic sphere and a planar SiO$_{2}$ substrate coated with graphene at room temperature (see the scheme in \Fig{Fig_setup}) and compared their results with theory predictions using the standard Lifshitz theory.

A crucial ingredient of the Lifshitz theory is the knowledge of the dielectric and diamagnetic response of the involved materials, which, in general, must take into account the presence of both losses and non-locality. In the analysis of M. Liu \emph{et al.} \cite{PRL_Mohideen,PRB_Mohideen}, the authors used a Quantum Field Theory (QFT) model $J_{\mu} = -\Pi_{\mu\nu}A^{\nu}$ providing a non-local electromagnetic response $\Pi_{\mu\nu}$ for graphene which, by construction, takes into account the complete electric and magnetic response. That model does not include the losses. In graphene several lossy mechanisms are present: in pristine graphene there are losses due to unavoidable interactions with phonons and Coulomb interaction, and in real samples of graphene also the scattering from impurities must be taken into account \cite{RevModPhys.81.109}\cite{DasSarmaRMP2011}\cite{Gusynin2006}\cite{Marinko2009}\cite{RMPdissipation2022}\cite{BOOKAbrikosov}. Hence that lossless model, as it is, cannot be used for general purposes. On the other hand, in \cite{non-local_Graphene_Lilia_Pablo}, a Kubo model for the non-local electric conductivity of graphene $J_{\mu} = \sigma_{\mu\nu}E^{\nu}$ has been derived. This electric conductivity naturally incorporates the effect of losses in a phenomenological way. By construction, while it provides the full response to the electric field, it does not incorporate all the possible magnetic responses that, in the purely non-local regime, must be included to have a full electromagnetic response \cite{PabloMauroGauge2024}. We have recently discussed this two models \cite{PabloMauroComparisonKuboQFT2024} showing that both comes from exactly the same Hamiltonian, and are based on exactly the same polarization $\Pi_{\mu\nu}$. We also showed that by phenomenologically introducing loss mechanisms into the polarization tensor, one can derive a general model that simultaneously captures all magnetic non-local effects (as in the QFT model) and accounts for losses (as in the Kubo model). This most complete model can now be used in all general purposes electromagnetic phenomena involving graphene. In \cite{PabloMauroComparisonKuboQFT2024} has been also shown that the QFT and Kubo models converge to the same local limit (once the proper losses have been added to the QFT model). Since experimental results are recently available for the Casimir-Lifshitz interaction in presence of graphene, it is interesting, for this specific electromagnetic phenomenon, to study the actual role played by losses, non-locality, and by a full magnetic response. To this purposes we will compare the experimental results \cite{PRL_Mohideen,PRB_Mohideen} with three models: (i) the Quantum Field Theory (QFT) based model \cite{Klimchitskaya2016,Klimchitskaya2017,Klimchitskaya2018} derived from the direct relation $J_{i} = - \Pi_{ij}A^{j}$ and used in \cite{PRL_Mohideen,PRB_Mohideen}, that is non-local, includes the magnetic response, but is lossless \cite{Klimchitskaya2016,Klimchitskaya2017,Klimchitskaya2018} (in this study we include the effect of losses for this model), (ii) the Kubo (K) formula for the electric conductivity, derived from the microscopic Ohm law $J_{i} = \sigma_{ij}E^{j}$, that is non-local and includes lossy mechanisms, but that considers only the electric response in a complete way, missing some non-local magnetic response terms, and (iii) the common local limit (L) of the Kubo model, having a very simple expression and eventually including losses (in the zero mass-gap limit that expression reduces to the known Falkovsky model \cite{Falkovsky2007b}). This investigation shows that this third simple local model (L), even without including losses, is largely enough to explain the experimental results, and that hence neither losses or non-locality (and therefore the full non-local response to magnetic fields) plays, by far, any effective role in the experimental range covered by the available experiments of the Casimir-Lifshitz effect.

We investigate also the sensitivity of the CLF gradient to the use of a Plasma or a Drude model for the involved metallic materials (gold and graphene) showing that a possible distinction between these two models cannot be explored with the available experiments.

The article is organized as follows. In \Sect{sect:Casimir} and \Sect{sect:Fresnel} we review the Lifshitz formula for calculating the CLF and Fresnel reflection matrices to be used.
In \Sect{Section_EM_response}, we review the three models for the electric conductivity of graphene we will compare. In \Sect{sect:experiment} and \Sect{sect:Drude_vs_Plasma}, we review the experiment published in \cite{PRL_Mohideen} and \cite{PRB_Mohideen}, check the Drude and Plasma prescriptions, the effect of losses and of non-locality. We finish with the conclusions in \Sect{sect:conclusions}.

\section{Casimir-Lifshitz force}\label{sect:Casimir}
For the geometric configuration of the experiment \cite{PRL_Mohideen,PRB_Mohideen} (see scheme in \Fig{Fig_setup}), the spatial gradient of the CLF $G(d) = \partial_{d}F(d)$ between a gold covered sphere of radius $R$ and a graphene covered SiO$_{2}$ plate separated by a distance $d$ can be safely expressed within the Proximity Force Approximation \cite{bordag2009advances} ($d/R < 0.012$):
\begin{eqnarray}\label{PFA_G}
G(d) = 2k_{B}TR{\dsum_{n=0}^{\infty}}'\hspace{-0.1cm}\dint_{0}^{\infty}\hspace{-0.3cm}\dd k_{\surf} \kappa_{z}k_{\surf}\Tr{ \mathbb{N}_{12}\left(\mathbbm{1} - \mathbb{N}_{12}\right)^{-1} }.
\end{eqnarray}
Here $\mathbb{N}_{12} = \mathbb{R}_{1}\mathbb{R}_{2}e^{-2d\kappa_{z} }$, $\mathbb{R}_{\alpha} = \mathbb{R}_{\alpha}(\kappa_{n},{k}_{\surf})$ is the Fresnel reflection matrix of the $\alpha$-th body, $\kappa_{z} = \sqrt{ \kappa_{n}^{2} + {k}_{\surf}^{2} }$, with $\kappa_{n} = \xi_n/c=\frac{2\pi}{\beta\hbar c}n$, $\xi_n$ are the Matsubara frequencies, $\beta = (k_{B}T)^{-1}$, and ${k}_{\surf}$ is the component of the EM wavevector parallel to the surface of the plate. The prime symbol indicates that the $n=0$ term has a $1/2$ weight.

In the following we will also use two limits of \Eq{PFA_G}, and in particular the zero temperature (T=0) limit:
\begin{eqnarray}\label{PFA_G_T=0}
G_{0}(d) = \frac{\hbar c R}{\pi} \!\!\int_{0}^{\infty}\!\!\!\!\!\dd\kappa\hspace{-0.1cm}\dint_{0}^{\infty}\hspace{-0.3cm}\dd k_{\surf} \kappa_{z}k_{\surf}\Tr{ \mathbb{N}_{12}\left(\mathbbm{1} - \mathbb{N}_{12}\right)^{-1} },
\end{eqnarray}
where $\kappa_{z} = \sqrt{ \kappa^{2} + k_{\surf}^{2} }$, and the high temperature/large distance classical limit, corresponding to the $n=0$ Matsubara term in \eq{PFA_G}: 
\begin{eqnarray}\label{PFA_G_cl}
G_{\rm{cl}}(d) = k_{B}TR\dint_{0}^{\infty}\dd k_{\surf} k_{\surf}^{2}\Tr{ \mathbb{N}_{12}\left(\mathbbm{1} - \mathbb{N}_{12}\right)^{-1} }.
\end{eqnarray}

\section{Fresnel Reflection matrices}\label{sect:Fresnel}
In this section we provide the expression for the Fresnel reflection matrices $\mathbb{R}_{1}$ and $\mathbb{R}_{2}$, corresponding to an SiO$_2$ plate coated with graphene and to a gold plate, respectively. These well be then used for the calculation of the CLF gradient $G(d)$.
\subsection{SiO\texorpdfstring{$_2$}{2} plate coated with graphene}
In general, the Fresnel matrix coefficients for a planar structure can be decomposed in four blocks:
\begin{eqnarray}
\mathbb{R} & = & \left(\begin{array}{cc}
R_{\rm{ss}} & R_{\rm{sp}}\\
R_{\rm{ps}} & R_{\rm{pp}}
\end{array}\right).
\end{eqnarray}
We consider here the particular case of a structure made by an infinite half space having relative dielectric susceptibility $\epsilon_{1}(\kappa)$ (in this case $\epsilon_{1}(\kappa)$ is the SiO$_2$ dielectric function) and relative diamagnetic susceptibility $\mu_{1} = 1$, coated with graphene sheet having longitudinal and transversal conductivity given by $\sigma_{L}(\kappa,k_{\surf})$ and $\sigma_{T}(\kappa,k_{\surf})$, respectively (in this case, we consider that the Hall conductivity is zero $\sigma_{H}(\kappa,k_{\surf}) = 0$ because in our case there is no induced time reversal symmetry breaking). Here we consider the conductivity and the dielectric functions along the imaginary frequency axis $\omega = \ii\xi = \ii c\kappa$. The corresponding terms of the reflection matrix, for imaginary frequencies, are  \cite{MAntezza17,Rodriguez-Lopez2020}
\begin{eqnarray}
R_{\rm{ss}}^{\rm{Gr}} &\equiv  R_{TE}^{\rm{Gr}} = & \dfrac{ \kappa_{z} - \kappa_{1,z} - 2\kappa\bar{\sigma}_{T} }{ \kappa_{z} + \kappa_{1,z} + 2\kappa\bar{\sigma}_{T} }, \label{RTE}\\
R_{\rm{pp}}^{\rm{Gr}} &\equiv R_{TM}^{\rm{Gr}} = &  \dfrac{ \epsilon_{1}\kappa_{z} - \kappa_{1,z} + 2\frac{\kappa_{z}\kappa_{1,z}}{\kappa}\bar{\sigma}_{L} }{ \epsilon_{1}\kappa_{z} + \kappa_{1,z} + 2\frac{\kappa_{z}\kappa_{1,z}}{\kappa}\bar{\sigma}_{L} },
\label{RTM}
\end{eqnarray}
and $R_{\rm{sp}}^{\rm{Gr}} = R_{\rm{ps}}^{\rm{Gr}} =  0$. Here $\bar{\sigma}_{L/T} = \frac{2\pi}{c}\sigma_{L/T}(\kappa,k_{\surf})$, $\kappa_{z} = \sqrt{ \kappa^{2} + k_{\surf}^{2} }$ and $\kappa_{1,z} = \sqrt{ \epsilon_{1}\kappa^{2} + k_{\surf}^{2} }$. 

In terms of the polarization operator, we have \cite{PRL_Mohideen,PRB_Mohideen}
\begin{eqnarray}
R_{\rm{ss}}^{\rm{Gr}} &\equiv  R_{TE}^{\rm{Gr}} = & \dfrac{ \kappa_{z} - \kappa_{1,z} - 2\frac{\bar{\Pi}_{T}}{c} }{ \kappa_{z} + \kappa_{1,z} + 2\frac{\bar{\Pi}_{T}}{c} }, \label{RTE_PI}\\
R_{\rm{pp}}^{\rm{Gr}} &\equiv R_{TM}^{\rm{Gr}} = &  \dfrac{ \epsilon_{1}\kappa_{z} - \kappa_{1,z} + 2\frac{\kappa_{z}\kappa_{1,z}}{c\kappa^{2}}\bar{\Pi}_{L} }{ \epsilon_{1}\kappa_{z} + \kappa_{1,z} + 2\frac{\kappa_{z}\kappa_{1,z}}{c\kappa^{2}}\bar{\Pi}_{L} },
\label{RTM_PI}
\end{eqnarray}
and $R_{\rm{sp}}^{\rm{Gr}} = R_{\rm{ps}}^{\rm{Gr}} =  0$. Here $\bar{\Pi}_{L/T} = \frac{2\pi}{c}\Pi_{L/T}(\kappa,k_{\surf})$. Note the similarities between Eqs.~(\ref{RTE},\ref{RTM}) and Eqs.~(\ref{RTE_PI},\ref{RTM_PI}), algebraically we can obtain one from the other by the substitution $\bar{\sigma}_{P} = \bar{\Pi}_{P}/(c\kappa)$, but physically we are not allowed to do that, because $\bar{\sigma}_{P}$ and $\bar{\Pi}_{P}$ have different physical meaning, being the former an electric response and the later an electromagnetic response.

\subsection{Gold plate}
In the case of the gold plate, one can still use \eq{RTE}-\eq{RTM} with $\bar{\sigma}_{L/T} =0$ and $\epsilon_{2}(\kappa)$ being the gold dielectric function: 
\begin{eqnarray}
R_{\text{ss}}^{\rm{Au}} & = & \dfrac{ \kappa_{z} - \kappa_{2,z} }{ \kappa_{z} + \kappa_{2,z} }, \label{RTEg}  \\
R_{\text{pp}}^{\rm{Au}} & = & \dfrac{ \epsilon_{2}\kappa_{z} - \kappa_{2,z} }{ \epsilon_{2}\kappa_{z} + \kappa_{2,z} },
\label{RTMg}
\end{eqnarray}
 with $\kappa_{2,z} = \sqrt{ \epsilon_{2}\kappa^{2} + k_{\surf}^{2} }$. Since in the following we will discuss the different predictions coming from using a Drude or a Plasma model at low frequencies, we derive below the corresponding zero frequency limit of the Fresnel reflection matrices of gold using the two different models.
 
\subsubsection{Drude model for gold}
If we take into account the effect of losses for gold at low frequencies (as one is supposed to do), the Drude model has to be used, hence including the effect of the mean life-time of the electronic quasiparticles with the parameter $\Gamma_{\rm{Au}} = \tau^{-1}$:
\begin{eqnarray}
\epsilon_{2}(\ii\xi) = 1 + \dfrac{\omega_{P}^{2}}{\xi(\xi + \Gamma_{\rm{Au}})},\label{DrudeGoldxi}
\end{eqnarray}
 where $\omega = \ii\xi$. In this case, at low frequencies the Fresnel reflection matrix for gold becomes
\begin{eqnarray}
\dlim_{\xi\to0}\mathbb{R}^{\rm{Au}} = \left(\begin{array}{cc}
0 & 0 \\
0 & 1 
\end{array}
\right).
\end{eqnarray}
 
\subsubsection{Plasma model for gold}
On the contrary, if one assumes that losses play no role at low frequencies ($\Gamma_{\rm{Au}}=0$), the Plasma model has to be used:
\begin{eqnarray}
\epsilon_{2}(\ii\xi) = 1 + \dfrac{\omega_{P}^{2}}{\xi^{2}},
\label{PlasmaGoldxi}
\end{eqnarray}
and in this case, at low frequencies, the Fresnel reflection matrix for gold becomes:
\begin{eqnarray}\label{Fresnel_Gold_n=0}
\dlim_{\xi\to0}\mathbb{R}^{\rm{Au}} = \left(\begin{array}{cc}
\dfrac{c k_{\surf}-\sqrt{ c^{2} k_{\surf}^{2} + \omega_{P}^{2} }}{c k_{\surf} + \sqrt{ c^{2} k_{\surf}^{2} + \omega_{P}^{2} }} & 0 \\
0 & 1
\end{array}\right),
\end{eqnarray}
hence it explicitly depends on $k_{\surf}$ and $\omega_{P}$.

\section{Electromagnetic response of graphene}\label{Section_EM_response}
To calculate the Fresnel reflection matrices (\ref{RTE},\ref{RTM}) for the graphene-coated plate, one needs to use a model for the graphene conductivity. In particular, the electronic current $\mean{J_{i}(\omega,\bm{k})}$ is obtained from the application of the Kubo formalism \cite{Kubo1957} to the microscopic Ohm law
\begin{eqnarray}\label{Ohm_Law}
\mean{J_{i}(\omega,\bm{k})} = \sigma_{ij}(\omega,\bm{k})E^{j}(\omega,\bm{k}),
\end{eqnarray}
where $\sigma_{ij}(\omega,\bm{k})$ is the electronic conductivity tensor, and $E^{j}(\omega,\bm{k})$ is the total electric field.

There are several different models for the electric conductivity of graphene (hydrodynamic-based models, Kubo formalism, tight-binding prescriptions, full ab-initio, QFT descriptions...). For the electric conductivity, we focus on two models used in the framework of CLF due to their compromise between simplicity, extended use in the literature and generality for the relevant experiments: A general non-local Kubo model $\sigma^{\rm{K}}$ obtained by the direct use of the Kubo formula \cite{non-local_Graphene_Lilia_Pablo}, and its local limit $\sigma^{\rm{L}}$ (which, in the zero mass limit $\Delta=0$ is the Falkovsky model \cite{Falkovsky2007b})

On the other hand, if the total electronic current $\mean{J_{i}^{total}(\omega,\bm{k})}$ is obtained from the full electromagnetic field as
\begin{eqnarray}\label{EM_response}
\mean{J_{i}^{total}(\omega,\bm{k})} = - \Pi_{ij}(\omega,\bm{k})A^{j}(\omega,\bm{k}),
\end{eqnarray}
we calculate the Fresnel reflection matrices using (\ref{RTE_PI},\ref{RTM_PI}) for the graphene-coated plate, where $\Pi_{ij}(\omega,\bm{k})$ is the Polarization operator tensor, and $A^{j}(\omega,\bm{k})$ is the em potential vector.
We use in our comparison the Quantum Field Theory based (QFT) and lossless model for $\Pi^{\rm{QFT}}$ \cite{Bordag2015} used in \cite{PRL_Mohideen,PRB_Mohideen}. In \cite{PabloMauroComparisonKuboQFT2024} it was shown that the Local model is obtained as the local limit ($\bm{k}_{\surf}\to\bm{0}$) of $\sigma^{\rm{K}}$ and of $\Pi^{\rm{QFT}}$ non-local models, that $\Pi^{\rm{QFT}}$ could be used to obtain $\sigma^{\rm{K}}$ once we properly add losses (see \Eq{Luttinger_sub}), and another "Non Regularized" (NR) electric conductivity $\sigma^{\rm{NR}}$ as well (see \Eq{NRmodel}). As written, $\sigma^{\rm{NR}}$ leads to transversal electric currents without losses that cannot be corrected by a simple addition of electronic dissipation. However, from \Eq{EM_response}, this anomalous term can be interpreted as a magnetization instead of a part of the electric conductivity $\sigma^{\rm{NR}}$, which has to be correctly regularized to be interpreted as an electric conductivity, leading to $\sigma^{\rm{K}}$.

In \cite{non-local_Graphene_Lilia_Pablo}, it was proven that the spatial components of the conductivity tensor $\sigma_{ij}$ for 2D Dirac materials can be conveniently given by separating between longitudinal $\sigma_{L}$, transverse $\sigma_{T}$, and Hall $\sigma_{H}$, contributions \cite{Zeitlin1995,Fialkovsky2011,Dorey1992,PabloMauroComparisonKuboQFT2024}
\begin{eqnarray}\label{General_non_local_conductivity_form}
\sigma_{ij}(\omega, \bm{k}_{\surf}) & =  &\frac{k_{i}k_{j}}{k_{\surf}^{2}}\sigma_{L}(\omega, k_{\surf})
  + 
\left( \delta_{ij} - \frac{k_{i}k_{j}}{k_{\surf}^{2}} \right)\sigma_{T}(\omega, k_{\surf})\nonumber\\
& & + \epsilon_{ij}\sigma_{H}(\omega, k_{\surf}).
\end{eqnarray}
A similar separation can be obtained for $\Pi_{ij}(\omega,\bm{k}_{\surf})$ \cite{PabloMauroComparisonKuboQFT2024}. Here, $\bm{k}_{\surf} = \left(k_{1},k_{2}\right)$, $k_{\surf}^{2} = \sqrt{k_{1}^{2} + k_{2}^{2}}$, $\delta_{ij}$ is the Kronecker delta function, $\epsilon_{ij}$ is the 2D Levi-Civita symbol. The symbols for the dependence on temperature $T$,  chemical potential $\mu$, and Dirac mass $\Delta$ have been omitted for brevity. All conductivity expressions in this paper explicitly depend on those parameters.
In \cite{PabloMauroComparisonKuboQFT2024} it has been shown that the non-local Kubo model (simply called "Kubo" conductivity $\sigma_{ij}^{\rm{K}}$) and the QFT model are based on the same Polarization Operator $\Pi_{ij}(\omega,\bm{k}_{\surf})$, their differences come from the incorporation of a dissipation term that accounts for the losses in the Kubo model $\Gamma = \tau^{-1}$, while the QFT model is explicitly a dissipation-less model without any losses term, however, we expand this model to incorporate losses here. In addition to that, in the Kubo model, the Kubo formalism \cite{Kubo1957} is applied to the microscopic Ohm law (\Eq{Ohm_Law}), therefore, the conductivity is obtained from the Luttinger formula \cite{Luttinger1968,Rammer_2007} (see Eq. (81) in \cite{PabloMauroComparisonKuboQFT2024})
\begin{eqnarray}\label{Luttinger_sub}
\sigma_{ij}^{{\rm K}}(\omega,\bm{k}_{\surf},\Gamma)\! =\! \dfrac{\Pi_{ij}(\omega\!+\!\ii\Gamma,\bm{k}_{\surf}) - \dlim_{\omega\to0}\Pi_{ij}(\omega\!+\!\ii\Gamma,\bm{k}_{\surf})}{-\ii\omega}.
\end{eqnarray}

Taking the data for the mean life-time of the electronic quasiparticle as $\tau\backsim 6\times 10^{-13}\text{ s}$ \cite{DasSarmaRMP2011}\cite{Marinko2009}, we represent the effect of losses in the electronic conductivity of graphene as $\hbar\Gamma_{\rm{Gr}} = 10^{-3}\eV$. 

It is worth stressing that the subtraction of the $\dlim_{\omega\to0}\Pi_{ij}(\omega+\ii\Gamma,\bm{k}_{\surf})$ term in \eq{Luttinger_sub} is not an ad-hoc prescription introduced by hands to cures the nonphysical plasma divergence at short frequencies. It is a necessary consequence of causality and of Ohm law, and its detailed analytical re-derivation can be found in Section III of \cite{PabloMauroComparisonKuboQFT2024}. This well known Luttinger subtraction is widely derived and used in classical papers in standard textbooks \cite{Luttinger1968,Rammer_2007}.

It is worth stressing that both $\sigma$ and $\Pi$ remains complex quantities even in absence of losses \cite{Mostdispersionrelations}, i.e. when $\Gamma\to0$. In general, in the lossless case, the response function are complex due to causality, this naturally enforces the Kramers-Kronig relations. A complex response function does not implies that losses are taking into account.

The explicit analytical form of $\sigma_{ij}^{{\rm K}}$ for real and complex frequencies in the zero temperature limit can be found in \cite{non-local_Graphene_Lilia_Pablo}\cite{PabloMauroComparisonKuboQFT2024}. The extension to finite temperature can be obtained by the Maldague formula \cite{Giuliani,Maldague}
\begin{eqnarray}\label{Maldague_Formula}
\sigma_{ij}(\omega, \bm{k}_{\surf}, \Gamma, \mu, T) = \int_{-\infty}^{\infty}\dd \Mu\dfrac{\sigma_{ij}(\omega, \bm{k}_{\surf}, \Gamma,\Mu,0)}{4 k_{\rm B}T\cosh^{2}\left(\frac{\Mu-\mu}{2 k_{\rm B}T}\right)} ,
\end{eqnarray}
where $\sigma_{ij}(\omega, \bm{k}_{\surf},\Gamma,\Mu,0)$ is the zero-temperature conductivity where the chemical potential is set to $\Mu$.

\begin{figure}[H]
\centering
\includegraphics[width=\linewidth]{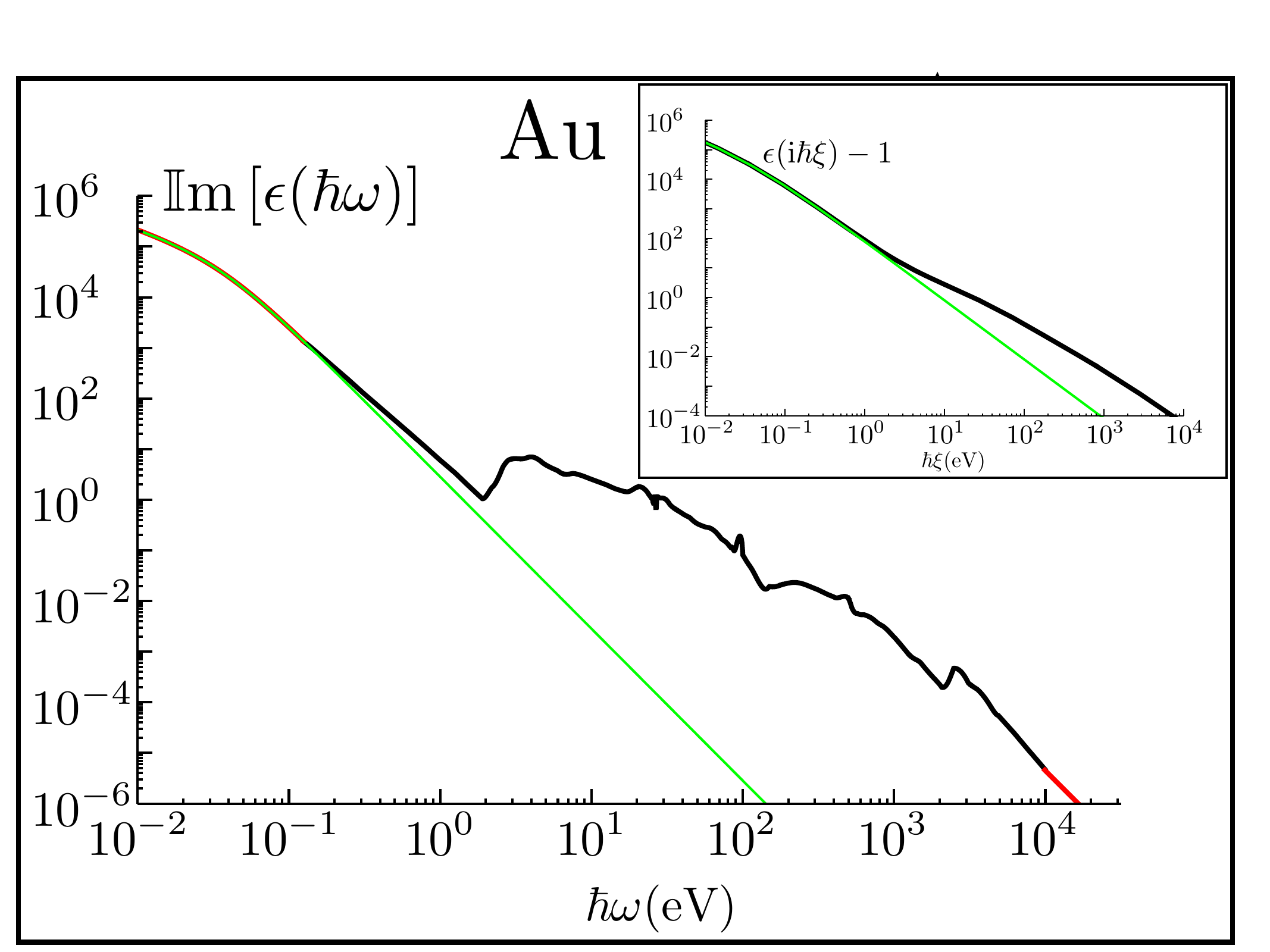}
\caption{ Log-log plot of the imaginary part of the dielectric susceptibility of gold(in black)  on real frequencies take from \cite{Palik_Handbook}. The extrapolations at low and high frequencies are plotted in red,  the Drude model for gold $\epsilon = 1 - {\omega_{P}^{2}}/{[\omega(\omega + \ii \Gamma_{\rm{Au}})]}$ is plotted in green. In the insert, the dielectric susceptibility of gold for imaginary frequencies ($\epsilon(\ii\hbar\xi) - 1$) obtained from the application of the Kramers-Krönig formula to real data following \cite{PhysRevB.77.035439} is plotted in black, while the Drude model of gold is plotted in green.}
\label{Fig_epsilonGold}
\end{figure}

In the QFT model, instead of the microscopic Ohm Law, the more general linear relationship of the electric conductivity with the vector potential is assumed $\mean{J_{i}(\omega,\bm{k})} = -\Pi_{ij}(\omega,\bm{k})A^{j}(\omega,\bm{k})$ \cite{Fialkovsky2012}, and the absence of losses is imposed. Using this result, sometimes a non-properly regularized electric conductivity, that we name "Non Regularized" conductivity $\sigma_{ij}^{\rm NR}$, is defined as \cite{Fialkovsky2012}\cite{Klimchitskaya2016} (see Eq. (56) in \cite{PabloMauroComparisonKuboQFT2024})
\begin{eqnarray}\label{NRmodel}
\sigma_{ij}^{{\rm NR}}(\omega,\bm{k}_{\surf}) = \dfrac{\Pi_{ij}^{{\rm QFT}}(\omega,\bm{k}_{\surf})}{-\ii\omega}.
\end{eqnarray}
The explicit analytical form of $\sigma_{ij}^{{\rm NR}}$ can be found in \cite{PabloMauroComparisonKuboQFT2024}\cite{Klimchitskaya2016}. As losses are neglected, this NR electric conductivity transform the Drude peaks of the electric conductivity tensor into Plasma peaks and, in addition to that, it predicts a new dissipation-less current coming from the inter-band contribution of $\sigma_{ij}^{\rm{NR}}(\omega,\bm{k}_{\surf})$ which has a Plasma behavior (\Eq{Anomalous_Plasmaterm_PiT_QFTb} of Appendix \ref{Appendix_QFTb}) \cite{PabloMauroComparisonKuboQFT2024}. We argue that this new term cannot be understood as part of the electric conductivity tensor, i.e. this current is not induced by an electric field, but by a magnetic field, therefore, we use, as in \cite{PRL_Mohideen,PRB_Mohideen}, directly the representation of the polarization tensor $\Pi_{ij}^{\rm{QFT}}(\omega,\bm{k}_{\surf})$ instead, given in  \cite{Bordag2015}\cite{PRL_Mohideen,PRB_Mohideen} and many others papers, and the reflection matrix terms (\Eqs{RTE_PI} and \eq{RTM_PI}) in terms of $\Pi_{ij}^{\rm{QFT}}(\omega,\bm{k}_{\surf}$ instead of (\Eqs{RTE} and \eq{RTM}) in terms of $\sigma_{ij}^{{\rm NR}}(\omega,\bm{k}_{\surf})$, which would give exactly the same result, but with the wrong physical interpretation coming from $\sigma_{ij}^{{\rm NR}}(\omega,\bm{k}_{\surf})$.

In addition to that, we include the effect of losses as
\begin{eqnarray}\label{QFT_response_with_dissipation}
\mean{J_{i}^{total}(\omega,\bm{k})} = -\Pi_{ij}(\omega + \ii\Gamma,\bm{k})A^{j}(\omega,\bm{k})    
\end{eqnarray}
in order to study the effect of losses in the QFT formalism, that might be included due to reasons already discussed in \cite{PabloMauroComparisonKuboQFT2024}.

\begin{figure}[H]
\centering
\includegraphics[width=1\linewidth]{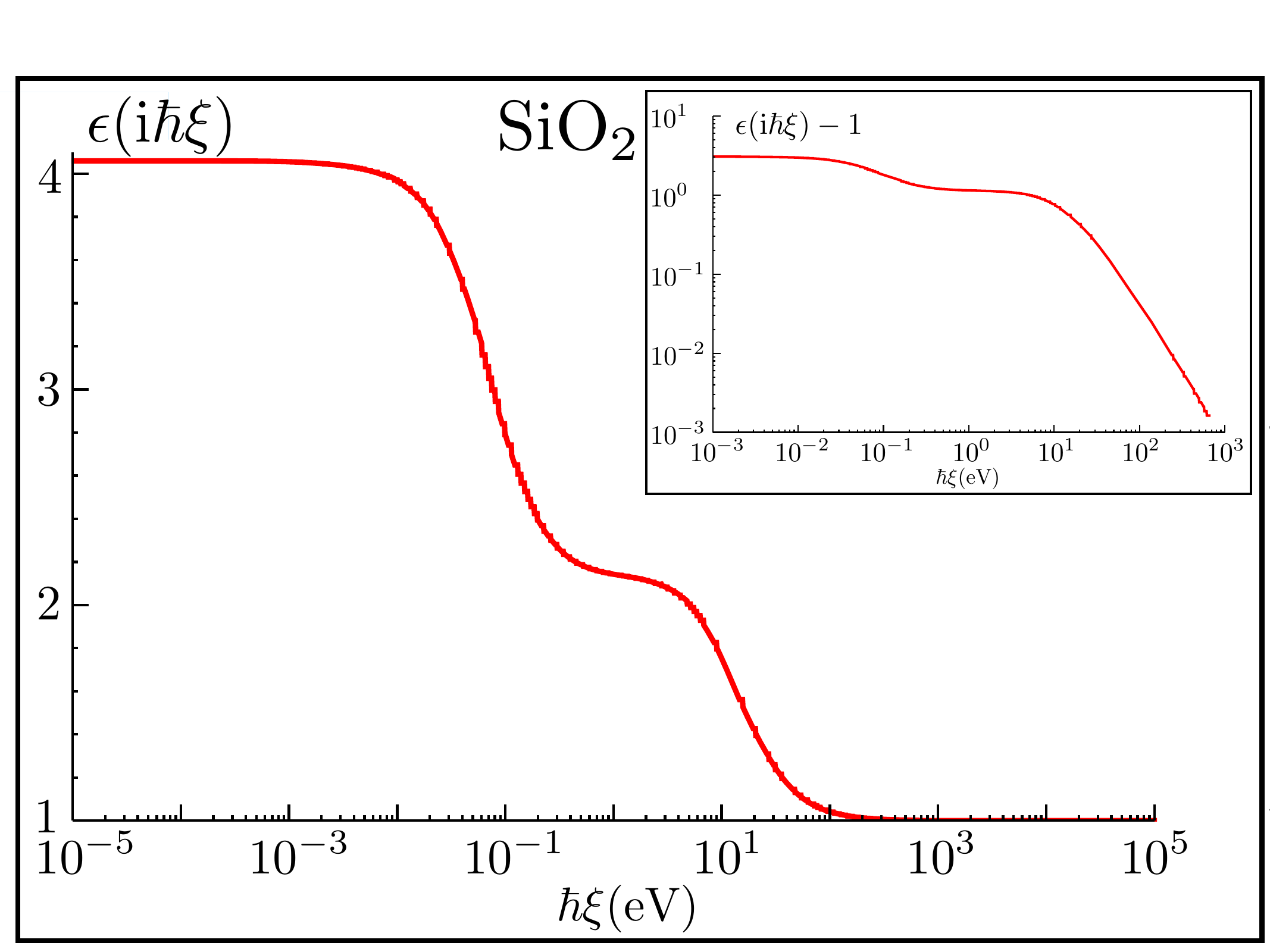}
\caption{  Log-Plot of the dielectric susceptibility of SiO$_{2}$ for imaginary frequencies, taken from \cite{Palik_Handbook}. In the insert, the dielectric susceptibility of SiO$_{2}$ ($\epsilon(\ii\hbar\xi) - 1$) for imaginary frequencies is represented in a double logarithmic plot.}
\label{Fig_epsilonSiO2}
\end{figure}

However, both Kubo and QFT models converge to the same electric conductivity in the local limit. This local limit ($\bm{k}_{\surf}\to\bm{0}$) of the conductivity for one massive Dirac cone can be analytically written in the $T\to0$ limit as (see Eq. (93) of \cite{non-local_Graphene_Lilia_Pablo})
\begin{eqnarray}\label{local_sigma_realw}
\sigma_{xx}^{\rm{L}}(\omega,\Gamma, \mu,T=0) & = & \ii\frac{\sigma_0}{\pi} 
\Bigg[ \frac{\mu^{2} - \Delta^{2}}{\abs{\mu}}\frac{1}{\Omega}\Theta\left(\abs{\mu} - \abs{\Delta}\right) \nonumber\\
&&\!\!\!\!\!\!\!\!\! \!\!\!\! \!\!\!\! \!\!\!\! \!\!\!\! \!\!\!\!   +\frac{\Delta^{2}}{M\Omega} - \frac{\Omega^{2}+4\Delta^{2}}{2\ii\Omega^{2}}   \tan^{-1}\left(\frac{\ii\Omega}{2M}\right) \Bigg], \\
\sigma_{xy}^{\rm{L}} (\omega,\Gamma, \mu,T=0) &=&  \frac{2\sigma_{0}}{\pi}\frac{\eta\Delta}{\ii\Omega}\tan^{-1}\left(\frac{\ii\Omega}{2M}\right). 
\nonumber
\end{eqnarray}
Note that $\sigma_{0}=\frac{\alpha c}{4}$ is the universal conductivity of graphene ($\alpha = \frac{e^{2}}{\hbar c}$ is the fine structure constant), $\Omega = \hbar\omega + \ii\hbar\Gamma$ and $M = \text{Max}\left[\abs{\Delta}, \abs{\mu}\right]$ and $\sigma^{\rm{L}}_{ij}(\omega,\Gamma, \mu,T=0) = \Pi^{\rm{L}}_{ij}(\omega,\Gamma, \mu,T=0)/(-\ii\omega)$. These results are per Dirac cone and they are consistent with the ones found in \cite{Ludwig1994}\cite{Gusynin2006}\cite{Falkovsky2007}\cite{Fialkovsky2008}\cite{MacDonald2006}\cite{Bordag2009}\cite{Klimchitskaya2018}\cite{PhysRevB.88.045442}\cite{WangKong2010}. The first term in $\sigma_{xx}$ corresponds to intra-band transitions, and the last two terms to inter-band transitions. Note that, in the local limit $\bm{k}_{\surf}=\bm{0}$ one obtains $\sigma_{xx}(\omega,\bm{0})=\sigma_{yy}(\omega,\bm{0})=\sigma_{L}(\omega,\bm{0})=\sigma_{T}(\omega,\bm{0})$, and $\sigma_{xy}(\omega,\bm{0})=-\sigma_{yx}(\omega,\bm{0})=\sigma_H(\omega, \bm{0})$. By using the Maldague formula \Eq{Maldague_Formula}, we can obtain the local conductivity for any temperature using the conductivity given in \Eq{local_sigma_realw}. From this result, in the $\Delta\to0$ limit, the Falkovsky model \cite{Falkovsky2007b} can be derived, as shown in \cite{PabloMauroComparisonKuboQFT2024}.

\begin{widetext}

\begin{figure}[H]
\centering
\begin{subfigure}[b]{0.49\textwidth}
   \centering
   \includegraphics[width=\textwidth]{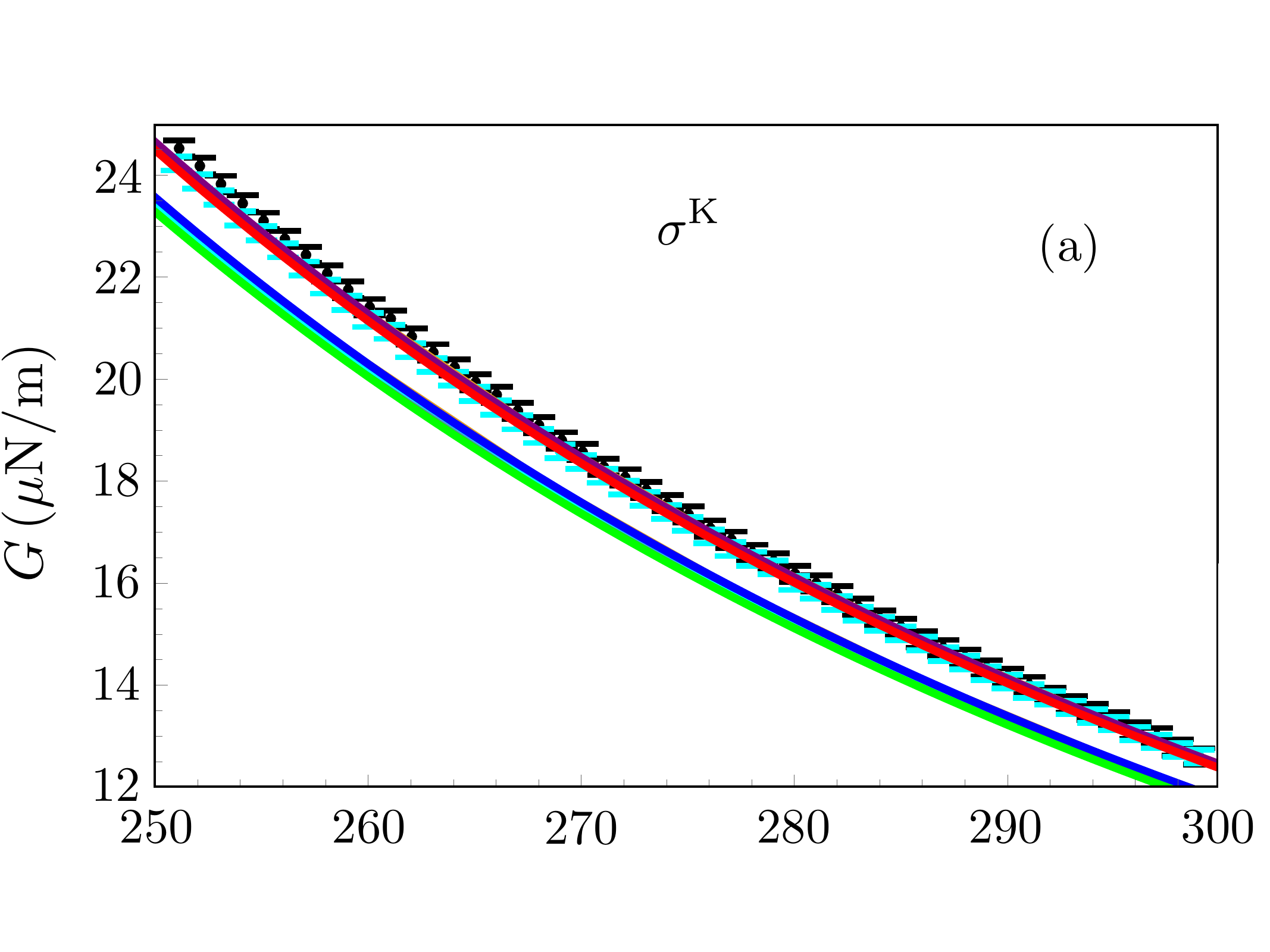}
\end{subfigure}
\hfill
\begin{subfigure}[b]{0.49\textwidth}
   \centering
   \includegraphics[width=\textwidth]{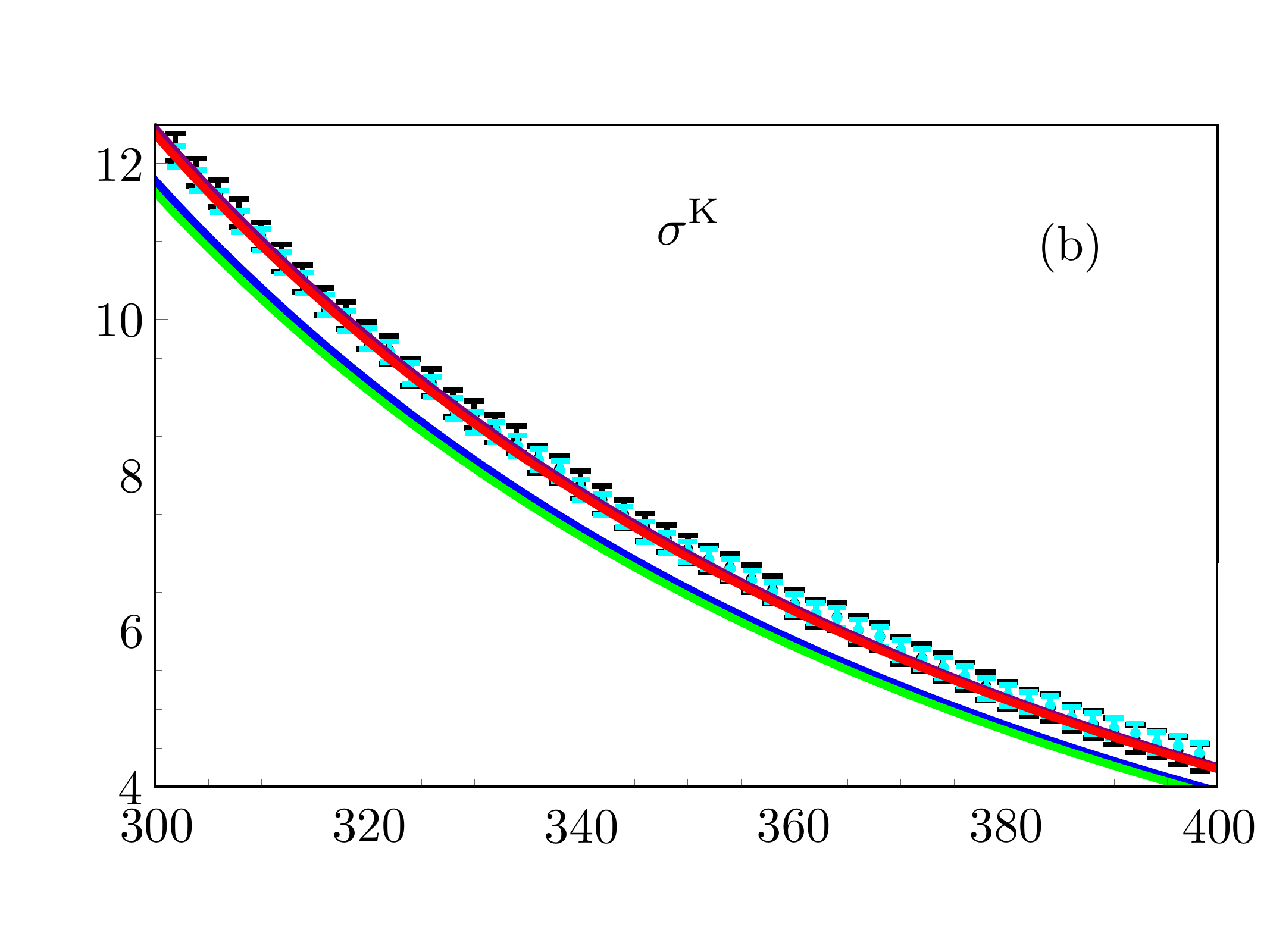}
\end{subfigure}
\hfill
\begin{subfigure}[b]{0.49\textwidth}
   \centering
   \includegraphics[width=\textwidth]{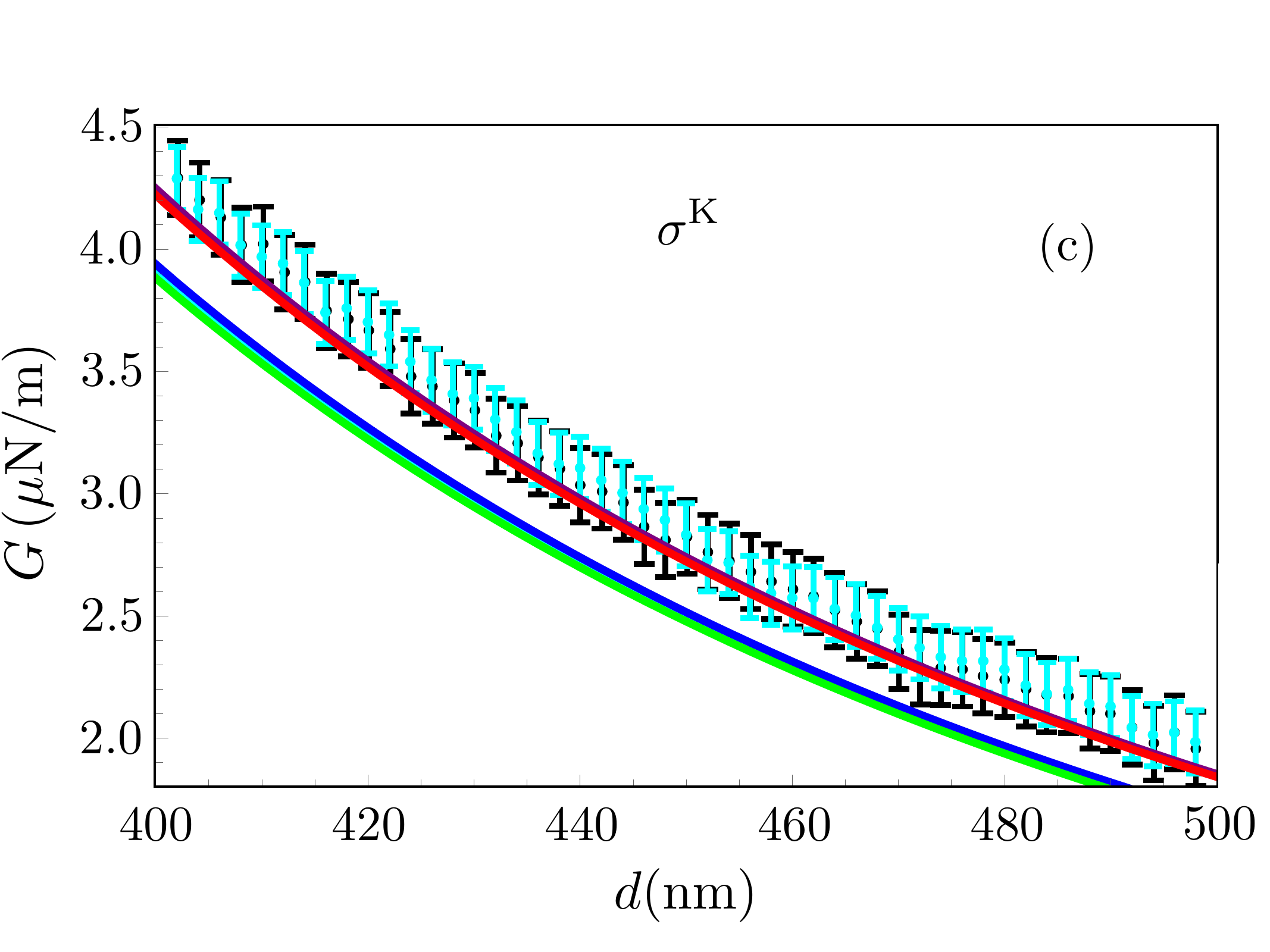}
\end{subfigure}
\hfill
\begin{subfigure}[b]{0.49\textwidth}
   \centering
   \includegraphics[width=\textwidth]{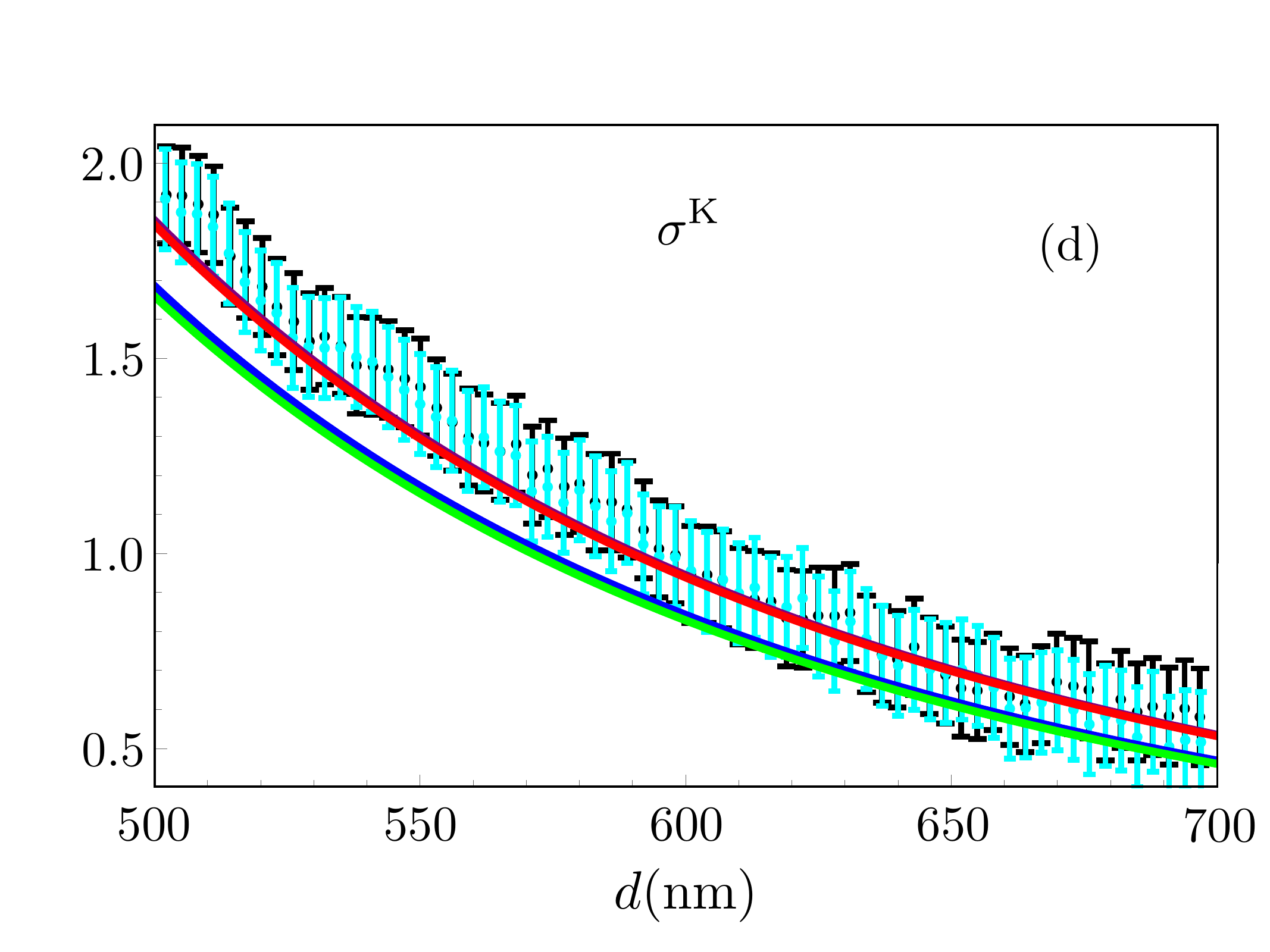}
\end{subfigure}
\caption{  Plots of the CLF gradient $G_{\rm{r}}(d)$ of \eq{Grough} and  as a function of the distance $d$. The experimental data are represented with the black (results of the PRL \cite{PRL_Mohideen}) and cyan (results of the PRB \cite{PRB_Mohideen}) points respectively, the numerical results at $T=294\KK$ are in purple ($\Delta = 0\eV$ and $\mu=0.25\eV$) and in red ($\Delta = 0.1\eV$ and $\mu=0.23\eV$), while the results at $T=0\KK$ are in blue ($\Delta = 0\eV$ and $\mu=0.25\eV$) and green ($\Delta = 0.1\eV$ and $\mu=0.23\eV$). For the numerical evaluation we used here the general non-local Kubo model $\sigma^{\rm{K}}$ for graphene \cite{non-local_Graphene_Lilia_Pablo}\cite{PabloMauroComparisonKuboQFT2024} with losses $\hbar\Gamma_{\rm{Gr}} = 10^{-3}\eV$. For the $n=0$ Matsubara term we used the Drude prescription for gold  \eq{DrudeGoldxi}, with $\hbar\omega_{P} = 9\eV$ and $\hbar\Gamma_{\rm{Au}} = 35\times 10^{-3}\eV$.}
\label{Fig_Comparison_NOLocal_Model}
\end{figure}
\end{widetext}

\section{Numerical calculation of the CLF gradient and comparison with experiments}\label{sect:experiment}
In this section, we compare the CLF gradient calculated using three different conductivity models (detailed in the previous section) with the experiment \cite{PRL_Mohideen, PRB_Mohideen}. 
The system consists of an Au-coated microsphere and a graphene sheet deposited on a silica glass (SiO$_{2}$) plate, for which we will use exactly the same system parameters used in \cite{PRL_Mohideen, PRB_Mohideen}. As done in \cite{PRL_Mohideen, PRB_Mohideen}, we will correct the CLF expressions to take into account the contribution of the roughness of the two surfaces: 
\begin{eqnarray} \label{Grough}
G_{\rm{r}}(d) = \left[ 1 + 10\dfrac{\delta_{s}^{2} + \delta_{g}^{2} }{d^{2}} \right]G(d)
\end{eqnarray}
where we will use  $\delta_{s} = 0.9\nm$ and $\delta_{g} = 1.5\text{nm}$ for the roughness of the metallic sphere and of the graphene, respectively. We note that the roughness correction in this experiment are of $\approx0.05\%$ at $d = 250\nm$ and of $\approx0.006\%$ at $d = 700\nm$.
The microsphere of diameter $2R = 120.7\pm0.1\mu\text{m}$ is made of hollow glass, and it is coated with a layer of $L = 120\pm 3\nm$ of thickness of Au.  The experiment was carried out at $T=294\pm0.5\KK$, the chemical potential of graphene was measured as $\mu = 0.24\pm0.01\text{eV}$. The SiO$_{2}$ substrate induces a non-topological mass gap to the graphene in the interval $m_{G} = (0.01-0.2) \eV$, corresponding to a Dirac mass $\Delta = m_{G}/2 = (0.005-0.1)\eV$. More experimental details are given in \cite{PRL_Mohideen, PRB_Mohideen}.

In our calculations, we use the dielectric permittivity of gold and SiO$_{2}$ tabulated in \cite{Palik_Handbook}, and applied a Kramers-Krönig transformation to obtain the results for imaginary frequencies \cite{PhysRevB.77.035439}\cite{Bimonte2010KK}, obtaining the results shown in \Fig{Fig_epsilonGold} for gold and in \Fig{Fig_epsilonSiO2} for SiO$_{2}$. For very low frequencies ($n=0$ Matsubara frequency), we have used for gold the Drude model \eq{DrudeGoldxi} with with $\hbar\omega_{P} = 9\eV$ and $\hbar\Gamma_{\rm{Au}} = 35\times 10^{-3}\eV$.

We compare the experimental results of \cite{PRL_Mohideen,PRB_Mohideen} for the CLF gradient with calculations using three different models: $\sigma^{\rm{K}}$ \cite{non-local_Graphene_Lilia_Pablo}, $\sigma^{\rm{L}}$ \cite{non-local_Graphene_Lilia_Pablo}\cite{Falkovsky2007b} and $\Pi^{\rm{QFT}}$ \cite{PRL_Mohideen}\cite{Bordag2015}. See also \cite{PabloMauroComparisonKuboQFT2024} for a comparison between the 3 different models.
Let us start discussing first the results obtained using $\sigma^{\rm{K}}$. In \Fig{Fig_Comparison_NOLocal_Model} the experimental results (extracted from the figures in \cite{PRL_Mohideen}) are plotted together with the theory using the non-local Kubo model $\sigma^{\rm{K}}$ \cite{non-local_Graphene_Lilia_Pablo}. The experimental data, with its error bars are represented as black points and bars, respectively, and they are compared with the numerical results for four different cases at $T=294\KK$ (purple ($\Delta = 0\eV$ and $\mu=0.25\eV$) and red ($\Delta = 0.1\eV$ and $\mu=0.23\eV$)) and at $T=0\KK$ (blue ($\Delta = 0\eV$ and $\mu=0.25\eV$) and green ($\Delta = 0.1\eV$ and $\mu=0.23\eV$)).

\begin{figure}[H]
\centering
\includegraphics[width=\linewidth]{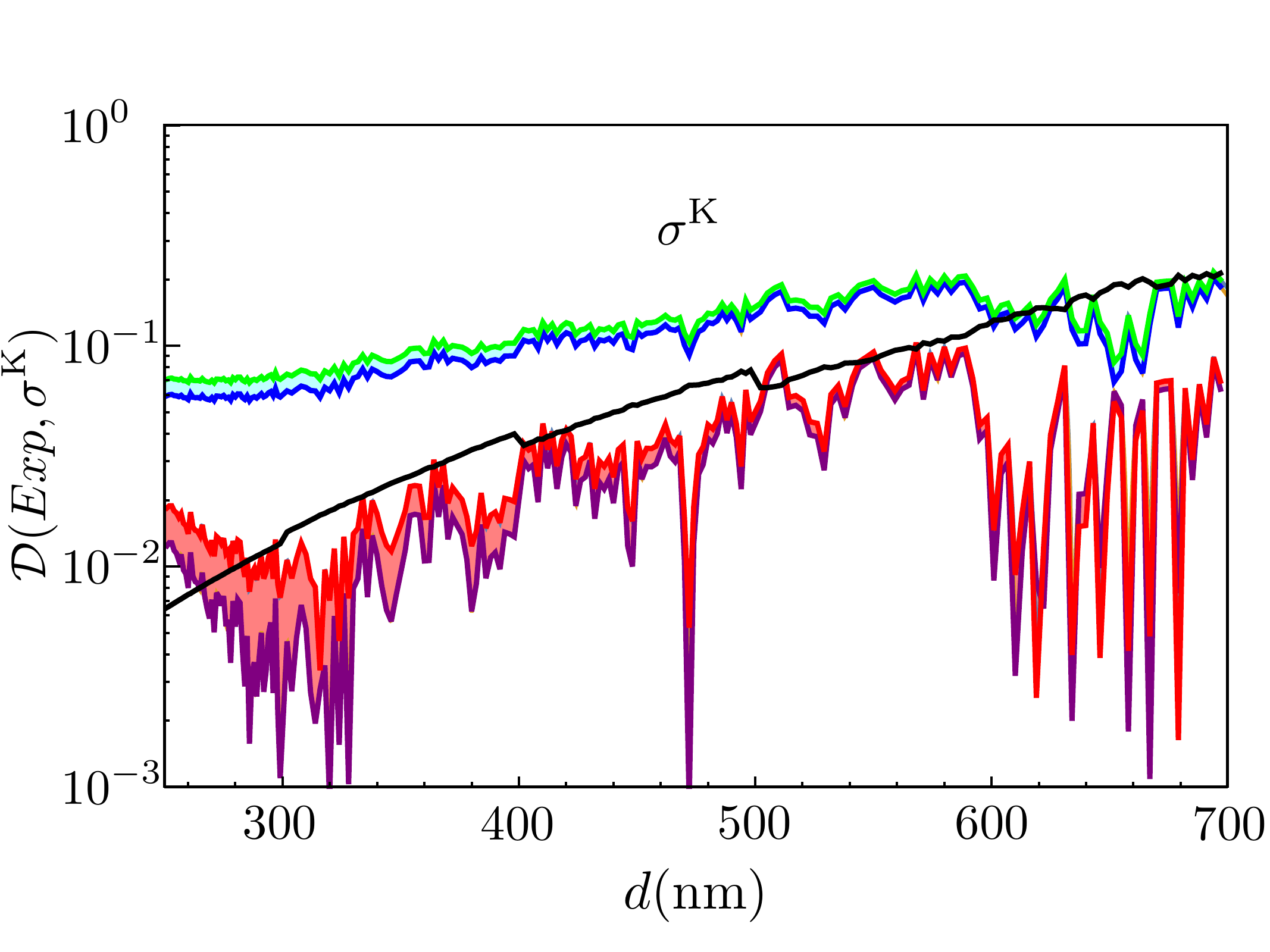}
\caption{ Logarithmic Plot of the relative difference $\mathcal{D}(Exp,\sigma^{\rm{K}})$ (\Eq{def_rel_diff})  of the theoretical predictions using the non-local Kubo model $\sigma^{\rm{K}}$ with the experimental result of \cite{PRL_Mohideen}. The numerical results at $T=294\KK$ are in purple ($\Delta = 0\eV$) and in red ($\Delta = 0.1\eV$), while the results at $T=0\KK$ are in blue ($\Delta = 0\eV$) and green ($\Delta = 0.1\eV$), the same as in \Fig{Fig_Comparison_NOLocal_Model}. The black curve is the error bar compared with the experimental value of each experimental point $\mathcal{D}(Exp,Exp\text{ lower bar})$. Theoretical values below this black curve show cases when the theoretical results are inside the experimental error-bars. We can observe a small discrepancy at short distances.}
\label{Fig_RelError_NoLocalPRL}
\end{figure}

\begin{figure}[H]
\centering
\includegraphics[width=\linewidth]{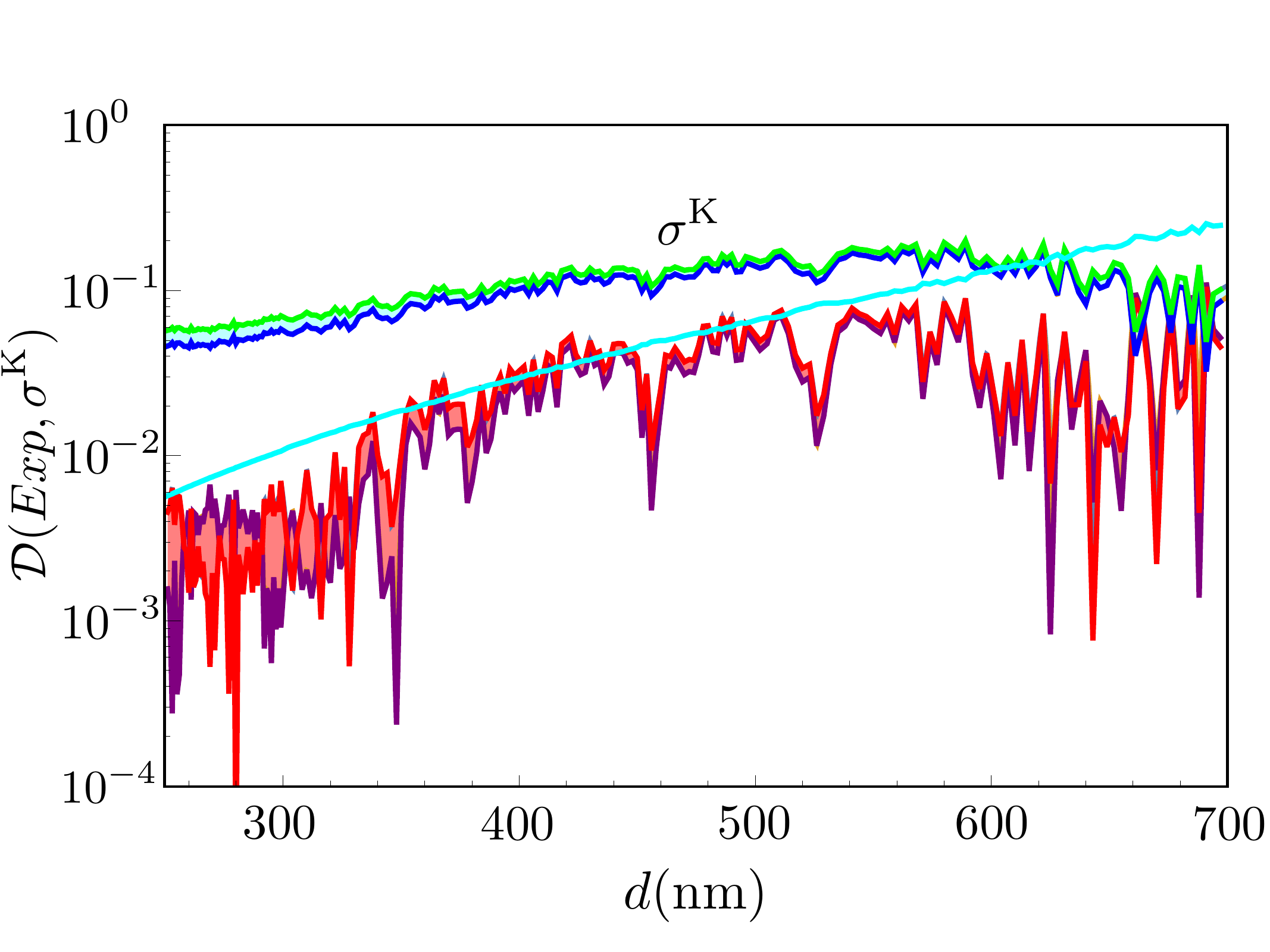}
\caption{ Logarithmic Plot of the relative difference $\mathcal{D}(Exp,\sigma^{\rm{K}})$ (\Eq{def_rel_diff})  of the theoretical predictions using the non-local Kubo model $\sigma^{\rm{K}}$ with the experimental result of \cite{PRB_Mohideen}. The numerical results at $T=294\KK$ are in purple ($\Delta = 0\eV$) and in red ($\Delta = 0.1\eV$), while the results at $T=0\KK$ are in blue ($\Delta = 0\eV$) and green ($\Delta = 0.1\eV$), the same as in \Fig{Fig_Comparison_NOLocal_Model}. The cyan curve is the error bar compared with the experimental value of each experimental point $\mathcal{D}(Exp,Exp\text{ lower bar})$. Theoretical values below this cyan curve show cases when the theoretical results are inside the experimental error-bars. We can observe that the small discrepancy at short distances that appeared in \Fig{Fig_RelError_NoLocalPRL} for the results of the PRL \cite{PRL_Mohideen} does not appear for the results of the PRB \cite{PRB_Mohideen}.}
\label{Fig_RelError_NoLocalPRB}
\end{figure}

We can see that the theory predictions at $T=294\KK$ are qualitatively in good agreement with the experimental results. To be more quantitative we represent in \Fig{Fig_RelError_NoLocalPRL} and \Fig{Fig_RelError_NoLocalPRB} the relative difference
\begin{eqnarray}\label{def_rel_diff}
\mathcal{D}(Exp,Theo) = \Abs{ \dfrac{Exp - Theo}{Exp} }
\end{eqnarray}
between the experimental result and each one of the four theory curves showed in  \Fig{Fig_Comparison_NOLocal_Model}, using the same color code. The black curve in \Fig{Fig_RelError_NoLocalPRL} is the relative difference between the lower experimental error bar and the experimental result in \Fig{Fig_Comparison_NOLocal_Model} for the results of the PRL \cite{PRL_Mohideen}, while the cyan curve in \Fig{Fig_RelError_NoLocalPRB} is the relative difference between the lower experimental error bar and the experimental result in \Fig{Fig_Comparison_NOLocal_Model} for the results of the PRB \cite{PRB_Mohideen}.

\begin{figure}[H]
\centering
\includegraphics[width=\linewidth]{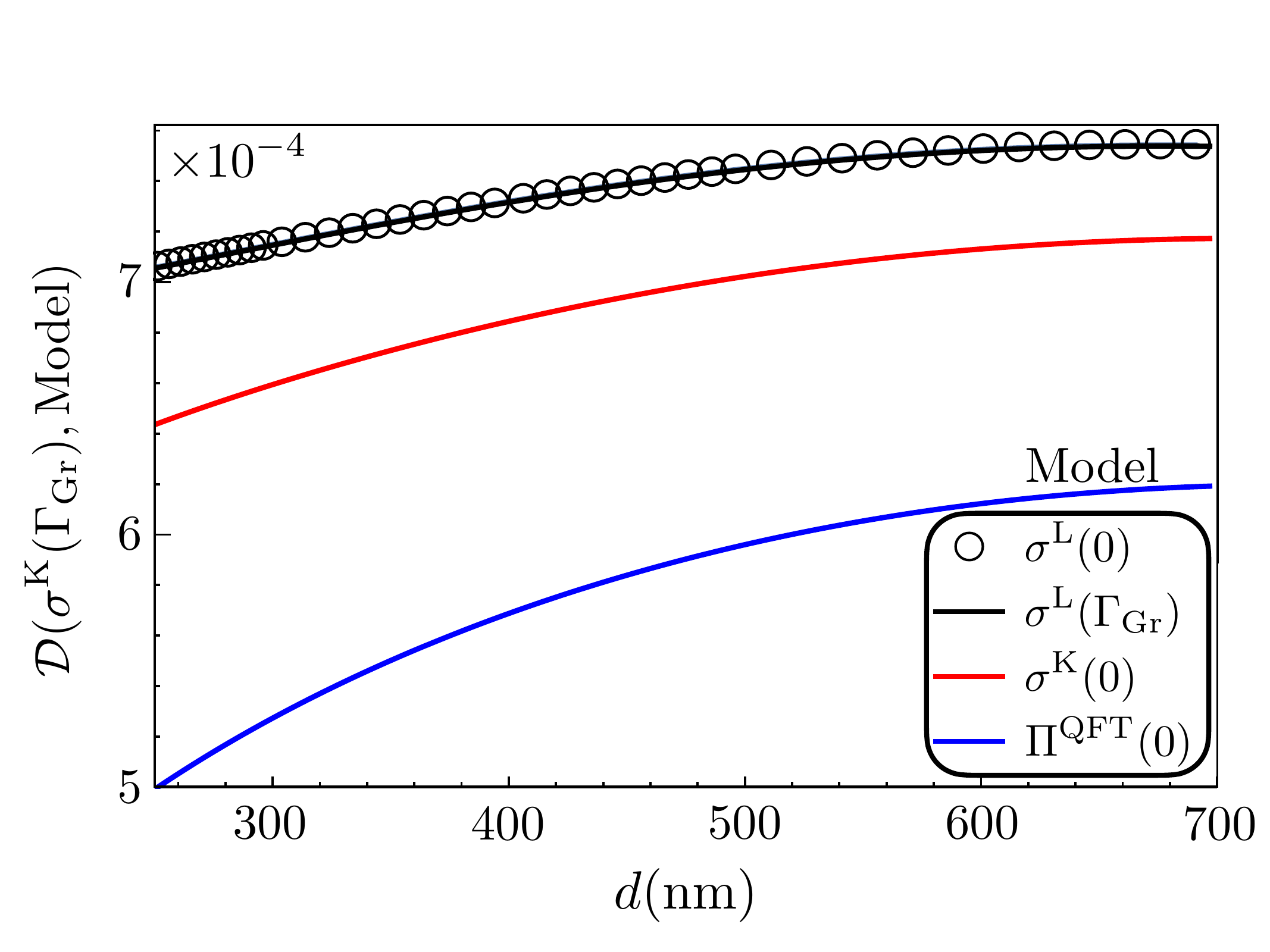}
\caption{ Relative difference $\mathcal{D}(\sigma^{\rm{K}}(\Gamma_{\rm{Gr}}),\text{Model})$ \Eq{def_rel_diff} with $\text{Model}=\{\sigma^{\rm{L}},\sigma^{\rm{K}},\Pi^{\rm{QFT}}\}$ of the CLF gradient $G(d)$ using: $\sigma^{\rm{L}}$ (with losses $\Gamma_{\rm{L}}=\Gamma_{\rm{Gr}}$, black full line, and without losses $\Gamma_{\rm{L}}= 0$, black circles), $\sigma^{\rm{K}}$ without losses $\Gamma_{\rm{K}}= 0$ (red line) and $\Pi^{\rm{QFT}}$ (blue line) with $\Gamma_{\rm{NR}} = 0$, all compared with the CLF gradient using the non-local Kubo model $\sigma^{\rm{K}}$ with losses $\Gamma_{\rm{K}} = \Gamma_{\rm{Gr}}$. Here $T=294\KK$, $\Delta = 0\eV$ and $\mu=0.25\eV$.
}
\label{Fig_RelError_NonLocality_with_NLKubo}
\end{figure}

All points below the black curve in \Fig{Fig_RelError_NoLocalPRL} and below the cyan curve in \Fig{Fig_RelError_NoLocalPRB} are consistent numerical predictions for the experimentally measured derivative of the Force experienced between the gold sphere and the graphene-covered SiO$_{2}$ plate. We can observe that the effect of temperature is distinguished by the experiment, the blue and green curves (predictions performed in the $T\to0$ limit, with \Eq{PFA_G_T=0}) do not fit the experimental results for distances $d<600\nm$, while the predictions performed taking into account the temperature of the experiment at $T=294\KK$ (red and purple curves, with \Eq{PFA_G})  fit the experimental results. We also observe that the realization of more experimental measurement in \cite{PRB_Mohideen} improves the statistical error and make the small discordance of the model with the experiment for small distances of \cite{PRL_Mohideen}, visible in \Fig{Fig_RelError_NoLocalPRL}, dissappear in \Fig{Fig_RelError_NoLocalPRB}.

Let us now analyze and compare, in \Fig{Fig_RelError_NonLocality_with_NLKubo}, the CLF gradient theory prediction using the other conductivity models. We keep the same experimental conditions for the three cases ($T=294\KK$, and graphene with $\Delta = 0\eV$ and $\mu=0.25\eV$). For gold, we use the Drude prescription here, the losses of gold are set to $\hbar\Gamma_{\rm{Au}} = 35\times 10^{-3}\eV$, and the losses of graphene are set to $\hbar\Gamma = 10^{-3}\eV$. The results obtained with the Non-Local Kubo conductivity $\sigma^{\rm{K}}$ without losses, Local conductivity $\sigma^{\rm{L}}$ (with and without losses) and QFT polarization operator $\Pi^{\rm{QFT}}$ models are practically identical, having a relative difference smaller than $10^{-3}$ with respect to predictions using $\sigma^{\rm{K}}$ with losses $\hbar\Gamma = 10^{-3}\eV$.
We can conclude that the experiment is performed in a region of parameters where the local model $\sigma^{\rm{L}}$ gives results very close to the one obtained with the non-local Kubo $\sigma^{\rm{K}}$ and to the one obtained with the full QFT polarization operator $\Pi^{\rm{QFT}}$, hence the experiment only test the graphene electric conductivity in the local regime.

Due to the equivalence of the three models in this given (local) experimental regime, the corresponding figures on the theory-experiment comparison 
obtained using the $\sigma^{\rm{L}}$ and $\Pi^{\rm{QFT}}$ models will be strictly indistinguishable from \Fig{Fig_Comparison_NOLocal_Model} and \Fig{Fig_RelError_NoLocalPRL}, as it is explicitly shown in the Appendices \ref{Appendix_Comparison_Local} and \ref{Appendix_Comparison_NR}.

\subsection{Effect of losses}
\begin{figure}[H]
\centering
\includegraphics[width=\linewidth]{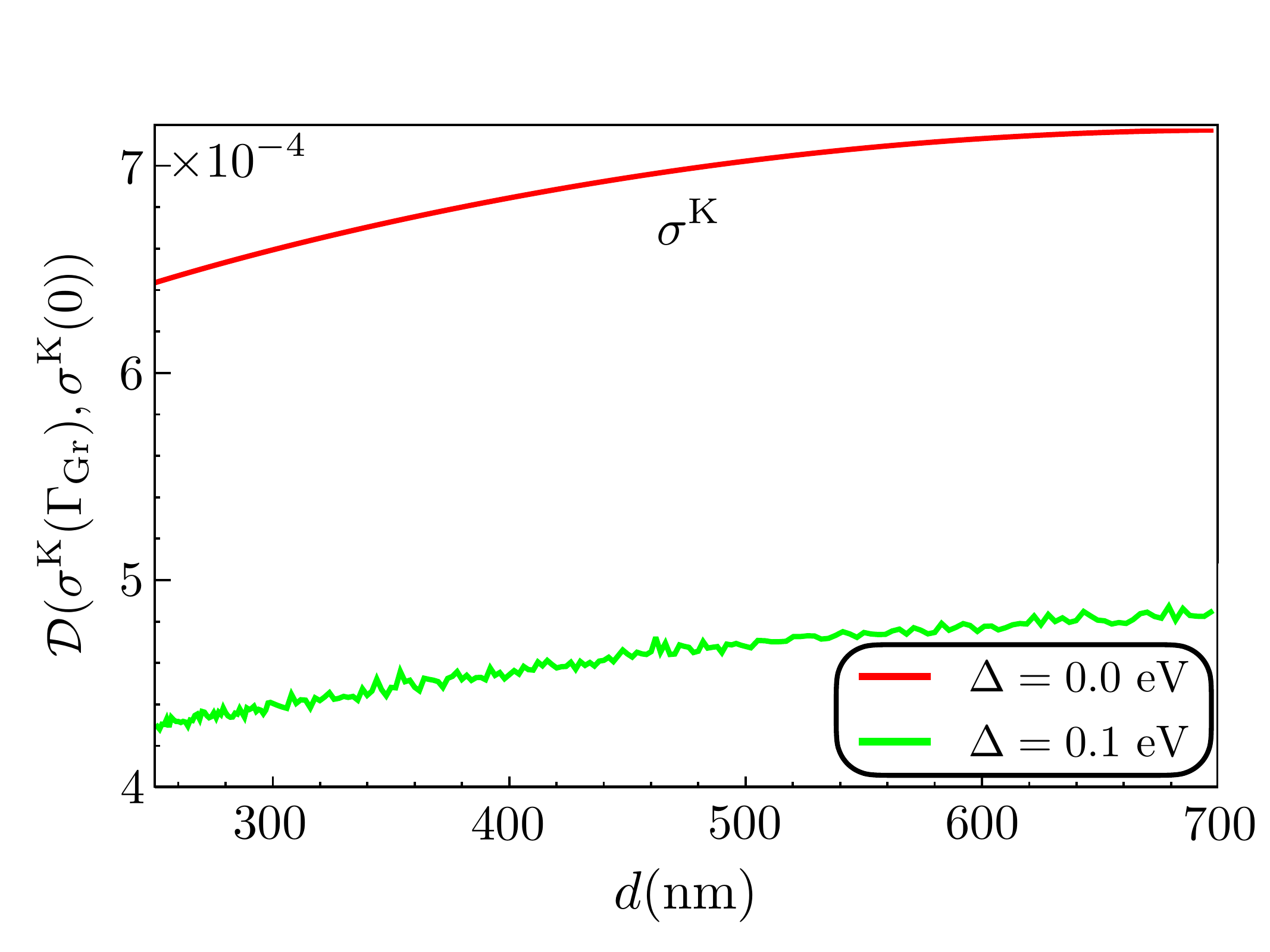}
\caption{Relative difference $\mathcal{D}(\sigma^{\rm{K}}(\Gamma_{\rm{Gr}}),\sigma^{\rm{K}}(0))$ \Eq{def_rel_diff} of $\sigma^{\rm{K}}$ without losses ($\Gamma = 0$ for graphene) compared with the case with losses ($\Gamma = \Gamma_{\rm{Gr}}$ for graphene). The red curve is the case with $T=294\KK$, $\Delta = 0\eV$ and $\mu=0.25\eV$, while the green curve is the case with $T=294\KK$, $\Delta = 0.1\eV$ and $\mu=0.23\eV$. For the $n=0$ Matsubara term we have used the Drude prescription for gold, with $\hbar\Gamma_{\rm{Au}} = 35\times 10^{-3}\eV$.}
\label{Fig_RelError_Losses_NLKubo}
\end{figure}

In this subsection we discuss the effect that losses in graphene \cite{PabloMauroComparisonKuboQFT2024} have in the experiment \cite{PRL_Mohideen,PRB_Mohideen} keeping the losses of gold as $\hbar\Gamma_{\rm{Au}} = 35\times 10^{-3}\eV$. Losses in graphene are a small non-zero quantity, and we assume that their effect on the electric conductivity can be well approached with the finite lifetime approximation by a constant imaginary dissipation rate $\Gamma = \tau^{-1}$ \cite{DasSarmaRMP2011}\cite{Adam2007}\cite{BOOKAbrikosov}\cite{Szunyogh1999}\cite{yanagisawa2004}, where $\tau\backsim 6\times 10^{-13}\text{ s}$ is estimated in \cite{DasSarmaRMP2011}\cite{Marinko2009}. The model we have that can handle losses is the Kubo model \cite{non-local_Graphene_Lilia_Pablo}, either in the local limit given in \Eq{local_sigma_realw} or the general non-local case given in \Eq{static_limit_sigma}. Then we compare the experimental results with a dissipation time of $\tau\backsim 6\times 10^{-13}\text{ s}$ with the zero-dissipation result $\tau \to\infty$. The results are shown in \Fig{Fig_RelError_Losses_NLKubo}, where we observe that, for the conditions of the experiment and considered distances, the relative error is lower than $10^{-3}$, and therefore, the effects of losses of graphene cannot be observed in the experiment. 

\section{Drude vs. Plasma}\label{sect:Drude_vs_Plasma}
In this section we compare the results by using the Drude and Plasma prescriptions to calculate the Casimir effect. In experiments on the CLF, it has been shown that the use of a Plasma prescription in the Casimir-Lifshitz theory (use of the zero losses limit of the metallic objects of the systems in the zero frequency term) provides a better theory experimental agreement than with a Drude model (i.e. use of non-zero losses for metals in the zero frequency term) \cite{PhysRevB.93.184434}\cite{universe7040093}\cite{Klimchitskaya_2022}. This contrasts with the natural choice of the Drude model for normal metals, where in the limit of zero frequency, losses are present if a DC electric field is applied, and some experiments fit better with this prescription \cite{Sushkov_2011}.

As the use of a Plasma or Drude model could in principle modify the theoretical prediction of the results of the experiment, we apply a detailed study here of the classical limit ($n=0$ Matsubara term) of the CLF gradient $G_{\rm{cl}}$ \Eq{PFA_G_cl}, which is the term affected by the choice between the two prescriptions.

\begin{figure}[H]
\centering
\includegraphics[width=\linewidth]{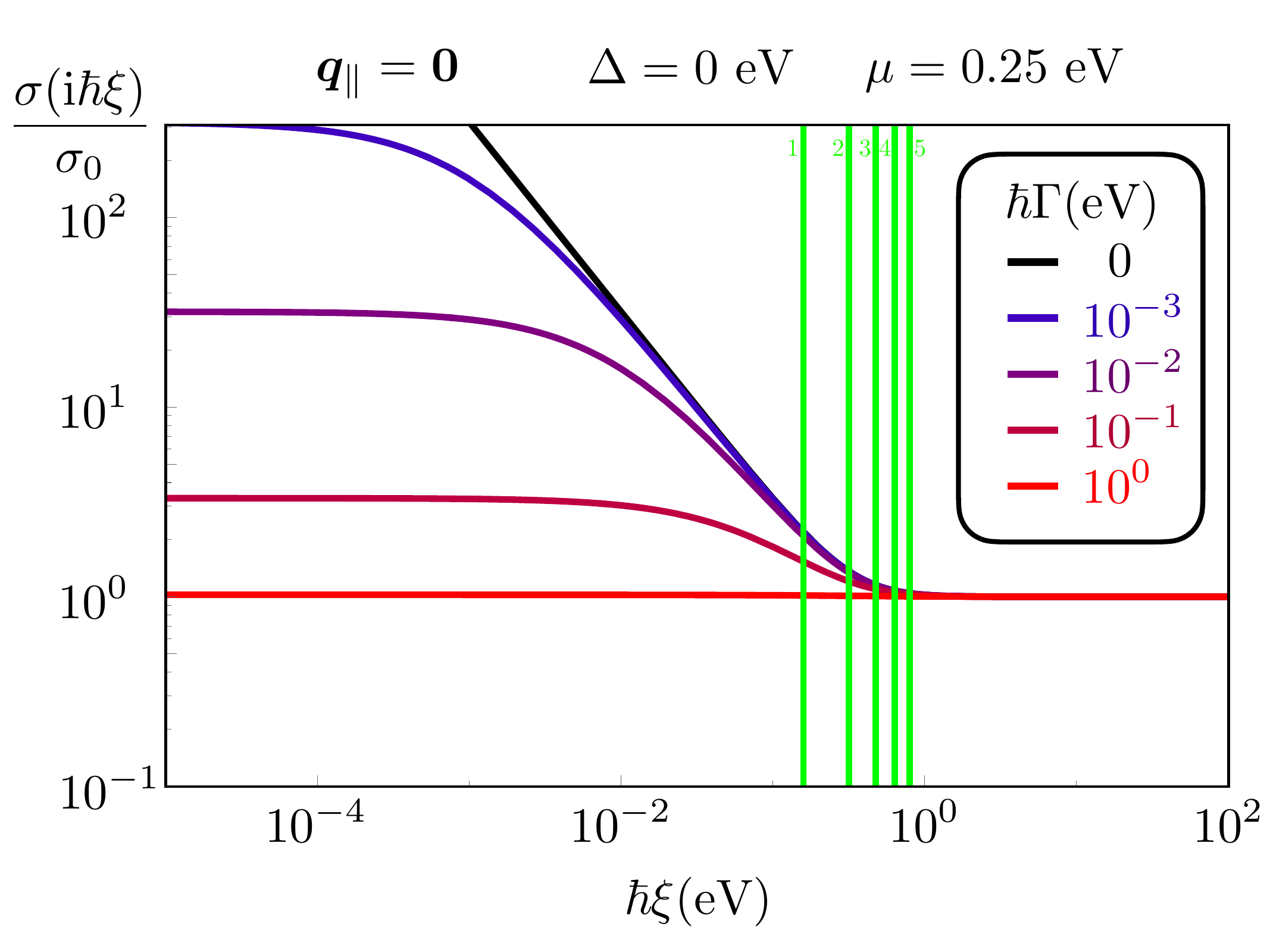}
\caption{ (Color online) Double-logarithmic plot of the electric conductivity of graphene $\sigma_{xx}(\ii\hbar\xi,\bm{q}_{\surf}=\bm{0},\Delta = 0,\mu=0.25\text{ eV},\Gamma)$ divided by the universal conductivity of graphene $\sigma_{0} = \frac{\alpha c}{4}$ for different losses $\Gamma$. The lossless case is the black curve, where we observe the divergence of the electric conductivity at $\omega=0$, and the blue, purple, red and orange represent the cases with $\hbar\Gamma$ equal to $10^{-3}\text{ eV}$, $10^{-2}\text{ eV}$, $10^{-1}\text{ eV}$ and $10^{0}\text{ eV}$ respectively. The green lines represent the five first non-zero Matsubara frequencies $\hbar\kappa_{n} = \frac{2\pi}{\beta c}n$ from $n=1$ to $n=5$ at the temperature of the experiment.}
\label{Fig_Effect_of_Losses_in_sigma_xx_Local}
\end{figure}

It is important to note that, for the electric conductivity at imaginary frequencies, the losses modify the electric conductivity more for smaller frequencies and, up to a certain threshold, the inclusion of losses do not modify at all the electric conductivity. For this reason, it is expected that only the electric conductivity at zero Matsubara frequency and, maybe, at first Matsubara frequency is noticeably modified by the inclusion of losses. In \Fig{Fig_Effect_of_Losses_in_sigma_xx_Local} we represent $\sigma_{xx}(\ii\hbar\xi,\bm{q}_{\surf}=\bm{0},\Delta = 0,\mu=0.25\text{ eV},\Gamma)$ for graphene for different losses. The lossless case is the black curve, where we observe the divergence of the electric conductivity at $\omega=0$, and the blue, purple, red and orange represent the cases with $\hbar\Gamma$ equal to $10^{-3}\text{ eV}$, $10^{-2}\text{ eV}$, $10^{-1}\text{ eV}$ and $10^{0}\text{ eV}$ respectively. Note that, for $\Gamma\to\infty$, we have $\dlim_{\Gamma\to\infty}\sigma_{xx}(\ii\hbar\xi,\bm{q}_{\surf}=\bm{0},\Delta = 0,\mu=0.25\text{ eV},\Gamma) = \frac{\alpha c}{4}$. The green lines represent the five first non-zero Matsubara frequencies. Then, it is clear that only the two or three first Matsubara frequencies (including the $n=0$ case) would be modified by the variation of losses and could modify the Casimir effect at finite temperature, but in practice we have observed that, numerically, we simply cannot distinguish the integrand for any non-zero Matsubara frequency with and without losses for the parameters of the experiment. Therefore, in what follows, we only are going to consider the effect of losses in the $n=0$ Matsubara term. As a conclusion, we could use a model to estimate the losses of graphene, like the one in \cite{PhysRevB.80.245435} where, assuming an electronic movility of $\mu = 10^{4}\text{ cm}^{2}/(\text{V s})$ we have $\tau^{-1} = \hbar\mu\sqrt{\pi n}/(e v_{F}) = 2.39\times 10^{-13}\text{ s}$ and $\hbar\Gamma_{\rm{Gr}} = \hbar\tau^{-1}= 2.75\times 10^{-3}\eV$, or we could take the data from the mean life-time of the electronic quasiparticle of experiments \cite{DasSarmaRMP2011}\cite{Marinko2009} as $\hbar\Gamma_{\rm{Gr}} = 10^{-3}\eV$, the numerical results for $G(d)$ are the same.

In the following we will need to calculate the $\xi\rightarrow0$ of graphene conductivity, and we can have typically three behaviors:
\begin{eqnarray}
\bar{\sigma}_{P}(\xi,\bm{k}_{\surf}) & \underset{\xi\to 0}{\approx} & \dfrac{\bar{\sigma}_{P,-1}(\bm{k}_{\surf})}{\xi} + \mathcal{O}\left[\xi^{0}\right]\label{def_sigma_m1}\\
\bar{\sigma}_{P}(\xi,\bm{k}_{\surf}) & \underset{\xi\to 0}{\approx} & \bar{\sigma}_{P,0}(\bm{k}_{\surf}) + \mathcal{O}\left[\xi^{1}\right]\label{def_sigma_0}\\
\bar{\sigma}_{P}(\xi,\bm{k}_{\surf}) & \underset{\xi\to 0}{\approx} & \bar{\sigma}_{P,1}(\bm{k}_{\surf})\xi + \mathcal{O}\left[\xi^{2}\right]\label{def_sigma_p1}
\end{eqnarray}
defining the constant $\bar{\sigma}_{P,0}$, $\bar{\sigma}_{P,-1}$ and $\bar{\sigma}_{P,1}$, where $P=\{\rm{L,T}\}$ are the two polarizations. We will use an equivalent definition for the Taylor series terms of $\bar{\Pi}_{P}(\xi,\bm{k}_{\surf})$.

\subsection{Local model}
\subsubsection{Drude metal model}
When losses are considered in the local model for the conductivity of graphene (\Eq{local_sigma_realw}), one can show that the $\xi\to0$ behavior of $\sigma_{P}(\xi)$ is given by \Eq{def_sigma_0} as
\begin{eqnarray}
\bar{\sigma}_{P,0}^{\rm{L}} & = & \frac{\alpha}{2} 
\Bigg[ \frac{\mu^{2} - \Delta^{2}}{\abs{\mu}}\frac{1}{\hbar\Gamma}\Theta\left(\abs{\mu} - \abs{\Delta}\right) \nonumber\\
& & + \frac{\Delta^{2}}{M\hbar\Gamma} + \frac{ 4\Delta^{2} - \hbar^{2}\Gamma^{2}}{2\hbar^{2}\Gamma^{2}} \tan^{-1}\left(\frac{\hbar\Gamma}{2M}\right) \Bigg].
\end{eqnarray}
for any polarization $P=\{\rm{L,T}\}$ at $T=0$. To obtain the result for any finite temperature we should use the Maldague formula \Eq{Maldague_Formula}. Therefore, the Fresnel reflection matrix for SiO$_{2}$ covered by graphene $\mathbb{R}^{\rm{Gr}}$ and for gold $\mathbb{R}^{\rm{Au}}$ with losses are equal to
\begin{eqnarray}\label{n=0_Drude_Fresnel_matrix}
\dlim_{\xi\to0}\mathbb{R}^{\rm{Gr}} = \dlim_{\xi\to0}\mathbb{R}^{\rm{Au}} = \left(\begin{array}{cc}
0 & 0 \\
0 & 1
\end{array}\right).
\end{eqnarray}
In this case, the $n=0$ Matsubara CLF gradient is exactly
\begin{eqnarray}\label{PFA_G_cl_Drude_Local}
G_{\rm{cl}}(d)
= G_{\rm{cl},\rm{pp}}
& = & k_{B}TR\dint_{0}^{\infty}\dd k_{\surf} k_{\surf}^{2}\dfrac{e^{-2dk_{\surf}}}{1 - e^{-2dk_{\surf}}}\nonumber\\
& = & - \frac{k_{B}TR \zeta(3)}{4 d^{3}},
\end{eqnarray}
which is the Drude result for the thermal CLF gradient.

\subsubsection{Plasma model}
\begin{figure}[H]
\centering
\includegraphics[width=\linewidth]{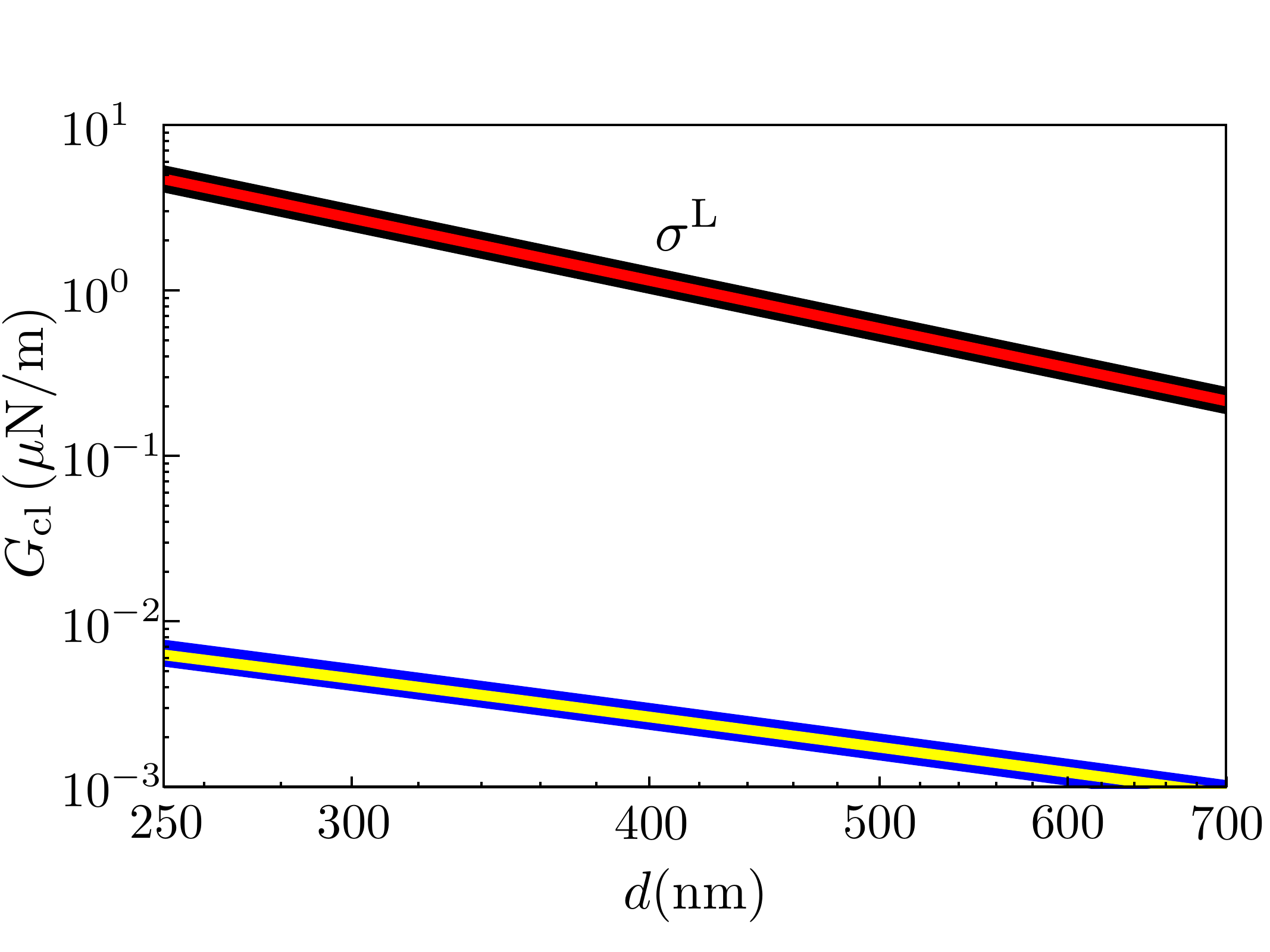}
\caption{ Double logarithmic plot of the $n=0$ Matsubara contribution of the CLF gradient $G_{\rm{cl}}$ calculated with the local model $\sigma^{\rm{L}}$ as a function of the distance $d$. The Black thick curve and the red curve are the results of the Plasma (\Eq{PFA_G_cl_Local}, using Eqs.~\eqref{PFA_G_cl_Localpp} and \eqref{PFA_G_cl_Localss_app}) and Drude (\Eq{PFA_G_cl_Drude_Local}) prescriptions, respectively. The blue curve is the exact result for the $\rm{ss}$ component of the Plasma prescription defined in \Eq{PFA_G_cl_Localss}, and the yellow curve is the approximation of \Eq{PFA_G_cl_Localss_app} for intermediate distances, when $c/\omega_{P} < d < d_{0}$.}
\label{Fig_LLP_Drude_Plasma_Local}
\end{figure}

If we artificially neglect losses in the local model for the conductivity of graphene \Eq{local_sigma_realw}, we obtain a Plasma model
\begin{eqnarray}
\bar{\sigma}_{P}^{\rm{L}}(\xi) \underset{\xi\to0}{\approx} \dfrac{\bar{\sigma}_{P,-1}^{\rm{L}}}{\xi}
\end{eqnarray}
with
\begin{eqnarray}
\bar{\sigma}_{P,-1}^{\rm{L}} = \dfrac{\alpha c}{4\pi\hbar}\dfrac{\mu^{2} - \Delta^{2}}{\abs{\mu}}\Theta\left(\abs{\mu} - \abs{\Delta}\right)
\end{eqnarray}
for any polarization $P=\{\rm{L,T}\}$ at $T=0$. To obtain the result for any finite temperature we should use the Maldague formula \Eq{Maldague_Formula}. Therefore, the Fresnel reflection matrix tends to
\begin{eqnarray}\label{Rmatrix_localcase_w0}
\dlim_{\xi\to0}\mathbb{R}^{\rm{Gr}} = \left(\begin{array}{cc}
\dfrac{ - \bar{\sigma}_{T,-1}^{\rm{L}}}{c k_{\surf} + \bar{\sigma}_{T,-1}^{\rm{L}} } & 0 \\
0 & 1
\end{array}\right)
\end{eqnarray}
In the Plasma prescription, the zero frequency limit of the Fresnel matrix for gold $\mathbb{R}^{\rm{Au}}$ is given by \Eq{Fresnel_Gold_n=0}. In this case, the CLF derivative is the sum of the contribution of the 2 unmixed polarizations
\begin{eqnarray}\label{PFA_G_cl_Local}
\!\!\!\!\!\!\!\!\!\!\!\!\!\!\!\!\!\!\!\!G_{\rm{cl}}(d) &=&  G_{\rm{cl},\rm{ss}} + G_{\rm{cl},\rm{pp}} \nonumber \\
&&\!\!\!\!\!\!\!\!\!\!\!\!\!\!\!=k_{B}TR\dint_{0}^{\infty}\dd k_{\surf} k_{\surf}^{2}\Tr{\dfrac{\mathbb{R}^{\rm{Au}}\mathbb{R}^{\rm{Gr}}e^{-2dk_{\surf}}}{ \mathbbm{1} - \mathbb{R}^{\rm{Au}}\mathbb{R}^{\rm{Gr}}e^{-2dk_{\surf}}}},
\end{eqnarray}
where the $\rm{pp}$ contribution equals to the Drude result given in \Eq{PFA_G_cl_Drude_Local}
\begin{eqnarray}\label{PFA_G_cl_Localpp}
G_{\rm{cl},\rm{pp}}(d) = - \frac{k_{B}TR \zeta(3)}{4 d^{3}}.
\end{eqnarray}
For the $\rm{ss}$ polarization, we have
\begin{eqnarray}\label{PFA_G_cl_Localss}
\!\!\!G_{\rm{cl},\rm{ss}}(d)
= k_{B}TR\!\!\dint_{0}^{\infty}\hspace{-0.25cm}\dd k_{\surf} k_{\surf}^{2}\Tr{\dfrac{R_{\rm{ss}}^{\rm{Au}}R_{\rm{ss}}^{\rm{Gr}}e^{-2dk_{\surf}}}{1 - R_{\rm{ss}}^{\rm{Au}}R_{\rm{ss}}^{\rm{Gr}}e^{-2dk_{\surf}}}}\!\!,
\end{eqnarray}
for which one we can distinguishes three regions:
\begin{eqnarray}
G_{\rm{cl},\rm{ss}}(d)  & \underset{d > d_{0}}{\approx} & - \frac{k_{B}TR \zeta(3)}{4 d^{3}},\\
G_{\rm{cl},\rm{ss}}(d) & \underset{d\ll\frac{c}{\omega_{P}}}{\approx} & -\frac{k_{B}TR\omega_{P}^{2}}{2 c^{2}d_{0}} \log\left(\frac{\sqrt{2}e^{\gamma}\omega_{P}d}{c}\right)\!\!,\\
G_{\rm{cl},\rm{ss}}(d) & \underset{\frac{c}{\omega_{P}}< d < d_{0}}{\approx} & - \frac{k_{B}TR c^{2}}{8\pi d_{0}d^{2}}\left( 1 - \dfrac{2c}{d\omega_{P}} \right),\label{PFA_G_cl_Localss_app}
\end{eqnarray}
where
\begin{eqnarray}\label{Def_d_0}
d_{0} = \dfrac{2\hbar c}{\alpha}\dfrac{\abs{\mu}}{\mu^{2} - \Delta^{2}} = 2.57\times 10^{-4}\text{ m}.
\end{eqnarray}
For gold, we have $c/\omega_{P}\approx 22\nm$, therefore, for the considered experiment, the $c/\omega_{P} < d < d_{0}$ is the relevant limit and the $\rm{ss}$ polarization has a negligible contribution that, in principle, will be very difficult to be measured, as can be observed in \Fig{Fig_LLP_Drude_Plasma_Local}. We conclude that, for the considered distances, Drude and Plasma prescriptions practically coincide when the Local model $\sigma^{\rm{L}}$ is used. To obtain an experimental signal that distinguishes between Drude and Plasma models in this experiment, we need to maximize the contribution of $G_{\rm{cl},\rm{ss}}(d)$ so that it becomes comparable to $G_{\rm{cl},\rm{pp}}(d)$ given in \Eq{PFA_G_cl_Localpp} by reducing $d_{0}$ given in \Eq{Def_d_0} as much as possible, reducing $\Delta$ and increasing $\mu$ as much as possible. However, even for $\Delta = 0\eV$ and unreasonable high chemical potential $\mu=1\eV$, we get $d_{0} = 5.4\times 10^{-5}\text{ m}$, far beyond the maximum experimental distance considered in the experiment.

\subsection{Non-Local Kubo model}
\subsubsection{Drude metal model}
When losses $\hbar\Gamma>0$ are considered in the non-local Kubo model, one can find that the $\xi\to0$ limit of $\bar{\sigma}_{P}^{\rm{K}}(\xi)$ tends to $\bar{\sigma}_{P,0}^{\rm{K}}$ of \Eq{def_sigma_0} given by \Eq{static_limit_sigma}. It is easy to check that $0 < \bar{\sigma}_{P,0}^{\rm{K}} < \infty$ for any polarization $P=\{\rm{L,T}\}$ and temperature $T$. Therefore, the Fresnel reflection matrix for SiO$_{2}$ covered by graphene and for gold are given by \Eq{n=0_Drude_Fresnel_matrix}. As a result, the zero Matsubara CLF gradient is the usual Drude result given in \Eq{PFA_G_cl_Drude_Local}.

\subsubsection{Plasma model}
If we artificially neglect the effect of losses in the non-local Kubo model for the conductivity of graphene making $\hbar\Gamma=0$, we obtain \Eq{static_limit_sigma} in the $T=0$ limit, and we have to apply the Maldague formula (\Eqref{Maldague_Formula}) to those conductivities to obtain the corresponding finite $T$ result.

\begin{figure}[H]
\centering
\includegraphics[width=\linewidth]{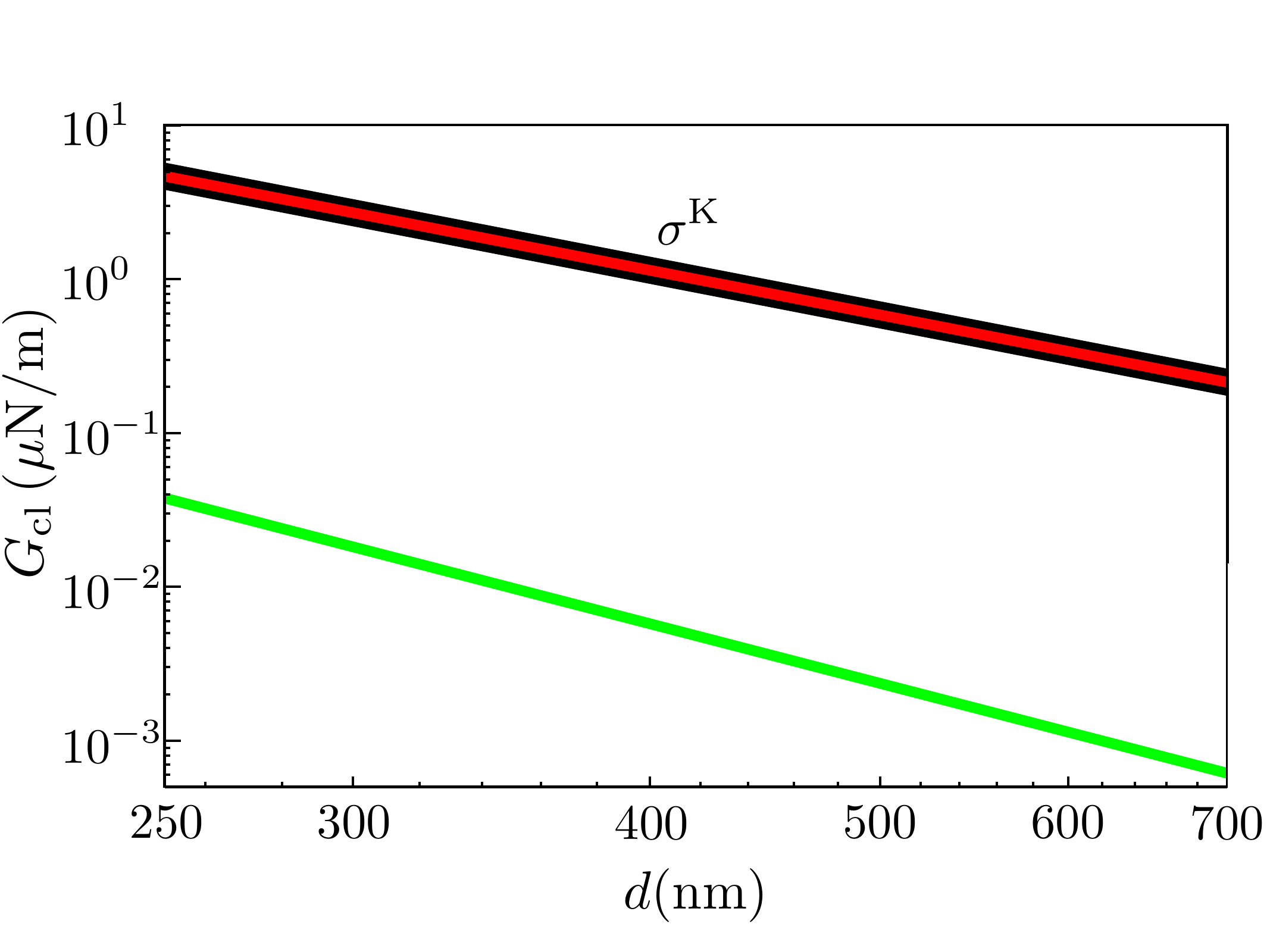}
\caption{ Double logarithmic plot of the $n=0$ Matsubara term of the CLF gradient $G_{\rm{cl}}$ calculated with the Kubo model $\sigma^{\rm{K}}$ as a function of the distance $d$. The Black thick curve is the numerical results using the Plasma prescription (\Eq{PFA_G_cl_Local} using \Eq{Fresnel_Gold_n=0} for $\mathbb{R}^{\rm{Au}}$ and \Eq{NonLocal_Plasma_Fresnel_Matriz_n=0} for $\mathbb{R}^{\rm{Gr}}$). The red curve is the numerical results using the Drude prescription, given by \Eq{PFA_G_cl_Drude_Local}.
The green curve is the difference between the Plasma (black) and Drude (red) results.}
\label{Fig_LLP_Drude_Plasma_NLKubo}
\end{figure}

In this case, the longitudinal conductivity behaves like \Eq{def_sigma_p1}, while the transversal conductivity behaves like \Eq{def_sigma_0}, as a consequence, the zero frequency limit of the Fresnel reflection matrix is
\begin{eqnarray}\label{NonLocal_Plasma_Fresnel_Matriz_n=0}
\dlim_{\xi\to0}\mathbb{R}^{\rm{Gr}} = \left(\begin{array}{cc}
0 & 0 \\
0 & \dfrac{ \epsilon_{1} - 1 + 2c k_{\surf}\bar{\sigma}_{L,1}^{\rm{K}} }{ \epsilon_{1} + 1 + 2c k_{\surf}\bar{\sigma}_{L,1}^{\rm{K}} }
\end{array}\right)
\end{eqnarray}
where $\dlim_{\xi\to0}\epsilon_{SiO_{2}}(\xi) = \epsilon_{1}$, and we have used that  $\bar{\sigma}_{L,1}^{\rm{K}} > 0$ and $0 \leq \bar{\sigma}_{T,0}^{\rm{K}} < \infty$ for all $k_{\surf} > 0$.
It is worth noting that for $k_{\surf}=0$, one has $\mathbb{R}_{ss}^{\rm{Gr}}(\xi\to0,k_{\surf}\to0)\neq 0$ (given by \Eq{Rmatrix_localcase_w0}), but this anomalous term is irrelevant for the computation of the CLF, because: i) this result does not generate a pole in the integrand, and ii) it is only obtained in a set of zero measure, therefore, it cannot contribute to the calculation of the CLF gradient and we must use \Eq{NonLocal_Plasma_Fresnel_Matriz_n=0} in the Lifshitz formula. By considering the lowest order expansion in $k_{\surf}\to0$, we have
\begin{eqnarray}\label{K_sigma_L_n=0}
\bar{\sigma}_{L,1}(k_{\surf}) & \underset{k_{\surf}\to 0}{\approx} & \alpha\dfrac{4\abs{\mu} - \abs{\Delta}}{\hbar v_{F}^{2}k_{\surf}^{2}} + \mathcal{O}\left[\dfrac{1}{k_{\surf}}\right],
\end{eqnarray}
where we have assumed that $\abs{\mu} > \abs{\Delta}$. As a consequence, the classical limit of the CLF gradient with the plasma prescription is approached by
\begin{eqnarray}\label{PFA_G_cl_NLK_pp}
\!\!\!\!\!\!\!\!\!\!\!\!\!\!G_{\rm{cl}}(d) \approx && \nonumber\\
&&\!\!\!\!\!\!\!\!\!\!\!\!\!\!\!\!\!\!\!\!\!\!\!\!- \frac{k_{B}TR \zeta(3)}{4 d^{3}}\left( 1 - \dfrac{6\hbar v_{F}^{2}}{\alpha c (4\abs{\mu} - \abs{\Delta})d} + \mathcal{O}\left[\dfrac{1}{d^{2}}\right]\right).
\end{eqnarray}
This result is always strictly smaller, but very close to the one obtained with the non-local Drude model \Eq{PFA_G_cl_Drude_Local}. So we conclude that also when we use the non-local Kubo model, the effect of Plasma prescription is irrelevant in this experiment, as can be observed in \Fig{Fig_LLP_Drude_Plasma_NLKubo}. This tiny correction of $G_{\rm{cl}}(d)$ is difficult to observe even numerically, so there is no hope we could distinguish it in the experiment.

\subsection{QFT model}
\subsubsection{Drude model}
When losses $\hbar\Gamma>0$ are considered in the QFT model, one can find that the $\xi\to0$ limit of $\bar{\Pi}_{P}^{\rm{QFT}}(\xi)$ tends to $\bar{\Pi}_{P,0}^{\rm{QFT}}$ of \Eq{def_sigma_0} given by \Eq{Def:Lim_xito0_Pi_L} and \Eq{Def:Lim_xito0_Pi_T}. It is easy to check that $0 < \bar{\Pi}_{P,0}^{\rm{QFT}} < \infty$ for any polarization $P=\{\rm{L,T}\}$ and temperature $T$. Therefore, the Fresnel reflection matrix for SiO$_{2}$ covered by graphene and for gold are given by \Eq{n=0_Drude_Fresnel_matrix}. As a result, the zero Matsubara CLF gradient is the usual Drude result given in \Eq{PFA_G_cl_Drude_Local}.
\subsubsection{Plasma model}
\begin{figure}[H]
\centering
\includegraphics[width=\linewidth]{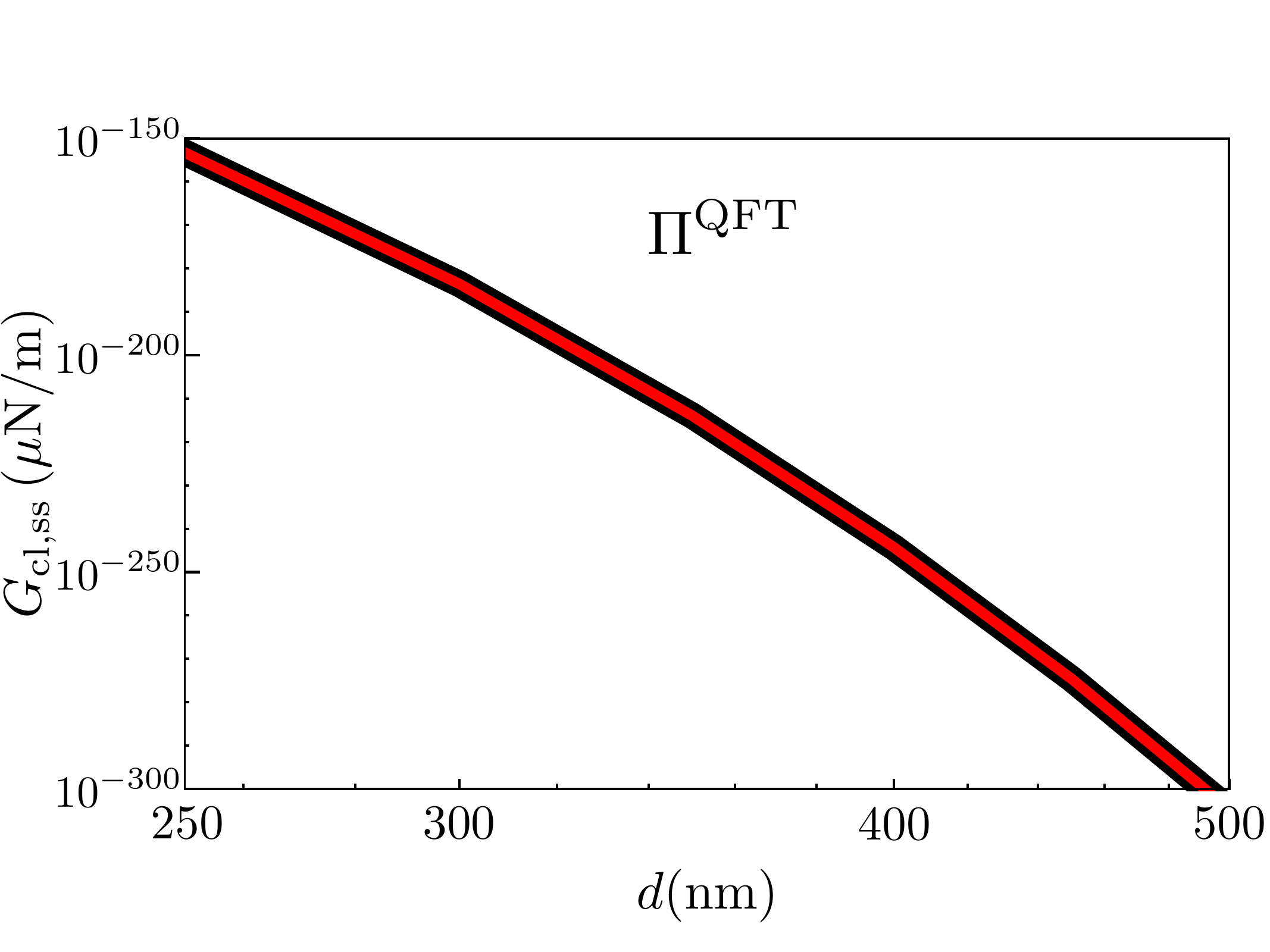}
\caption{ Double logarithmic plot of the $n=0$ Matsubara term of the CLF gradient $G_{\rm{cl},\rm{ss}}$ calculated with the QFT model $\Pi^{\rm{QFT}}$ as a function of the distance $d$. The black thick curve is the numerical result obtained using \Eq{PFA_G_cl_Local} with \Eq{Fresnel_Gold_n=0} for $\mathbb{R}^{\rm{Au}}$ and \Eq{Rss_Gr_NR} and \Eq{Rpp_Gr_NR} for $\mathbb{R}^{\rm{Gr}}$, and the red curve is the approximation given in \Eq{PFA_G_cl_NR_ss}.}
\label{Fig_LLP_Drude_Plasma_NR_exp}
\end{figure}

In the case of the QFT model in the dissipation-less limit $\Gamma\to0$, one can find that the $\xi\to0$ limit of the Longitudinal Polarization operator can be approached using \Eq{Def:PiL_QFTb}, from where we obtain that (\Eq{Def:Lim_xito0_Pi_L})
\begin{eqnarray}
\bar{\Pi}_{L}^{\rm{QFT}}(\xi) \underset{\xi\to 0}{\approx} \bar{\Pi}_{L,2}^{\rm{QFT}}\xi^{2},
\end{eqnarray}
with $\bar{\Pi}_{L,2}^{\rm{QFT}}$ given in \Eq{Def:Lim_xito0_PiL_QFTb}, and the Transversal conductivity, using \Eq{Anomalous_Plasmaterm_PiT_QFTb} can be approached in the $\xi\to0$ limit as (\Eq{Def:Lim_xito0_Pi_T})
\begin{eqnarray}\label{Def:Lim_xito0_sigma_T2}
\bar{\Pi}_{T}^{\rm{QFT}}(\xi)
\underset{\xi\to0}{\approx} \bar{\Pi}_{T,0}^{\rm{QFT}},
\end{eqnarray}
with $\Pi_{T,0}^{\rm{QFT}}$ given in \Eq{Def:Plasma_Pi_T0}. When $\abs{\mu} > \abs{\Delta}$, at room temperature, one can show that those results can be safely approximated by
\begin{eqnarray}
\bar{\Pi}_{T,0}^{\rm{QFT}}
& \approx & \alpha v_{F}k_{\surf}\frac{\pi}{2}\Theta\left( k_{\surf} - K \right),\\
\bar{\Pi}_{L,2}^{\rm{QFT}}
& \underset{k_{\surf}\to 0}{\approx} & \alpha\dfrac{4\abs{\mu}-\abs{\Delta}}{\hbar v_{F}^{2}k_{\surf}^{2}},\label{NR_Pi_L_n=0}
\end{eqnarray}
with $K = 2\sqrt{\mu^{2} - \Delta^{2}}/(\hbar v_{F})$. Therefore, for SiO$_{2}$ covered by graphene, the zero frequency Fresnel coefficients are
\begin{eqnarray}
\lim_{\xi\to0}R_{\rm{ss}}^{\rm{Gr}} & = & \dfrac{-\bar{\Pi}_{T,0}^{\rm{QFT}}}{ c k_{\surf} + \bar{\Pi}_{T,0}^{\rm{QFT}}} \approx \dfrac{-\frac{\pi\alpha v_{F}}{2c}}{ 1 + \frac{\pi\alpha v_{F}}{2c}}\Theta\left( k_{\surf} - K \right)\label{Rss_Gr_NR}\\
\lim_{\xi\to0}R_{\rm{pp}}^{\rm{Gr}} & = & \dfrac{ \epsilon_{1} - 1 + 2c k_{\surf}\bar{\Pi}_{L,2}^{\rm{QFT}}}{ \epsilon_{1} + 1 + 2c k_{\surf}\bar{\Pi}_{L,2}^{\rm{QFT}} }\label{Rpp_Gr_NR}
\end{eqnarray}
where we have used that, in this model $\Pi_{T,0}^{\rm{QFT}}\neq 0$ only for $k_{\surf} > K = 2\sqrt{ \mu^{2} - \Delta^{2} }/(\hbar v_{F})$. As a result, the integral contained in the $n=0$ Matsubara term of the CLF gradient only has contributions of those large wavevectors. For those cases, the Fresnel coefficients of gold can be approximated as
\begin{eqnarray}
\lim_{\xi\to0}R_{\rm{ss}}^{\rm{Au}} & \underset{k_{\surf}>K}{\approx} & - \left(\frac{\omega_{P}}{2c k_{\surf}}\right)^{2}\\
\lim_{\xi\to0}R_{\rm{pp}}^{\rm{Au}} & = & 1
\end{eqnarray}
and the $n=0$ Matsubara term of the CLF gradient due to the $\rm{ss}$ Fresnel coefficients can be  approximated by
\begin{eqnarray}\label{PFA_G_cl_NR_ss}
G_{\rm{cl},\rm{ss}}
& \approx & k_{B}TR\frac{\omega_{P}^{2}\pi\alpha v_{F}}{16c^{3}}\dfrac{e^{-2dK}}{d}.
\end{eqnarray}
In the experiment \cite{PRL_Mohideen}, $\Delta = 0.1\eV$ and $\mu = 0.23\eV$, therefore we have $K^{-1}=d_{0} = \dfrac{\hbar v_{F}}{2\sqrt{\mu^{2} - \Delta^{2}}} \approx 1.6\nm$ (and $K^{-1}\approx 1.3\nm$ for $\Delta = 0$ and $\mu=0.25\eV$). In \Fig{Fig_LLP_Drude_Plasma_NR_exp}, the contribution of $G_{\rm{cl},\rm{ss}}$ to the experimental result for the conditions and distances of the experiment \cite{PRL_Mohideen} can be observed. This $n=0$ Matsubara term is exponentially neglected for the experimentally relevant distances, therefore, even when there is a magnetization contribution to the $n=0$ Matsubara term, in practice it cannot contribute to the final result. $G_{\rm{cl},\rm{ss}}$ is much smaller (more than $10^{8}$ times) than $G_{\rm{cl},\rm{pp}}$ given in \Eq{PFA_G_cl_NLK_pp} for all distances, so it is a complete unobservable effect.

The $\rm{pp}$ term of the Fresnel reflection matrix is modified into \Eq{Rpp_Gr_NR}. This reflection coefficient is the same as the one obtained with the non-local Kubo model $\sigma^{\rm{K}}$ when the Plasma prescription is used and also the expression for $\bar{\sigma}_{L,1}^{\rm{K}} = \bar{\Pi}_{L,2}^{\rm{QFT}}$ is the same (compare \Eq{NR_Pi_L_n=0} with \Eq{K_sigma_L_n=0}), then we obtain that $G_{\rm{cl},\rm{pp}}(d)$ is equal to \Eq{PFA_G_cl_NLK_pp}.

This result cannot be distinguished in the experiment from the usual Drude $n=0$ Matsubara term of the CLF gradient, as shown in \Fig{Fig_LLP_Drude_Plasma_NLKubo} when losses are considered. We conclude that, with the QFT model, the Plasma and Drude prescriptions provides the same numerical result, to all practical extents.
In conclusion, for any of the three models used for the graphene conductivity, and for both Drude or Plasma prescriptions, the $n=0$ Matsubara term always provides a numerical result indistinguishable from the theoretical Drude result given in \eq{PFA_G_cl_Drude_Local}, i.e. the usual Drude result for the thermal CLF. Hence it is not possible, in this experiment, to distinguish the Drude from the Plasma prescription.

\section{Conclusion}\label{sect:conclusions}
We have compared the CLF gradient experimental results \cite{PRL_Mohideen, PRB_Mohideen} between a gold sphere and a graphene-coated plate with the Lifshitz theory using three different models for the electric response of graphene.
In particular, we have considered the general non-local Kubo model for the electric response, the local limit of the Kubo model for the electric response, and the lossless model used in \cite{PRL_Mohideen,PRB_Mohideen} for the full em response. In \cite{PabloMauroComparisonKuboQFT2024} it has been shown that the Kubo and QFT models coincide in the local limiting case, where they exactly provide the local Kubo expression (once also losses are included) for the electric conductivity as the full em response of graphene.
We show that the non-local Kubo and the local Kubo models provide practically identical predictions for the CLF gradient when calculated using the parameters of the experiment \cite{PRL_Mohideen, PRB_Mohideen}. Hence that experiment is sensible only to local effects of graphene. The QFT model also produces results practically identical to the two other cases, and this is since it is applied in a region where the (essentially non-local) magnetic response does not contribute appreciatively to the total result, i.e. the local regime. There, the three models provide results for the CLF gradient with a relative difference smaller than $10^{-3}$. 
Thanks to the natural presence of losses in the Kubo model, we have also tested the effect of losses in graphene response, and we have shown that the CLF gradient is practically insensitive to graphene losses for the parameters used in experiment \cite{PRL_Mohideen,PRB_Mohideen}. This also explains why, again, the QFT lossless model still provides results very similar to that of the Kubo model for the given experimental parameters.
Finally, we have investigated the effect of using a Plasma model for both gold and graphene, in comparison with the correct Drude/Kubo models including losses at zero frequency. We find that, for the experimental parameters \cite{PRL_Mohideen,PRB_Mohideen}, the Plasma models provide results practically identical to the Drude ones, and that the magnetic response derived from the QFT model, that has been sometimes interpreted as a Plasma term in the electric conductivity of graphene $\sigma^{^{\rm{NR}}}$ (\Eq{NRmodel}) when it is not properly regularized as a true electric conductivity $\sigma^{^{\rm{K}}}$ (\Eq{Luttinger_sub}) does not contribute the the experimentally measured CLF gradient.

We have summarized the obtained results in Tab.~\ref{tab_Comparison_dissipation}
\begin{table}[H]
\centering
\begin{tabular}{V{3} c|c|c|c V{3}}
\hlineB{4}
& Drude & Plasma & Distinguisibility\\
 & $\Gamma > 0$ & $\Gamma = 0$ & Drude of Plasma\\
\hlineB{4}
Local & \Eq{PFA_G_cl_Drude_Local} & \Eq{PFA_G_cl_Localpp} + \Eq{PFA_G_cl_Localss} & $d>d_{0}$ \\
\hline
Kubo & \Eq{PFA_G_cl_Drude_Local} & \Eq{PFA_G_cl_NLK_pp} $\,\phantom{ + \Eq{PFA_G_cl_NLK_pp}}$ & Impossible\\
\hline
QFT & \Eq{PFA_G_cl_Drude_Local} & \Eq{PFA_G_cl_NLK_pp} + \Eq{PFA_G_cl_NR_ss}  & Impossible\\
\hlineB{4}
\end{tabular}
\caption{ Table results for the three considered models with and without dissipation. }
\label{tab_Comparison_dissipation}
\end{table}

In conclusion, we show that the extremely simple local Kubo model given by $\sigma_{L}(\omega,\bm{0})=\sigma_{T}(\omega,\bm{0})=\sigma_{xx}(\omega,\bm{0})$ of \Eq{local_sigma_realw} together with \Eq{Maldague_Formula}, explicitly depending on Dirac mass, chemical potential, losses and temperature, is largely enough to fully describe  the experimental results in \cite{PRL_Mohideen}, and can be safely and effectively used for future comparisons with all classical experimental configurations. In any case, for general purposes, we also show how to include losses in the Polarization tensor expression, providing a complete non-local and lossy em response function for graphene.

For the future, it would be interesting to compare the Kubo model with ab-initio ones \cite{PhysRevB.103.125421}, the contribution of the magnetic induced current to fluctuation-induced interactions and to study the effects of graphene in NEMS/MEMS based experimental structures \cite{Nature_2021} where non-locality might possibly play some role in really extreme conditions.

\begin{acknowledgments}
P. R.-L. acknowledges support from Ministerio de Ciencia e Innovaci\'on (Spain), Agencia Estatal de Investigaci\'on, under project NAUTILUS (PID2022-139524NB-I00), from AYUDA PUENTE, URJC, from QuantUM program of the University of Montpellier and the hospitality of the Theory of Light-Matter and Quantum Phenomena group at the Laboratoire Charles Coulomb, University of Montpellier, where part of this work was done.
M.A. acknowledges the QuantUM program of the University of Montpellier, the grant ”CAT”, No. A-HKUST604/20, from the ANR/RGC Joint Research Scheme sponsored by the French National Research Agency (ANR) and the Research Grants Council (RGC) of the Hong Kong Special Administrative Region.
The authors acknowledge Jian-Sheng Wang for comments and suggestions.
\end{acknowledgments}





\begin{thebibliography}{60}%
\makeatletter
\providecommand \@ifxundefined [1]{%
 \@ifx{#1\undefined}
}%
\providecommand \@ifnum [1]{%
 \ifnum #1\expandafter \@firstoftwo
 \else \expandafter \@secondoftwo
 \fi
}%
\providecommand \@ifx [1]{%
 \ifx #1\expandafter \@firstoftwo
 \else \expandafter \@secondoftwo
 \fi
}%
\providecommand \natexlab [1]{#1}%
\providecommand \enquote  [1]{``#1''}%
\providecommand \bibnamefont  [1]{#1}%
\providecommand \bibfnamefont [1]{#1}%
\providecommand \citenamefont [1]{#1}%
\providecommand \href@noop [0]{\@secondoftwo}%
\providecommand \href [0]{\begingroup \@sanitize@url \@href}%
\providecommand \@href[1]{\@@startlink{#1}\@@href}%
\providecommand \@@href[1]{\endgroup#1\@@endlink}%
\providecommand \@sanitize@url [0]{\catcode `\\12\catcode `\$12\catcode
  `\&12\catcode `\#12\catcode `\^12\catcode `\_12\catcode `\%12\relax}%
\providecommand \@@startlink[1]{}%
\providecommand \@@endlink[0]{}%
\providecommand \url  [0]{\begingroup\@sanitize@url \@url }%
\providecommand \@url [1]{\endgroup\@href {#1}{\urlprefix }}%
\providecommand \urlprefix  [0]{URL }%
\providecommand \Eprint [0]{\href }%
\providecommand \doibase [0]{https://doi.org/}%
\providecommand \selectlanguage [0]{\@gobble}%
\providecommand \bibinfo  [0]{\@secondoftwo}%
\providecommand \bibfield  [0]{\@secondoftwo}%
\providecommand \translation [1]{[#1]}%
\providecommand \BibitemOpen [0]{}%
\providecommand \bibitemStop [0]{}%
\providecommand \bibitemNoStop [0]{.\EOS\space}%
\providecommand \EOS [0]{\spacefactor3000\relax}%
\providecommand \BibitemShut  [1]{\csname bibitem#1\endcsname}%
\let\auto@bib@innerbib\@empty
\bibitem [{\citenamefont {Dzyaloshinskii}\ \emph {et~al.}(1961)\citenamefont
  {Dzyaloshinskii}, \citenamefont {Lifshitz},\ and\ \citenamefont
  {Pitaevskii}}]{DLP_1961}%
  \BibitemOpen
  \bibfield  {author} {\bibinfo {author} {\bibfnamefont {I.~E.}\ \bibnamefont
  {Dzyaloshinskii}}, \bibinfo {author} {\bibfnamefont {E.~M.}\ \bibnamefont
  {Lifshitz}},\ and\ \bibinfo {author} {\bibfnamefont {L.~P.}\ \bibnamefont
  {Pitaevskii}},\ }\bibfield  {title} {\bibinfo {title} {General theory of van
  der waals' forces},\ }\href {https://doi.org/10.1070/PU1961v004n02ABEH003330}
  {\bibfield  {journal} {\bibinfo  {journal} {Soviet Physics Uspekhi}\ }\textbf
  {\bibinfo {volume} {4}},\ \bibinfo {pages} {153} (\bibinfo {year}
  {1961})}\BibitemShut {NoStop}%
\bibitem [{\citenamefont {Woods}\ \emph {et~al.}(2016)\citenamefont {Woods},
  \citenamefont {Dalvit}, \citenamefont {Tkatchenko}, \citenamefont
  {Rodriguez-Lopez}, \citenamefont {Rodriguez},\ and\ \citenamefont
  {Podgornik}}]{RevModPhys.88.045003}%
  \BibitemOpen
  \bibfield  {author} {\bibinfo {author} {\bibfnamefont {L.~M.}\ \bibnamefont
  {Woods}}, \bibinfo {author} {\bibfnamefont {D.~A.~R.}\ \bibnamefont
  {Dalvit}}, \bibinfo {author} {\bibfnamefont {A.}~\bibnamefont {Tkatchenko}},
  \bibinfo {author} {\bibfnamefont {P.}~\bibnamefont {Rodriguez-Lopez}},
  \bibinfo {author} {\bibfnamefont {A.~W.}\ \bibnamefont {Rodriguez}},\ and\
  \bibinfo {author} {\bibfnamefont {R.}~\bibnamefont {Podgornik}},\ }\bibfield
  {title} {\bibinfo {title} {Materials perspective on casimir and van der waals
  interactions},\ }\href {https://doi.org/10.1103/RevModPhys.88.045003}
  {\bibfield  {journal} {\bibinfo  {journal} {Rev. Mod. Phys.}\ }\textbf
  {\bibinfo {volume} {88}},\ \bibinfo {pages} {045003} (\bibinfo {year}
  {2016})}\BibitemShut {NoStop}%
\bibitem [{\citenamefont {Abbas}\ \emph {et~al.}(2017)\citenamefont {Abbas},
  \citenamefont {Guizal},\ and\ \citenamefont {Antezza}}]{MAntezza17}%
  \BibitemOpen
  \bibfield  {author} {\bibinfo {author} {\bibfnamefont {C.}~\bibnamefont
  {Abbas}}, \bibinfo {author} {\bibfnamefont {B.}~\bibnamefont {Guizal}},\ and\
  \bibinfo {author} {\bibfnamefont {M.}~\bibnamefont {Antezza}},\ }\bibfield
  {title} {\bibinfo {title} {Strong thermal and electrostatic manipulation of
  the casimir force in graphene multilayers},\ }\href
  {https://doi.org/10.1103/PhysRevLett.118.126101} {\bibfield  {journal}
  {\bibinfo  {journal} {Phys. Rev. Lett.}\ }\textbf {\bibinfo {volume} {118}},\
  \bibinfo {pages} {126101} (\bibinfo {year} {2017})}\BibitemShut {NoStop}%
\bibitem [{\citenamefont {Rodriguez-Lopez}\ \emph {et~al.}(2017)\citenamefont
  {Rodriguez-Lopez}, \citenamefont {Kort-Kamp}, \citenamefont {Dalvit},\ and\
  \citenamefont {Woods}}]{Rodriguez-Lopez2017}%
  \BibitemOpen
  \bibfield  {author} {\bibinfo {author} {\bibfnamefont {P.}~\bibnamefont
  {Rodriguez-Lopez}}, \bibinfo {author} {\bibfnamefont {W.~J.~M.}\ \bibnamefont
  {Kort-Kamp}}, \bibinfo {author} {\bibfnamefont {D.~A.~R.}\ \bibnamefont
  {Dalvit}},\ and\ \bibinfo {author} {\bibfnamefont {L.~M.}\ \bibnamefont
  {Woods}},\ }\bibfield  {title} {\bibinfo {title} {Casimir force phase
  transitions in the graphene family},\ }\href
  {https://doi.org/10.1038/ncomms14699} {\bibfield  {journal} {\bibinfo
  {journal} {Nature Communications}\ }\textbf {\bibinfo {volume} {8}},\
  \bibinfo {pages} {14699} (\bibinfo {year} {2017})}\BibitemShut {NoStop}%
\bibitem [{\citenamefont {G\'omez-Santos}(2009)}]{GomezSantos2009}%
  \BibitemOpen
  \bibfield  {author} {\bibinfo {author} {\bibfnamefont {G.}~\bibnamefont
  {G\'omez-Santos}},\ }\bibfield  {title} {\bibinfo {title} {Thermal van der
  waals interaction between graphene layers},\ }\href
  {https://doi.org/10.1103/PhysRevB.80.245424} {\bibfield  {journal} {\bibinfo
  {journal} {Phys. Rev. B}\ }\textbf {\bibinfo {volume} {80}},\ \bibinfo
  {pages} {245424} (\bibinfo {year} {2009})}\BibitemShut {NoStop}%
\bibitem [{\citenamefont {Bimonte}\ \emph {et~al.}(2017)\citenamefont
  {Bimonte}, \citenamefont {Klimchitskaya},\ and\ \citenamefont
  {Mostepanenko}}]{Bimonte2017}%
  \BibitemOpen
  \bibfield  {author} {\bibinfo {author} {\bibfnamefont {G.}~\bibnamefont
  {Bimonte}}, \bibinfo {author} {\bibfnamefont {G.~L.}\ \bibnamefont
  {Klimchitskaya}},\ and\ \bibinfo {author} {\bibfnamefont {V.~M.}\
  \bibnamefont {Mostepanenko}},\ }\bibfield  {title} {\bibinfo {title} {Thermal
  effect in the casimir force for graphene and graphene-coated substrates:
  Impact of nonzero mass gap and chemical potential},\ }\href
  {https://doi.org/10.1103/PhysRevB.96.115430} {\bibfield  {journal} {\bibinfo
  {journal} {Phys. Rev. B}\ }\textbf {\bibinfo {volume} {96}},\ \bibinfo
  {pages} {115430} (\bibinfo {year} {2017})}\BibitemShut {NoStop}%
\bibitem [{\citenamefont {Bordag}\ \emph
  {et~al.}(2009{\natexlab{a}})\citenamefont {Bordag}, \citenamefont
  {Fialkovsky}, \citenamefont {Gitman},\ and\ \citenamefont
  {Vassilevich}}]{Bordag2009}%
  \BibitemOpen
  \bibfield  {author} {\bibinfo {author} {\bibfnamefont {M.}~\bibnamefont
  {Bordag}}, \bibinfo {author} {\bibfnamefont {I.~V.}\ \bibnamefont
  {Fialkovsky}}, \bibinfo {author} {\bibfnamefont {D.~M.}\ \bibnamefont
  {Gitman}},\ and\ \bibinfo {author} {\bibfnamefont {D.~V.}\ \bibnamefont
  {Vassilevich}},\ }\bibfield  {title} {\bibinfo {title} {Casimir interaction
  between a perfect conductor and graphene described by the dirac model},\
  }\href {https://doi.org/10.1103/PhysRevB.80.245406} {\bibfield  {journal}
  {\bibinfo  {journal} {Phys. Rev. B}\ }\textbf {\bibinfo {volume} {80}},\
  \bibinfo {pages} {245406} (\bibinfo {year} {2009}{\natexlab{a}})}\BibitemShut
  {NoStop}%
\bibitem [{\citenamefont {Drosdoff}\ and\ \citenamefont
  {Woods}(2010)}]{Woods2010}%
  \BibitemOpen
  \bibfield  {author} {\bibinfo {author} {\bibfnamefont {D.}~\bibnamefont
  {Drosdoff}}\ and\ \bibinfo {author} {\bibfnamefont {L.~M.}\ \bibnamefont
  {Woods}},\ }\bibfield  {title} {\bibinfo {title} {Casimir forces and graphene
  sheets},\ }\href {https://doi.org/10.1103/PhysRevB.82.155459} {\bibfield
  {journal} {\bibinfo  {journal} {Phys. Rev. B}\ }\textbf {\bibinfo {volume}
  {82}},\ \bibinfo {pages} {155459} (\bibinfo {year} {2010})}\BibitemShut
  {NoStop}%
\bibitem [{\citenamefont {Wang}\ and\ \citenamefont
  {Antezza}(2024)}]{Arxivwang2024photon}%
  \BibitemOpen
  \bibfield  {author} {\bibinfo {author} {\bibfnamefont {J.-S.}\ \bibnamefont
  {Wang}}\ and\ \bibinfo {author} {\bibfnamefont {M.}~\bibnamefont {Antezza}},\
  }\bibfield  {title} {\bibinfo {title} {Photon mediated transport of energy,
  linear momentum, and angular momentum in fullerene and graphene systems
  beyond local equilibrium},\ }\href
  {https://doi.org/10.1103/PhysRevB.109.125105} {\bibfield  {journal} {\bibinfo
   {journal} {Phys. Rev. B}\ }\textbf {\bibinfo {volume} {109}},\ \bibinfo
  {pages} {125105} (\bibinfo {year} {2024})}\BibitemShut {NoStop}%
\bibitem [{\citenamefont {Rodriguez-Lopez}\ \emph {et~al.}(2024)\citenamefont
  {Rodriguez-Lopez}, \citenamefont {Le}, \citenamefont {Bondarev},
  \citenamefont {Antezza},\ and\ \citenamefont {Woods}}]{PabloRL2024}%
  \BibitemOpen
  \bibfield  {author} {\bibinfo {author} {\bibfnamefont {P.}~\bibnamefont
  {Rodriguez-Lopez}}, \bibinfo {author} {\bibfnamefont {D.-N.}\ \bibnamefont
  {Le}}, \bibinfo {author} {\bibfnamefont {I.~V.}\ \bibnamefont {Bondarev}},
  \bibinfo {author} {\bibfnamefont {M.}~\bibnamefont {Antezza}},\ and\ \bibinfo
  {author} {\bibfnamefont {L.~M.}\ \bibnamefont {Woods}},\ }\bibfield  {title}
  {\bibinfo {title} {Giant anisotropy and casimir phenomena: The case of carbon
  nanotube metasurfaces},\ }\href {https://doi.org/10.1103/PhysRevB.109.035422}
  {\bibfield  {journal} {\bibinfo  {journal} {Phys. Rev. B}\ }\textbf {\bibinfo
  {volume} {109}},\ \bibinfo {pages} {035422} (\bibinfo {year}
  {2024})}\BibitemShut {NoStop}%
\bibitem [{\citenamefont {Jeyar}\ \emph
  {et~al.}(2023{\natexlab{a}})\citenamefont {Jeyar}, \citenamefont {Luo},
  \citenamefont {Austry}, \citenamefont {Guizal}, \citenamefont {Zheng},
  \citenamefont {Chan},\ and\ \citenamefont {Antezza}}]{PhysRevA.108.062811}%
  \BibitemOpen
  \bibfield  {author} {\bibinfo {author} {\bibfnamefont {Y.}~\bibnamefont
  {Jeyar}}, \bibinfo {author} {\bibfnamefont {M.}~\bibnamefont {Luo}}, \bibinfo
  {author} {\bibfnamefont {K.}~\bibnamefont {Austry}}, \bibinfo {author}
  {\bibfnamefont {B.}~\bibnamefont {Guizal}}, \bibinfo {author} {\bibfnamefont
  {Y.}~\bibnamefont {Zheng}}, \bibinfo {author} {\bibfnamefont {H.~B.}\
  \bibnamefont {Chan}},\ and\ \bibinfo {author} {\bibfnamefont
  {M.}~\bibnamefont {Antezza}},\ }\bibfield  {title} {\bibinfo {title} {Tunable
  nonadditivity in the casimir-lifshitz force between graphene gratings},\
  }\href {https://doi.org/10.1103/PhysRevA.108.062811} {\bibfield  {journal}
  {\bibinfo  {journal} {Phys. Rev. A}\ }\textbf {\bibinfo {volume} {108}},\
  \bibinfo {pages} {062811} (\bibinfo {year} {2023}{\natexlab{a}})}\BibitemShut
  {NoStop}%
\bibitem [{\citenamefont {Jeyar}\ \emph
  {et~al.}(2023{\natexlab{b}})\citenamefont {Jeyar}, \citenamefont {Austry},
  \citenamefont {Luo}, \citenamefont {Guizal}, \citenamefont {Chan},\ and\
  \citenamefont {Antezza}}]{PhysRevB.108.115412}%
  \BibitemOpen
  \bibfield  {author} {\bibinfo {author} {\bibfnamefont {Y.}~\bibnamefont
  {Jeyar}}, \bibinfo {author} {\bibfnamefont {K.}~\bibnamefont {Austry}},
  \bibinfo {author} {\bibfnamefont {M.}~\bibnamefont {Luo}}, \bibinfo {author}
  {\bibfnamefont {B.}~\bibnamefont {Guizal}}, \bibinfo {author} {\bibfnamefont
  {H.~B.}\ \bibnamefont {Chan}},\ and\ \bibinfo {author} {\bibfnamefont
  {M.}~\bibnamefont {Antezza}},\ }\bibfield  {title} {\bibinfo {title}
  {Casimir-lifshitz force between graphene-based structures out of thermal
  equilibrium},\ }\href {https://doi.org/10.1103/PhysRevB.108.115412}
  {\bibfield  {journal} {\bibinfo  {journal} {Phys. Rev. B}\ }\textbf {\bibinfo
  {volume} {108}},\ \bibinfo {pages} {115412} (\bibinfo {year}
  {2023}{\natexlab{b}})}\BibitemShut {NoStop}%
\bibitem [{\citenamefont {Jeyar}\ \emph {et~al.}(2024)\citenamefont {Jeyar},
  \citenamefont {Luo}, \citenamefont {Guizal}, , \citenamefont {Chan},\ and\
  \citenamefont {Antezza}}]{YoussefPRB}%
  \BibitemOpen
  \bibfield  {author} {\bibinfo {author} {\bibfnamefont {Y.}~\bibnamefont
  {Jeyar}}, \bibinfo {author} {\bibfnamefont {M.}~\bibnamefont {Luo}}, \bibinfo
  {author} {\bibfnamefont {B.}~\bibnamefont {Guizal}}, , \bibinfo {author}
  {\bibfnamefont {H.~B.}\ \bibnamefont {Chan}},\ and\ \bibinfo {author}
  {\bibfnamefont {M.}~\bibnamefont {Antezza}},\ }\href
  {https://arxiv.org/abs/2405.14523} {\bibinfo {title} {Casimir-lifshitz force
  for graphene-covered gratings}} (\bibinfo {year} {2024}),\ \Eprint
  {https://arxiv.org/abs/2405.14523} {arXiv:2405.14523 [cond-mat.mes-hall]}
  \BibitemShut {NoStop}%
\bibitem [{\citenamefont {Liu}\ \emph {et~al.}(2021{\natexlab{a}})\citenamefont
  {Liu}, \citenamefont {Zhang}, \citenamefont {Klimchitskaya}, \citenamefont
  {Mostepanenko},\ and\ \citenamefont {Mohideen}}]{PRL_Mohideen}%
  \BibitemOpen
  \bibfield  {author} {\bibinfo {author} {\bibfnamefont {M.}~\bibnamefont
  {Liu}}, \bibinfo {author} {\bibfnamefont {Y.}~\bibnamefont {Zhang}}, \bibinfo
  {author} {\bibfnamefont {G.~L.}\ \bibnamefont {Klimchitskaya}}, \bibinfo
  {author} {\bibfnamefont {V.~M.}\ \bibnamefont {Mostepanenko}},\ and\ \bibinfo
  {author} {\bibfnamefont {U.}~\bibnamefont {Mohideen}},\ }\bibfield  {title}
  {\bibinfo {title} {Demonstration of an unusual thermal effect in the casimir
  force from graphene},\ }\href
  {https://doi.org/10.1103/PhysRevLett.126.206802} {\bibfield  {journal}
  {\bibinfo  {journal} {Phys. Rev. Lett.}\ }\textbf {\bibinfo {volume} {126}},\
  \bibinfo {pages} {206802} (\bibinfo {year} {2021}{\natexlab{a}})}\BibitemShut
  {NoStop}%
\bibitem [{\citenamefont {Liu}\ \emph {et~al.}(2021{\natexlab{b}})\citenamefont
  {Liu}, \citenamefont {Zhang}, \citenamefont {Klimchitskaya}, \citenamefont
  {Mostepanenko},\ and\ \citenamefont {Mohideen}}]{PRB_Mohideen}%
  \BibitemOpen
  \bibfield  {author} {\bibinfo {author} {\bibfnamefont {M.}~\bibnamefont
  {Liu}}, \bibinfo {author} {\bibfnamefont {Y.}~\bibnamefont {Zhang}}, \bibinfo
  {author} {\bibfnamefont {G.~L.}\ \bibnamefont {Klimchitskaya}}, \bibinfo
  {author} {\bibfnamefont {V.~M.}\ \bibnamefont {Mostepanenko}},\ and\ \bibinfo
  {author} {\bibfnamefont {U.}~\bibnamefont {Mohideen}},\ }\bibfield  {title}
  {\bibinfo {title} {Experimental and theoretical investigation of the thermal
  effect in the casimir interaction from graphene},\ }\href
  {https://doi.org/10.1103/PhysRevB.104.085436} {\bibfield  {journal} {\bibinfo
   {journal} {Phys. Rev. B}\ }\textbf {\bibinfo {volume} {104}},\ \bibinfo
  {pages} {085436} (\bibinfo {year} {2021}{\natexlab{b}})}\BibitemShut
  {NoStop}%
\bibitem [{\citenamefont {Castro~Neto}\ \emph {et~al.}(2009)\citenamefont
  {Castro~Neto}, \citenamefont {Guinea}, \citenamefont {Peres}, \citenamefont
  {Novoselov},\ and\ \citenamefont {Geim}}]{RevModPhys.81.109}%
  \BibitemOpen
  \bibfield  {author} {\bibinfo {author} {\bibfnamefont {A.~H.}\ \bibnamefont
  {Castro~Neto}}, \bibinfo {author} {\bibfnamefont {F.}~\bibnamefont {Guinea}},
  \bibinfo {author} {\bibfnamefont {N.~M.~R.}\ \bibnamefont {Peres}}, \bibinfo
  {author} {\bibfnamefont {K.~S.}\ \bibnamefont {Novoselov}},\ and\ \bibinfo
  {author} {\bibfnamefont {A.~K.}\ \bibnamefont {Geim}},\ }\bibfield  {title}
  {\bibinfo {title} {The electronic properties of graphene},\ }\href
  {https://doi.org/10.1103/RevModPhys.81.109} {\bibfield  {journal} {\bibinfo
  {journal} {Rev. Mod. Phys.}\ }\textbf {\bibinfo {volume} {81}},\ \bibinfo
  {pages} {109} (\bibinfo {year} {2009})}\BibitemShut {NoStop}%
\bibitem [{\citenamefont {Das~Sarma}\ \emph {et~al.}(2011)\citenamefont
  {Das~Sarma}, \citenamefont {Adam}, \citenamefont {Hwang},\ and\ \citenamefont
  {Rossi}}]{DasSarmaRMP2011}%
  \BibitemOpen
  \bibfield  {author} {\bibinfo {author} {\bibfnamefont {S.}~\bibnamefont
  {Das~Sarma}}, \bibinfo {author} {\bibfnamefont {S.}~\bibnamefont {Adam}},
  \bibinfo {author} {\bibfnamefont {E.~H.}\ \bibnamefont {Hwang}},\ and\
  \bibinfo {author} {\bibfnamefont {E.}~\bibnamefont {Rossi}},\ }\bibfield
  {title} {\bibinfo {title} {Electronic transport in two-dimensional
  graphene},\ }\href {https://doi.org/10.1103/RevModPhys.83.407} {\bibfield
  {journal} {\bibinfo  {journal} {Rev. Mod. Phys.}\ }\textbf {\bibinfo {volume}
  {83}},\ \bibinfo {pages} {407} (\bibinfo {year} {2011})}\BibitemShut
  {NoStop}%
\bibitem [{\citenamefont {Gusynin}\ \emph {et~al.}(2006)\citenamefont
  {Gusynin}, \citenamefont {Sharapov},\ and\ \citenamefont
  {Carbotte}}]{Gusynin2006}%
  \BibitemOpen
  \bibfield  {author} {\bibinfo {author} {\bibfnamefont {V.~P.}\ \bibnamefont
  {Gusynin}}, \bibinfo {author} {\bibfnamefont {S.~G.}\ \bibnamefont
  {Sharapov}},\ and\ \bibinfo {author} {\bibfnamefont {J.~P.}\ \bibnamefont
  {Carbotte}},\ }\bibfield  {title} {\bibinfo {title} {Unusual microwave
  response of dirac quasiparticles in graphene},\ }\href
  {https://doi.org/10.1103/PhysRevLett.96.256802} {\bibfield  {journal}
  {\bibinfo  {journal} {Phys. Rev. Lett.}\ }\textbf {\bibinfo {volume} {96}},\
  \bibinfo {pages} {256802} (\bibinfo {year} {2006})}\BibitemShut {NoStop}%
\bibitem [{\citenamefont {Jablan}\ \emph
  {et~al.}(2009{\natexlab{a}})\citenamefont {Jablan}, \citenamefont {Buljan},\
  and\ \citenamefont {Solja\ifmmode \check{c}\else
  \v{c}\fi{}i\ifmmode~\acute{c}\else \'{c}\fi{}}}]{Marinko2009}%
  \BibitemOpen
  \bibfield  {author} {\bibinfo {author} {\bibfnamefont {M.}~\bibnamefont
  {Jablan}}, \bibinfo {author} {\bibfnamefont {H.}~\bibnamefont {Buljan}},\
  and\ \bibinfo {author} {\bibfnamefont {M.}~\bibnamefont {Solja\ifmmode
  \check{c}\else \v{c}\fi{}i\ifmmode~\acute{c}\else \'{c}\fi{}}},\ }\bibfield
  {title} {\bibinfo {title} {Plasmonics in graphene at infrared frequencies},\
  }\href {https://doi.org/10.1103/PhysRevB.80.245435} {\bibfield  {journal}
  {\bibinfo  {journal} {Phys. Rev. B}\ }\textbf {\bibinfo {volume} {80}},\
  \bibinfo {pages} {245435} (\bibinfo {year} {2009}{\natexlab{a}})}\BibitemShut
  {NoStop}%
\bibitem [{\citenamefont {Hartnoll}\ and\ \citenamefont
  {Mackenzie}(2022)}]{RMPdissipation2022}%
  \BibitemOpen
  \bibfield  {author} {\bibinfo {author} {\bibfnamefont {S.~A.}\ \bibnamefont
  {Hartnoll}}\ and\ \bibinfo {author} {\bibfnamefont {A.~P.}\ \bibnamefont
  {Mackenzie}},\ }\bibfield  {title} {\bibinfo {title} {Colloquium: Planckian
  dissipation in metals},\ }\href
  {https://doi.org/10.1103/RevModPhys.94.041002} {\bibfield  {journal}
  {\bibinfo  {journal} {Rev. Mod. Phys.}\ }\textbf {\bibinfo {volume} {94}},\
  \bibinfo {pages} {041002} (\bibinfo {year} {2022})}\BibitemShut {NoStop}%
\bibitem [{\citenamefont {Abrikosov}\ \emph {et~al.}(1975)\citenamefont
  {Abrikosov}, \citenamefont {Dzyaloshinskii}, \citenamefont {Gorkov},\ and\
  \citenamefont {Silverman}}]{BOOKAbrikosov}%
  \BibitemOpen
  \bibfield  {author} {\bibinfo {author} {\bibfnamefont {A.~A.}\ \bibnamefont
  {Abrikosov}}, \bibinfo {author} {\bibfnamefont {I.}~\bibnamefont
  {Dzyaloshinskii}}, \bibinfo {author} {\bibfnamefont {L.~P.}\ \bibnamefont
  {Gorkov}},\ and\ \bibinfo {author} {\bibfnamefont {R.~A.}\ \bibnamefont
  {Silverman}},\ }\href {https://cds.cern.ch/record/107441} {\emph {\bibinfo
  {title} {{Methods of quantum field theory in statistical physics}}}}\
  (\bibinfo  {publisher} {Dover},\ \bibinfo {address} {New York, NY},\ \bibinfo
  {year} {1975})\BibitemShut {NoStop}%
\bibitem [{\citenamefont {Rodriguez-Lopez}\ \emph {et~al.}(2018)\citenamefont
  {Rodriguez-Lopez}, \citenamefont {Kort-Kamp}, \citenamefont {Dalvit},\ and\
  \citenamefont {Woods}}]{non-local_Graphene_Lilia_Pablo}%
  \BibitemOpen
  \bibfield  {author} {\bibinfo {author} {\bibfnamefont {P.}~\bibnamefont
  {Rodriguez-Lopez}}, \bibinfo {author} {\bibfnamefont {W.~J.~M.}\ \bibnamefont
  {Kort-Kamp}}, \bibinfo {author} {\bibfnamefont {D.~A.~R.}\ \bibnamefont
  {Dalvit}},\ and\ \bibinfo {author} {\bibfnamefont {L.~M.}\ \bibnamefont
  {Woods}},\ }\bibfield  {title} {\bibinfo {title} {Nonlocal optical response
  in topological phase transitions in the graphene family},\ }\href
  {https://doi.org/10.1103/PhysRevMaterials.2.014003} {\bibfield  {journal}
  {\bibinfo  {journal} {Phys. Rev. Mater.}\ }\textbf {\bibinfo {volume} {2}},\
  \bibinfo {pages} {014003} (\bibinfo {year} {2018})}\BibitemShut {NoStop}%
\bibitem [{\citenamefont {Rodriguez-Lopez}\ and\ \citenamefont
  {Antezza}(2024)}]{PabloMauroGauge2024}%
  \BibitemOpen
  \bibfield  {author} {\bibinfo {author} {\bibfnamefont {P.}~\bibnamefont
  {Rodriguez-Lopez}}\ and\ \bibinfo {author} {\bibfnamefont {M.}~\bibnamefont
  {Antezza}},\ }\bibfield  {title} {\bibinfo {title} {Future publication},\
  }\href@noop {} {\bibfield  {journal} {\bibinfo  {journal} {Future
  Publication}\ } (\bibinfo {year} {2024})}\BibitemShut {NoStop}%
\bibitem [{\citenamefont {Rodriguez-Lopez}\ \emph {et~al.}(2025)\citenamefont
  {Rodriguez-Lopez}, \citenamefont {Wang},\ and\ \citenamefont
  {Antezza}}]{PabloMauroComparisonKuboQFT2024}%
  \BibitemOpen
  \bibfield  {author} {\bibinfo {author} {\bibfnamefont {P.}~\bibnamefont
  {Rodriguez-Lopez}}, \bibinfo {author} {\bibfnamefont {J.-S.}\ \bibnamefont
  {Wang}},\ and\ \bibinfo {author} {\bibfnamefont {M.}~\bibnamefont
  {Antezza}},\ }\bibfield  {title} {\bibinfo {title} {Electric conductivity in
  graphene: Kubo model versus a nonlocal quantum field theory model},\ }\href
  {https://doi.org/10.1103/PhysRevB.111.115428} {\bibfield  {journal} {\bibinfo
   {journal} {Phys. Rev. B}\ }\textbf {\bibinfo {volume} {111}},\ \bibinfo
  {pages} {115428} (\bibinfo {year} {2025})}\BibitemShut {NoStop}%
\bibitem [{\citenamefont {Klimchitskaya}\ and\ \citenamefont
  {Mostepanenko}(2016)}]{Klimchitskaya2016}%
  \BibitemOpen
  \bibfield  {author} {\bibinfo {author} {\bibfnamefont {G.~L.}\ \bibnamefont
  {Klimchitskaya}}\ and\ \bibinfo {author} {\bibfnamefont {V.~M.}\ \bibnamefont
  {Mostepanenko}},\ }\bibfield  {title} {\bibinfo {title} {Conductivity of pure
  graphene: Theoretical approach using the polarization tensor},\ }\href
  {https://doi.org/10.1103/PhysRevB.93.245419} {\bibfield  {journal} {\bibinfo
  {journal} {Phys. Rev. B}\ }\textbf {\bibinfo {volume} {93}},\ \bibinfo
  {pages} {245419} (\bibinfo {year} {2016})}\BibitemShut {NoStop}%
\bibitem [{\citenamefont {Klimchitskaya}\ \emph {et~al.}(2017)\citenamefont
  {Klimchitskaya}, \citenamefont {Mostepanenko},\ and\ \citenamefont
  {Petrov}}]{Klimchitskaya2017}%
  \BibitemOpen
  \bibfield  {author} {\bibinfo {author} {\bibfnamefont {G.~L.}\ \bibnamefont
  {Klimchitskaya}}, \bibinfo {author} {\bibfnamefont {V.~M.}\ \bibnamefont
  {Mostepanenko}},\ and\ \bibinfo {author} {\bibfnamefont {V.~M.}\ \bibnamefont
  {Petrov}},\ }\bibfield  {title} {\bibinfo {title} {Conductivity of graphene
  in the framework of dirac model: Interplay between nonzero mass gap and
  chemical potential},\ }\href {https://doi.org/10.1103/PhysRevB.96.235432}
  {\bibfield  {journal} {\bibinfo  {journal} {Phys. Rev. B}\ }\textbf {\bibinfo
  {volume} {96}},\ \bibinfo {pages} {235432} (\bibinfo {year}
  {2017})}\BibitemShut {NoStop}%
\bibitem [{\citenamefont {Klimchitskaya}\ and\ \citenamefont
  {Mostepanenko}(2018)}]{Klimchitskaya2018}%
  \BibitemOpen
  \bibfield  {author} {\bibinfo {author} {\bibfnamefont {G.~L.}\ \bibnamefont
  {Klimchitskaya}}\ and\ \bibinfo {author} {\bibfnamefont {V.~M.}\ \bibnamefont
  {Mostepanenko}},\ }\bibfield  {title} {\bibinfo {title} {Kramers-kronig
  relations and causality conditions for graphene in the framework of the dirac
  model},\ }\href {https://doi.org/10.1103/PhysRevD.97.085001} {\bibfield
  {journal} {\bibinfo  {journal} {Phys. Rev. D}\ }\textbf {\bibinfo {volume}
  {97}},\ \bibinfo {pages} {085001} (\bibinfo {year} {2018})}\BibitemShut
  {NoStop}%
\bibitem [{\citenamefont {Falkovsky}\ and\ \citenamefont
  {Pershoguba}(2007)}]{Falkovsky2007b}%
  \BibitemOpen
  \bibfield  {author} {\bibinfo {author} {\bibfnamefont {L.~A.}\ \bibnamefont
  {Falkovsky}}\ and\ \bibinfo {author} {\bibfnamefont {S.~S.}\ \bibnamefont
  {Pershoguba}},\ }\bibfield  {title} {\bibinfo {title} {Optical far-infrared
  properties of a graphene monolayer and multilayer},\ }\href
  {https://doi.org/10.1103/PhysRevB.76.153410} {\bibfield  {journal} {\bibinfo
  {journal} {Phys. Rev. B}\ }\textbf {\bibinfo {volume} {76}},\ \bibinfo
  {pages} {153410} (\bibinfo {year} {2007})}\BibitemShut {NoStop}%
\bibitem [{\citenamefont {Bordag}\ \emph
  {et~al.}(2009{\natexlab{b}})\citenamefont {Bordag}, \citenamefont
  {Klimchitskaya}, \citenamefont {Mohideen},\ and\ \citenamefont
  {Mostepanenko}}]{bordag2009advances}%
  \BibitemOpen
  \bibfield  {author} {\bibinfo {author} {\bibfnamefont {M.}~\bibnamefont
  {Bordag}}, \bibinfo {author} {\bibfnamefont {G.}~\bibnamefont
  {Klimchitskaya}}, \bibinfo {author} {\bibfnamefont {U.}~\bibnamefont
  {Mohideen}},\ and\ \bibinfo {author} {\bibfnamefont {V.}~\bibnamefont
  {Mostepanenko}},\ }\href {https://books.google.es/books?id=CqE1f_s5PgYC}
  {\emph {\bibinfo {title} {Advances in the Casimir Effect}}},\ International
  Series of Monographs on Physics\ (\bibinfo  {publisher} {OUP Oxford},\
  \bibinfo {year} {2009})\BibitemShut {NoStop}%
\bibitem [{\citenamefont {Rodriguez-Lopez}\ \emph {et~al.}(2020)\citenamefont
  {Rodriguez-Lopez}, \citenamefont {Popescu}, \citenamefont {Fialkovsky},
  \citenamefont {Khusnutdinov},\ and\ \citenamefont
  {Woods}}]{Rodriguez-Lopez2020}%
  \BibitemOpen
  \bibfield  {author} {\bibinfo {author} {\bibfnamefont {P.}~\bibnamefont
  {Rodriguez-Lopez}}, \bibinfo {author} {\bibfnamefont {A.}~\bibnamefont
  {Popescu}}, \bibinfo {author} {\bibfnamefont {I.}~\bibnamefont {Fialkovsky}},
  \bibinfo {author} {\bibfnamefont {N.}~\bibnamefont {Khusnutdinov}},\ and\
  \bibinfo {author} {\bibfnamefont {L.~M.}\ \bibnamefont {Woods}},\ }\bibfield
  {title} {\bibinfo {title} {Signatures of complex optical response in casimir
  interactions of type i and ii weyl semimetals},\ }\href
  {https://doi.org/10.1038/s43246-020-0015-4} {\bibfield  {journal} {\bibinfo
  {journal} {Communications Materials}\ }\textbf {\bibinfo {volume} {1}},\
  \bibinfo {pages} {14} (\bibinfo {year} {2020})}\BibitemShut {NoStop}%
\bibitem [{\citenamefont {Kubo}(1957)}]{Kubo1957}%
  \BibitemOpen
  \bibfield  {author} {\bibinfo {author} {\bibfnamefont {R.}~\bibnamefont
  {Kubo}},\ }\bibfield  {title} {\bibinfo {title} {Statistical-mechanical
  theory of irreversible processes. i. general theory and simple applications
  to magnetic and conduction problems},\ }\href
  {https://doi.org/10.1143/JPSJ.12.570} {\bibfield  {journal} {\bibinfo
  {journal} {Journal of the Physical Society of Japan}\ }\textbf {\bibinfo
  {volume} {12}},\ \bibinfo {pages} {570} (\bibinfo {year} {1957})},\ \Eprint
  {https://arxiv.org/abs/https://doi.org/10.1143/JPSJ.12.570}
  {https://doi.org/10.1143/JPSJ.12.570} \BibitemShut {NoStop}%
\bibitem [{\citenamefont {Bordag}\ \emph {et~al.}(2015)\citenamefont {Bordag},
  \citenamefont {Klimchitskaya}, \citenamefont {Mostepanenko},\ and\
  \citenamefont {Petrov}}]{Bordag2015}%
  \BibitemOpen
  \bibfield  {author} {\bibinfo {author} {\bibfnamefont {M.}~\bibnamefont
  {Bordag}}, \bibinfo {author} {\bibfnamefont {G.~L.}\ \bibnamefont
  {Klimchitskaya}}, \bibinfo {author} {\bibfnamefont {V.~M.}\ \bibnamefont
  {Mostepanenko}},\ and\ \bibinfo {author} {\bibfnamefont {V.~M.}\ \bibnamefont
  {Petrov}},\ }\bibfield  {title} {\bibinfo {title} {Quantum field theoretical
  description for the reflectivity of graphene},\ }\href
  {https://doi.org/10.1103/PhysRevD.91.045037} {\bibfield  {journal} {\bibinfo
  {journal} {Phys. Rev. D}\ }\textbf {\bibinfo {volume} {91}},\ \bibinfo
  {pages} {045037} (\bibinfo {year} {2015})}\BibitemShut {NoStop}%
\bibitem [{\citenamefont {Zeitlin}(1995)}]{Zeitlin1995}%
  \BibitemOpen
  \bibfield  {author} {\bibinfo {author} {\bibfnamefont {V.}~\bibnamefont
  {Zeitlin}},\ }\bibfield  {title} {\bibinfo {title} {Qed2+1 with nonzero
  fermion density and the quantum hall effect},\ }\href
  {https://doi.org/https://doi.org/10.1016/0370-2693(95)00488-7} {\bibfield
  {journal} {\bibinfo  {journal} {Physics Letters B}\ }\textbf {\bibinfo
  {volume} {352}},\ \bibinfo {pages} {422} (\bibinfo {year}
  {1995})}\BibitemShut {NoStop}%
\bibitem [{\citenamefont {Fialkovsky}\ \emph {et~al.}(2011)\citenamefont
  {Fialkovsky}, \citenamefont {Marachevsky},\ and\ \citenamefont
  {Vassilevich}}]{Fialkovsky2011}%
  \BibitemOpen
  \bibfield  {author} {\bibinfo {author} {\bibfnamefont {I.~V.}\ \bibnamefont
  {Fialkovsky}}, \bibinfo {author} {\bibfnamefont {V.~N.}\ \bibnamefont
  {Marachevsky}},\ and\ \bibinfo {author} {\bibfnamefont {D.~V.}\ \bibnamefont
  {Vassilevich}},\ }\bibfield  {title} {\bibinfo {title} {Finite-temperature
  casimir effect for graphene},\ }\href
  {https://doi.org/10.1103/PhysRevB.84.035446} {\bibfield  {journal} {\bibinfo
  {journal} {Phys. Rev. B}\ }\textbf {\bibinfo {volume} {84}},\ \bibinfo
  {pages} {035446} (\bibinfo {year} {2011})}\BibitemShut {NoStop}%
\bibitem [{\citenamefont {Dorey}\ and\ \citenamefont
  {Mavromatos}(1992)}]{Dorey1992}%
  \BibitemOpen
  \bibfield  {author} {\bibinfo {author} {\bibfnamefont {N.}~\bibnamefont
  {Dorey}}\ and\ \bibinfo {author} {\bibfnamefont {N.}~\bibnamefont
  {Mavromatos}},\ }\bibfield  {title} {\bibinfo {title} {Qed3 and
  two-dimensional superconductivity without parity violation},\ }\href
  {https://doi.org/https://doi.org/10.1016/0550-3213(92)90632-L} {\bibfield
  {journal} {\bibinfo  {journal} {Nuclear Physics B}\ }\textbf {\bibinfo
  {volume} {386}},\ \bibinfo {pages} {614} (\bibinfo {year}
  {1992})}\BibitemShut {NoStop}%
\bibitem [{\citenamefont {Luttinger}(1968)}]{Luttinger1968}%
  \BibitemOpen
  \bibfield  {author} {\bibinfo {author} {\bibfnamefont {J.~M.}\ \bibnamefont
  {Luttinger}},\ }\bibinfo {title} {Transport theory},\ in\ \href
  {https://doi.org/10.1007/978-1-4899-6435-9_4} {\emph {\bibinfo {booktitle}
  {Mathematical Methods in Solid State and Superfluid Theory: Scottish
  Universities' Summer School}}},\ \bibinfo {editor} {edited by\ \bibinfo
  {editor} {\bibfnamefont {R.~C.}\ \bibnamefont {Clark}}\ and\ \bibinfo
  {editor} {\bibfnamefont {G.~H.}\ \bibnamefont {Derrick}}}\ (\bibinfo
  {publisher} {Springer US},\ \bibinfo {address} {Boston, MA},\ \bibinfo {year}
  {1968})\ pp.\ \bibinfo {pages} {157--193}\BibitemShut {NoStop}%
\bibitem [{\citenamefont {Rammer}(2007)}]{Rammer_2007}%
  \BibitemOpen
  \bibfield  {author} {\bibinfo {author} {\bibfnamefont {J.}~\bibnamefont
  {Rammer}},\ }\href@noop {} {\emph {\bibinfo {title} {Quantum Field Theory of
  Non-equilibrium States}}}\ (\bibinfo  {publisher} {Cambridge University
  Press},\ \bibinfo {year} {2007})\BibitemShut {NoStop}%
\bibitem [{\citenamefont {Klimchitskaya}\ and\ \citenamefont
  {Mostepanenko}(2023)}]{Mostdispersionrelations}%
  \BibitemOpen
  \bibfield  {author} {\bibinfo {author} {\bibfnamefont {G.~L.}\ \bibnamefont
  {Klimchitskaya}}\ and\ \bibinfo {author} {\bibfnamefont {V.~M.}\ \bibnamefont
  {Mostepanenko}},\ }\bibfield  {title} {\bibinfo {title} {Quantum field
  theoretical framework for the electromagnetic response of graphene and
  dispersion relations with implications to the casimir effect},\ }\href
  {https://doi.org/10.1103/PhysRevD.107.105007} {\bibfield  {journal} {\bibinfo
   {journal} {Phys. Rev. D}\ }\textbf {\bibinfo {volume} {107}},\ \bibinfo
  {pages} {105007} (\bibinfo {year} {2023})}\BibitemShut {NoStop}%
\bibitem [{\citenamefont {Giuliani}\ and\ \citenamefont
  {Vignale}(2005)}]{Giuliani}%
  \BibitemOpen
  \bibfield  {author} {\bibinfo {author} {\bibfnamefont {G.}~\bibnamefont
  {Giuliani}}\ and\ \bibinfo {author} {\bibfnamefont {G.}~\bibnamefont
  {Vignale}},\ }\href {https://doi.org/10.1017/CBO9780511619915} {\emph
  {\bibinfo {title} {Quantum Theory of the Electron Liquid}}}\ (\bibinfo
  {publisher} {Cambridge University Press},\ \bibinfo {year}
  {2005})\BibitemShut {NoStop}%
\bibitem [{\citenamefont {Maldague}(1978)}]{Maldague}%
  \BibitemOpen
  \bibfield  {author} {\bibinfo {author} {\bibfnamefont {P.~F.}\ \bibnamefont
  {Maldague}},\ }\bibfield  {title} {\bibinfo {title} {Many-body corrections to
  the polarizability of the two-dimensional electron gas},\ }\href
  {https://doi.org/https://doi.org/10.1016/0039-6028(78)90507-1} {\bibfield
  {journal} {\bibinfo  {journal} {Surface Science}\ }\textbf {\bibinfo {volume}
  {73}},\ \bibinfo {pages} {296} (\bibinfo {year} {1978})}\BibitemShut
  {NoStop}%
\bibitem [{\citenamefont {Palik}\ and\ \citenamefont
  {Prucha}(1997)}]{Palik_Handbook}%
  \BibitemOpen
  \bibfield  {author} {\bibinfo {author} {\bibfnamefont {E.~D.}\ \bibnamefont
  {Palik}}\ and\ \bibinfo {author} {\bibfnamefont {E.~J.}\ \bibnamefont
  {Prucha}},\ }\href {https://cds.cern.ch/record/396087} {\emph {\bibinfo
  {title} {Handbook of optical constants of solids}}}\ (\bibinfo  {publisher}
  {Academic Press},\ \bibinfo {address} {Boston, MA},\ \bibinfo {year}
  {1997})\BibitemShut {NoStop}%
\bibitem [{\citenamefont {Svetovoy}\ \emph {et~al.}(2008)\citenamefont
  {Svetovoy}, \citenamefont {van Zwol}, \citenamefont {Palasantzas},\ and\
  \citenamefont {De~Hosson}}]{PhysRevB.77.035439}%
  \BibitemOpen
  \bibfield  {author} {\bibinfo {author} {\bibfnamefont {V.~B.}\ \bibnamefont
  {Svetovoy}}, \bibinfo {author} {\bibfnamefont {P.~J.}\ \bibnamefont {van
  Zwol}}, \bibinfo {author} {\bibfnamefont {G.}~\bibnamefont {Palasantzas}},\
  and\ \bibinfo {author} {\bibfnamefont {J.~T.~M.}\ \bibnamefont {De~Hosson}},\
  }\bibfield  {title} {\bibinfo {title} {Optical properties of gold films and
  the casimir force},\ }\href {https://doi.org/10.1103/PhysRevB.77.035439}
  {\bibfield  {journal} {\bibinfo  {journal} {Phys. Rev. B}\ }\textbf {\bibinfo
  {volume} {77}},\ \bibinfo {pages} {035439} (\bibinfo {year}
  {2008})}\BibitemShut {NoStop}%
\bibitem [{\citenamefont {Fialkovsky}\ and\ \citenamefont
  {Vassilevich}(2012)}]{Fialkovsky2012}%
  \BibitemOpen
  \bibfield  {author} {\bibinfo {author} {\bibfnamefont {I.~V.}\ \bibnamefont
  {Fialkovsky}}\ and\ \bibinfo {author} {\bibfnamefont {D.~V.}\ \bibnamefont
  {Vassilevich}},\ }\bibfield  {title} {\bibinfo {title} {Quantum field theory
  in graphene},\ }\href {https://doi.org/10.1142/S0217751X1260007X} {\bibfield
  {journal} {\bibinfo  {journal} {International Journal of Modern Physics A}\
  }\textbf {\bibinfo {volume} {27}},\ \bibinfo {pages} {1260007} (\bibinfo
  {year} {2012})},\ \Eprint
  {https://arxiv.org/abs/https://doi.org/10.1142/S0217751X1260007X}
  {https://doi.org/10.1142/S0217751X1260007X} \BibitemShut {NoStop}%
\bibitem [{\citenamefont {Ludwig}\ \emph {et~al.}(1994)\citenamefont {Ludwig},
  \citenamefont {Fisher}, \citenamefont {Shankar},\ and\ \citenamefont
  {Grinstein}}]{Ludwig1994}%
  \BibitemOpen
  \bibfield  {author} {\bibinfo {author} {\bibfnamefont {A.~W.~W.}\
  \bibnamefont {Ludwig}}, \bibinfo {author} {\bibfnamefont {M.~P.~A.}\
  \bibnamefont {Fisher}}, \bibinfo {author} {\bibfnamefont {R.}~\bibnamefont
  {Shankar}},\ and\ \bibinfo {author} {\bibfnamefont {G.}~\bibnamefont
  {Grinstein}},\ }\bibfield  {title} {\bibinfo {title} {Integer quantum hall
  transition: An alternative approach and exact results},\ }\href
  {https://doi.org/10.1103/PhysRevB.50.7526} {\bibfield  {journal} {\bibinfo
  {journal} {Phys. Rev. B}\ }\textbf {\bibinfo {volume} {50}},\ \bibinfo
  {pages} {7526} (\bibinfo {year} {1994})}\BibitemShut {NoStop}%
\bibitem [{\citenamefont {Falkovsky}\ and\ \citenamefont
  {Varlamov}(2007)}]{Falkovsky2007}%
  \BibitemOpen
  \bibfield  {author} {\bibinfo {author} {\bibfnamefont {L.~A.}\ \bibnamefont
  {Falkovsky}}\ and\ \bibinfo {author} {\bibfnamefont {A.~A.}\ \bibnamefont
  {Varlamov}},\ }\bibfield  {title} {\bibinfo {title} {Space-time dispersion of
  graphene conductivity},\ }\href {https://doi.org/10.1140/epjb/e2007-00142-3}
  {\bibfield  {journal} {\bibinfo  {journal} {The European Physical Journal B}\
  }\textbf {\bibinfo {volume} {56}},\ \bibinfo {pages} {281} (\bibinfo {year}
  {2007})}\BibitemShut {NoStop}%
\bibitem [{\citenamefont {Falkovsky}(2008)}]{Fialkovsky2008}%
  \BibitemOpen
  \bibfield  {author} {\bibinfo {author} {\bibfnamefont {L.~A.}\ \bibnamefont
  {Falkovsky}},\ }\bibfield  {title} {\bibinfo {title} {Optical properties of
  doped graphene layers},\ }\href {https://doi.org/10.1134/S1063776108030175}
  {\bibfield  {journal} {\bibinfo  {journal} {Journal of Experimental and
  Theoretical Physics}\ }\textbf {\bibinfo {volume} {106}},\ \bibinfo {pages}
  {575} (\bibinfo {year} {2008})}\BibitemShut {NoStop}%
\bibitem [{\citenamefont {Sinitsyn}\ \emph {et~al.}(2006)\citenamefont
  {Sinitsyn}, \citenamefont {Hill}, \citenamefont {Min}, \citenamefont
  {Sinova},\ and\ \citenamefont {MacDonald}}]{MacDonald2006}%
  \BibitemOpen
  \bibfield  {author} {\bibinfo {author} {\bibfnamefont {N.~A.}\ \bibnamefont
  {Sinitsyn}}, \bibinfo {author} {\bibfnamefont {J.~E.}\ \bibnamefont {Hill}},
  \bibinfo {author} {\bibfnamefont {H.}~\bibnamefont {Min}}, \bibinfo {author}
  {\bibfnamefont {J.}~\bibnamefont {Sinova}},\ and\ \bibinfo {author}
  {\bibfnamefont {A.~H.}\ \bibnamefont {MacDonald}},\ }\bibfield  {title}
  {\bibinfo {title} {Charge and spin hall conductivity in metallic graphene},\
  }\href {https://doi.org/10.1103/PhysRevLett.97.106804} {\bibfield  {journal}
  {\bibinfo  {journal} {Phys. Rev. Lett.}\ }\textbf {\bibinfo {volume} {97}},\
  \bibinfo {pages} {106804} (\bibinfo {year} {2006})}\BibitemShut {NoStop}%
\bibitem [{\citenamefont {Xiao}\ and\ \citenamefont
  {Wen}(2013)}]{PhysRevB.88.045442}%
  \BibitemOpen
  \bibfield  {author} {\bibinfo {author} {\bibfnamefont {X.}~\bibnamefont
  {Xiao}}\ and\ \bibinfo {author} {\bibfnamefont {W.}~\bibnamefont {Wen}},\
  }\bibfield  {title} {\bibinfo {title} {Optical conductivities and signatures
  of topological insulators with hexagonal warping},\ }\href
  {https://doi.org/10.1103/PhysRevB.88.045442} {\bibfield  {journal} {\bibinfo
  {journal} {Phys. Rev. B}\ }\textbf {\bibinfo {volume} {88}},\ \bibinfo
  {pages} {045442} (\bibinfo {year} {2013})}\BibitemShut {NoStop}%
\bibitem [{\citenamefont {Tse}\ and\ \citenamefont
  {MacDonald}(2010)}]{WangKong2010}%
  \BibitemOpen
  \bibfield  {author} {\bibinfo {author} {\bibfnamefont {W.-K.}\ \bibnamefont
  {Tse}}\ and\ \bibinfo {author} {\bibfnamefont {A.~H.}\ \bibnamefont
  {MacDonald}},\ }\bibfield  {title} {\bibinfo {title} {Giant magneto-optical
  kerr effect and universal faraday effect in thin-film topological
  insulators},\ }\href {https://doi.org/10.1103/PhysRevLett.105.057401}
  {\bibfield  {journal} {\bibinfo  {journal} {Phys. Rev. Lett.}\ }\textbf
  {\bibinfo {volume} {105}},\ \bibinfo {pages} {057401} (\bibinfo {year}
  {2010})}\BibitemShut {NoStop}%
\bibitem [{\citenamefont {Bimonte}(2010)}]{Bimonte2010KK}%
  \BibitemOpen
  \bibfield  {author} {\bibinfo {author} {\bibfnamefont {G.}~\bibnamefont
  {Bimonte}},\ }\bibfield  {title} {\bibinfo {title} {Generalized
  kramers-kronig transform for casimir effect computations},\ }\href
  {https://doi.org/10.1103/PhysRevA.81.062501} {\bibfield  {journal} {\bibinfo
  {journal} {Phys. Rev. A}\ }\textbf {\bibinfo {volume} {81}},\ \bibinfo
  {pages} {062501} (\bibinfo {year} {2010})}\BibitemShut {NoStop}%
\bibitem [{\citenamefont {Adam}\ \emph {et~al.}(2007)\citenamefont {Adam},
  \citenamefont {Hwang}, \citenamefont {Galitski},\ and\ \citenamefont
  {Sarma}}]{Adam2007}%
  \BibitemOpen
  \bibfield  {author} {\bibinfo {author} {\bibfnamefont {S.}~\bibnamefont
  {Adam}}, \bibinfo {author} {\bibfnamefont {E.~H.}\ \bibnamefont {Hwang}},
  \bibinfo {author} {\bibfnamefont {V.~M.}\ \bibnamefont {Galitski}},\ and\
  \bibinfo {author} {\bibfnamefont {S.~D.}\ \bibnamefont {Sarma}},\ }\bibfield
  {title} {\bibinfo {title} {A self-consistent theory for graphene transport},\
  }\href {https://doi.org/10.1073/pnas.0704772104} {\bibfield  {journal}
  {\bibinfo  {journal} {Proceedings of the National Academy of Sciences}\
  }\textbf {\bibinfo {volume} {104}},\ \bibinfo {pages} {18392} (\bibinfo
  {year} {2007})},\ \Eprint
  {https://arxiv.org/abs/https://www.pnas.org/doi/pdf/10.1073/pnas.0704772104}
  {https://www.pnas.org/doi/pdf/10.1073/pnas.0704772104} \BibitemShut {NoStop}%
\bibitem [{\citenamefont {Szunyogh}\ and\ \citenamefont
  {Weinberger}(1999)}]{Szunyogh1999}%
  \BibitemOpen
  \bibfield  {author} {\bibinfo {author} {\bibfnamefont {L.}~\bibnamefont
  {Szunyogh}}\ and\ \bibinfo {author} {\bibfnamefont {P.}~\bibnamefont
  {Weinberger}},\ }\bibfield  {title} {\bibinfo {title} {Evaluation of the
  optical conductivity tensor in terms of contour integrations},\ }\href
  {https://doi.org/10.1088/0953-8984/11/50/333} {\bibfield  {journal} {\bibinfo
   {journal} {Journal of Physics: Condensed Matter}\ }\textbf {\bibinfo
  {volume} {11}},\ \bibinfo {pages} {10451} (\bibinfo {year}
  {1999})}\BibitemShut {NoStop}%
\bibitem [{\citenamefont {Yanagisawa}\ and\ \citenamefont
  {Shibata}(2004)}]{yanagisawa2004}%
  \BibitemOpen
  \bibfield  {author} {\bibinfo {author} {\bibfnamefont {T.}~\bibnamefont
  {Yanagisawa}}\ and\ \bibinfo {author} {\bibfnamefont {H.}~\bibnamefont
  {Shibata}},\ }\href@noop {} {\bibinfo {title} {Optical properties of
  unconventional superconductors}} (\bibinfo {year} {2004}),\ \Eprint
  {https://arxiv.org/abs/cond-mat/0408054} {arXiv:cond-mat/0408054}
  \BibitemShut {NoStop}%
\bibitem [{\citenamefont {Bimonte}\ \emph {et~al.}(2016)\citenamefont
  {Bimonte}, \citenamefont {L\'opez},\ and\ \citenamefont
  {Decca}}]{PhysRevB.93.184434}%
  \BibitemOpen
  \bibfield  {author} {\bibinfo {author} {\bibfnamefont {G.}~\bibnamefont
  {Bimonte}}, \bibinfo {author} {\bibfnamefont {D.}~\bibnamefont {L\'opez}},\
  and\ \bibinfo {author} {\bibfnamefont {R.~S.}\ \bibnamefont {Decca}},\
  }\bibfield  {title} {\bibinfo {title} {Isoelectronic determination of the
  thermal casimir force},\ }\href {https://doi.org/10.1103/PhysRevB.93.184434}
  {\bibfield  {journal} {\bibinfo  {journal} {Phys. Rev. B}\ }\textbf {\bibinfo
  {volume} {93}},\ \bibinfo {pages} {184434} (\bibinfo {year}
  {2016})}\BibitemShut {NoStop}%
\bibitem [{\citenamefont {Bimonte}\ \emph {et~al.}(2021)\citenamefont
  {Bimonte}, \citenamefont {Spreng}, \citenamefont {Maia~Neto}, \citenamefont
  {Ingold}, \citenamefont {Klimchitskaya}, \citenamefont {Mostepanenko},\ and\
  \citenamefont {Decca}}]{universe7040093}%
  \BibitemOpen
  \bibfield  {author} {\bibinfo {author} {\bibfnamefont {G.}~\bibnamefont
  {Bimonte}}, \bibinfo {author} {\bibfnamefont {B.}~\bibnamefont {Spreng}},
  \bibinfo {author} {\bibfnamefont {P.~A.}\ \bibnamefont {Maia~Neto}}, \bibinfo
  {author} {\bibfnamefont {G.-L.}\ \bibnamefont {Ingold}}, \bibinfo {author}
  {\bibfnamefont {G.~L.}\ \bibnamefont {Klimchitskaya}}, \bibinfo {author}
  {\bibfnamefont {V.~M.}\ \bibnamefont {Mostepanenko}},\ and\ \bibinfo {author}
  {\bibfnamefont {R.~S.}\ \bibnamefont {Decca}},\ }\bibfield  {title} {\bibinfo
  {title} {Measurement of the casimir force between 0.2 and 8 {$\mu$}m:
  Experimental procedures and comparison with theory},\ }\bibfield  {journal}
  {\bibinfo  {journal} {Universe}\ }\textbf {\bibinfo {volume} {7}},\ \href
  {https://doi.org/10.3390/universe7040093} {10.3390/universe7040093} (\bibinfo
  {year} {2021})\BibitemShut {NoStop}%
\bibitem [{\citenamefont {Klimchitskaya}\ and\ \citenamefont
  {Mostepanenko}(2022)}]{Klimchitskaya_2022}%
  \BibitemOpen
  \bibfield  {author} {\bibinfo {author} {\bibfnamefont {G.~L.}\ \bibnamefont
  {Klimchitskaya}}\ and\ \bibinfo {author} {\bibfnamefont {V.~M.}\ \bibnamefont
  {Mostepanenko}},\ }\bibfield  {title} {\bibinfo {title} {Current status of
  the problem of thermal casimir force},\ }\bibfield  {journal} {\bibinfo
  {journal} {International Journal of Modern Physics A}\ }\textbf {\bibinfo
  {volume} {37}},\ \href {https://doi.org/10.1142/s0217751x22410020}
  {10.1142/s0217751x22410020} (\bibinfo {year} {2022})\BibitemShut {NoStop}%
\bibitem [{\citenamefont {Sushkov}\ \emph {et~al.}(2011)\citenamefont
  {Sushkov}, \citenamefont {Kim}, \citenamefont {Dalvit},\ and\ \citenamefont
  {Lamoreaux}}]{Sushkov_2011}%
  \BibitemOpen
  \bibfield  {author} {\bibinfo {author} {\bibfnamefont {A.~O.}\ \bibnamefont
  {Sushkov}}, \bibinfo {author} {\bibfnamefont {W.~J.}\ \bibnamefont {Kim}},
  \bibinfo {author} {\bibfnamefont {D.~A.~R.}\ \bibnamefont {Dalvit}},\ and\
  \bibinfo {author} {\bibfnamefont {S.~K.}\ \bibnamefont {Lamoreaux}},\
  }\bibfield  {title} {\bibinfo {title} {Observation of the thermal casimir
  force},\ }\href {https://doi.org/10.1038/nphys1909} {\bibfield  {journal}
  {\bibinfo  {journal} {Nature Physics}\ }\textbf {\bibinfo {volume} {7}},\
  \bibinfo {pages} {230–233} (\bibinfo {year} {2011})}\BibitemShut {NoStop}%
\bibitem [{\citenamefont {Jablan}\ \emph
  {et~al.}(2009{\natexlab{b}})\citenamefont {Jablan}, \citenamefont {Buljan},\
  and\ \citenamefont {Solja\ifmmode \check{c}\else
  \v{c}\fi{}i\ifmmode~\acute{c}\else \'{c}\fi{}}}]{PhysRevB.80.245435}%
  \BibitemOpen
  \bibfield  {author} {\bibinfo {author} {\bibfnamefont {M.}~\bibnamefont
  {Jablan}}, \bibinfo {author} {\bibfnamefont {H.}~\bibnamefont {Buljan}},\
  and\ \bibinfo {author} {\bibfnamefont {M.}~\bibnamefont {Solja\ifmmode
  \check{c}\else \v{c}\fi{}i\ifmmode~\acute{c}\else \'{c}\fi{}}},\ }\bibfield
  {title} {\bibinfo {title} {Plasmonics in graphene at infrared frequencies},\
  }\href {https://doi.org/10.1103/PhysRevB.80.245435} {\bibfield  {journal}
  {\bibinfo  {journal} {Phys. Rev. B}\ }\textbf {\bibinfo {volume} {80}},\
  \bibinfo {pages} {245435} (\bibinfo {year} {2009}{\natexlab{b}})}\BibitemShut
  {NoStop}%
\bibitem [{\citenamefont {Zhu}\ \emph {et~al.}(2021)\citenamefont {Zhu},
  \citenamefont {Antezza},\ and\ \citenamefont {Wang}}]{PhysRevB.103.125421}%
  \BibitemOpen
  \bibfield  {author} {\bibinfo {author} {\bibfnamefont {T.}~\bibnamefont
  {Zhu}}, \bibinfo {author} {\bibfnamefont {M.}~\bibnamefont {Antezza}},\ and\
  \bibinfo {author} {\bibfnamefont {J.-S.}\ \bibnamefont {Wang}},\ }\bibfield
  {title} {\bibinfo {title} {Dynamical polarizability of graphene with spatial
  dispersion},\ }\href {https://doi.org/10.1103/PhysRevB.103.125421} {\bibfield
   {journal} {\bibinfo  {journal} {Phys. Rev. B}\ }\textbf {\bibinfo {volume}
  {103}},\ \bibinfo {pages} {125421} (\bibinfo {year} {2021})}\BibitemShut
  {NoStop}%
\bibitem [{\citenamefont {Wang}\ \emph {et~al.}(2021)\citenamefont {Wang},
  \citenamefont {Tang}, \citenamefont {Ng}, \citenamefont {Messina},
  \citenamefont {Guizal}, \citenamefont {Crosse}, \citenamefont {Antezza},
  \citenamefont {Chan},\ and\ \citenamefont {Chan}}]{Nature_2021}%
  \BibitemOpen
  \bibfield  {author} {\bibinfo {author} {\bibfnamefont {M.}~\bibnamefont
  {Wang}}, \bibinfo {author} {\bibfnamefont {L.}~\bibnamefont {Tang}}, \bibinfo
  {author} {\bibfnamefont {C.~Y.}\ \bibnamefont {Ng}}, \bibinfo {author}
  {\bibfnamefont {R.}~\bibnamefont {Messina}}, \bibinfo {author} {\bibfnamefont
  {B.}~\bibnamefont {Guizal}}, \bibinfo {author} {\bibfnamefont {J.~A.}\
  \bibnamefont {Crosse}}, \bibinfo {author} {\bibfnamefont {M.}~\bibnamefont
  {Antezza}}, \bibinfo {author} {\bibfnamefont {C.~T.}\ \bibnamefont {Chan}},\
  and\ \bibinfo {author} {\bibfnamefont {H.~B.}\ \bibnamefont {Chan}},\
  }\bibfield  {title} {\bibinfo {title} {Strong geometry dependence of the
  casimir force between interpenetrated rectangular gratings},\ }\href
  {https://doi.org/10.1038/s41467-021-20891-4} {\bibfield  {journal} {\bibinfo
  {journal} {Nature Communications}\ }\textbf {\bibinfo {volume} {12}},\
  \bibinfo {pages} {600} (\bibinfo {year} {2021})}\BibitemShut {NoStop}%
\end{thebibliography}
\providecommand{\noopsort}[1]{}\providecommand{\singleletter}[1]{#1}%

\appendix

\begin{widetext}
\section{Low frequency limit of the Kubo electric conductivity and of the QFT based model for em response for graphene.}
In section \ref{sect:Drude_vs_Plasma} we analyzed the difference between the Drude and Plasma prescription on the calculation of the CLF gradient. To this extent we need to apply such prescriptions to the $n=0$ Matsubara term, and hence we need to calculate the low frequency limit of the graphene non-local Kubo conductivity and of the QFT conductivity. In this appendix we derive such limits.

\subsection{Low frequency limit of non-local Kubo conductivity}\label{App_Kubo}
Here we derive the low frequency limit of the graphene non-local Kubo conductivity \cite{non-local_Graphene_Lilia_Pablo}\cite{PabloMauroComparisonKuboQFT2024}. For the Plasma prescription, we will also need to discard losses, hence we set $\xi\to0$ with $\Gamma\to0$ while for the Drude prescription we set $\xi\to0$ with $\Gamma\neq 0$. In the $T=0$ limit we have:
\begin{eqnarray}\label{static_limit_sigma}
\lim_{\Omega\to 0}\sigma_{T}^{\rm{K}}(\Omega,k_{\surf}) & = & \frac{\alpha c}{\pi}\frac{\tilde{K}^{2}}{\tilde{k}_{\surf}\sqrt{ \tilde{K}^{2} - \tilde{k}_{\surf}^{2}}}\Theta\left( \tilde{K} - \tilde{k}_{\surf} \right)\Theta(\abs{\mu} - \abs{\Delta})\nonumber\\
\sigma_{L}^{\rm{K}}(\Omega,k_{\surf}) & \underset{\Omega\to 0}{\approx} & 0 + \Omega\sigma_{L,1}(k_{\surf}) \nonumber\\
\lim_{\Omega\to 0}\sigma_{H}^{\rm{K}}(\Omega,k_{\surf}) & = & - \frac{\alpha c}{2\pi}\frac{\eta\Delta}{\tilde{k}_{\surf}}\left[ \Theta(\abs{\Delta} - \abs{\mu})\tan^{-1}\left(\frac{\tilde{k}_{\surf}}{2\abs{\Delta}}\right)
 + \Theta(\abs{\mu} - \abs{\Delta})\cos^{-1}\left(\frac{2\abs{\mu}}{\sqrt{ \tilde{k}_{\surf}^{2} + 4\Delta^{2} }}\right)
\right]\nonumber\\
\sigma_{L,1}(k_{\surf}) & = & \frac{\alpha c}{2\pi}\frac{\hbar}{2\tilde{k}_{\surf}}\left[
\frac{\abs{\Delta}}{\tilde{k}_{\surf}} + \frac{\tilde{k}_{\surf}^{2} - 4\Delta^{2}}{2\tilde{k}_{\surf}^{2}}\tan^{-1}\left(\frac{\tilde{k}_{\surf}}{2\abs{\Delta}}\right) + \Theta(\abs{\mu} - \abs{\Delta})\left( 2\frac{\abs{\mu} - \abs{\Delta}}{\tilde{k}_{\surf}} + \frac{4\Delta^{2} - \tilde{k}_{\surf}^{2}}{4\tilde{k}_{\surf}^{2}} X
\right) \right]\nonumber\\
X & = & \Theta\left( \tilde{k}_{\surf}^{2} - \tilde{K}^{2} \right)\tan^{-1}\left(2\frac{\abs{\mu}\tilde{k}_{\surf} - \abs{\Delta}\sqrt{ \tilde{k}_{\surf}^{2} - \tilde{K}^{2} }}{4\abs{\Delta}\abs{\mu} + \tilde{k}_{\surf}\sqrt{ \tilde{k}_{\surf}^{2} - \tilde{K}^{2} }}\right) + \Theta\left( \tilde{K}^{2} - \tilde{k}_{\surf}^{2} \right)\tan^{-1}\left(\frac{\tilde{k}_{\surf}}{2\abs{\Delta}}\right)
\end{eqnarray}
where $\Omega = \omega + \ii\Gamma$, $\tilde{k}_{\surf} = \hbar v_{F}k_{\surf}$ and $\tilde{K} = \hbar v_{F}K = 2\sqrt{\mu^{2} - \Delta^{2}}$. We have to apply the Maldague formula, given in \Eq{Maldague_Formula} to obtain the corresponding finite $T$ results.

\subsection{Low frequency limit of the QFT Polarization Operator}\label{Appendix_QFTb}
Here we derive the low frequency limit of the graphene polarization operator for the QFT model \cite{Bordag2015}. We start with the definition of the auxiliary functions
\begin{eqnarray}
\Psi(x, y) = 2\left[y\sqrt{ 1 + x^{2} - y^{2} } + \left( 1 - x^{2} \right) \tan^{-1}\left(\tfrac{\sqrt{ 1 + x^{2} - y^{2} } }{y}\right) \right],
\end{eqnarray}
\begin{eqnarray}
\Psi(x) = \dlim_{y\to x}\Psi(x, y) = 2\left[ x + (1 - x^{2})\tan^{-1}\left(\frac{1}{x}\right)\right],
\end{eqnarray}
From \cite{PabloMauroComparisonKuboQFT2024}, the longitudinal Polarization term is
\begin{eqnarray}
\Pi_{L}^{\rm{QFT}}
& = & 
\xi\frac{\alpha c\lambda\Psi\left(\delta\right)}{4\pi} + \xi\frac{\alpha c\Xi\tilde{\theta}_{z}}{\pi\tilde{k}_{\surf}^{2}} \int_{\delta}^{\infty}\dd u N_{\mu}\left(\tfrac{\tilde{\theta}_{z}}{2}u\right) \left(1-\Real{\frac{ 1 - u^{2} + 2\ii\lambda u }{\sqrt{ 1 - u^{2} + 2\ii\lambda u + \left( 1 - \lambda^{2}\right) \delta^{2} } } }\right),\label{Def:PiL_QFTb}\\
\Pi_{L}^{\rm{QFT}}
& \underset{\xi\to0}{\approx} & \xi\frac{\alpha c}{4\pi}\frac{\Xi}{\tilde{k}_{\surf}}
\left[ \Psi(x) + 4\int_{x}^{\infty}\dd u N_{\mu}\left(\tfrac{\tilde{k}_{\surf}}{2}u\right) \left(1-\Real{\frac{ 1 - u^{2} }{\sqrt{ 1 - u^{2} + x^{2} } } }\right)\right],\label{Def:Lim_xito0_PiL_QFTb}
\end{eqnarray}
where $\Xi = \hbar\xi$, $\tilde{k}_{\surf} = \hbar v_{F}k_{\surf}$, $\tilde{\theta}_{z} = \sqrt{ \Xi^{2} + \tilde{k}_{\surf}^{2} }$, $\delta = \tfrac{2\Delta}{\tilde{\theta}_{z}}$, $\lambda = \frac{\Xi}{\tilde{\theta}_{z}}$, $x = \dlim_{\hbar\xi\to0}\delta = \tfrac{2\Delta}{\tilde{k}_{\surf}}$, $y = \tfrac{2\abs{\mu}}{\tilde{k}_{\surf}}$, and
\begin{eqnarray}
N_{\mu}(\epsilon) = \frac{1}{e^{\beta\left(\epsilon + \mu\right)} + 1 } + \frac{1}{e^{\beta\left(\epsilon - \mu\right)} + 1 }.
\end{eqnarray}
The relevant contribution of the conductivity to the Fresnel reflection coefficients in the $\xi\to0$ limit is
\begin{eqnarray}\label{Def:Lim_xito0_Pi_L}
\Pi_{L}^{\rm{QFT}}(\xi)
& \underset{\xi\to0}{\approx} & \Pi_{L,2}^{\rm{QFT}}\xi^{2},
\end{eqnarray}
where, in the $T=0$ limit we have
\begin{eqnarray}\label{Def_Mostepanenko_Pi_L_xi_0}
\Pi_{L,2}^{\rm{QFT}}
& = & \frac{\alpha c}{4\pi}\frac{\hbar}{\tilde{k}_{\surf}}
\Bigg[ \Psi(x)\Theta(x-y) + \Theta(y-x)\Real{4y\left( 1-\sqrt{ 1 + x^{2} - y^{2} } \right) + \Psi(x,y)}\Bigg].
\end{eqnarray}
Following the same procedure, now we compute the transversal Polarization operator term \cite{PabloMauroComparisonKuboQFT2024}
\begin{eqnarray}\label{Anomalous_Plasmaterm_PiT_QFTb}
\Pi_{T}^{\rm{QFT}}
& = & \xi\frac{\alpha c\Psi(\delta)}{4\pi\lambda} - \xi\frac{\alpha c\Xi\tilde{\theta}_{z}}{\pi\tilde{k}_{\surf}^{2}} \int_{\delta}^{\infty}\dd u\,N_{\mu}\left(\tfrac{\tilde{\theta}_{z}}{2}u\right)\left( 1 -\Real{\frac{\left( 1 + \ii\lambda^{-1}u \right)^{2} + \left( \lambda^{-2} - 1 \right)\delta^{2} }{\sqrt{ 1 - u^{2} + 2\ii\lambda u + \delta^{2}\left( 1 - \lambda^{2} \right)}}}\right)\nonumber\\
\Pi_{T}^{\rm{QFT}} & \underset{\xi\to0}{\approx} & \xi\frac{\alpha c}{4\pi}\left[\frac{\tilde{k}_{\surf}}{\Xi}\Psi(x) + 4\int_{x}^{\infty}\dd u\,N_{\mu}\left(\tfrac{\tilde{k}_{\surf}}{2}u\right)\left( \frac{\tilde{k}_{\surf}}{\Xi}\Real{\frac{ x^{2} - u^{2} }{\sqrt{ 1 - u^{2} + x^{2} }}} + \Imag{\frac{u \left( 2 - u^{2} + x^{2} \right)}{\left( 1 - u^{2} + x^{2} \right)^{3/2}}} \right)\right].
\end{eqnarray}
We can obtain closed formulas for the different integrals in the $T\to0$ limit as
\begin{eqnarray}
\dlim_{T\to0}I_{T,1}
& = & \dlim_{T\to0}\int_{x}^{\infty}\dd u\,N_{\mu}\left(\tfrac{\tilde{k}_{\surf}}{2}u\right)\Real{\frac{ x^{2} - u^{2} }{\sqrt{ 1 - u^{2} + x^{2} }}}\nonumber\\
& = & \Theta(y - x)\frac{1}{4}\left[ \Theta\left( 1 + x^{2} - y^{2} \right) \Psi(x, y) - \Psi(x) \right],
\end{eqnarray}
\begin{eqnarray}
\dlim_{T\to0}I_{T,2}
& = & \dlim_{T\to0}\int_{x}^{\infty}\dd u\,N_{\mu}\left(\tfrac{\tilde{k}_{\surf}}{2}u\right)\Imag{ \frac{u\left( 2 - u^{2} + x^{2} \right)}{\left( 1 - u^{2} + x^{2} \right)^{3/2}} }\nonumber\\
& = & \frac{ \tilde{K}^{2} }{\tilde{k}_{\surf}\sqrt{ \tilde{K}^{2} - \tilde{k}_{\surf}^{2} } }\Theta\left( \tilde{K} - \tilde{k}_{\surf} \right)\Theta(\abs{\mu} - \abs{\Delta}).
\end{eqnarray}
Expression $I_{T,2}$ coincides with the zero temperature zero frequency limit of the transversal conductivity derived from the Kubo formalism, while $I_{T,1}$ is sometimes interpreted as a dissipation-less Plasma term of the transversal conductivity of the QFT model when the assumption $\sigma_{ij}^{\rm{NR}}(\omega,\bm{k}_{\surf}) = \Pi_{ij}^{\rm{QFT}}(\omega,\bm{k}_{\surf})/(-\ii\omega)$ is used to obtain the conductivity of the system instead of the use of the correct Luttinger formula (\Eq{Luttinger_sub}). Here we interpret this term as a magnetic response. This magnetic term is
\begin{eqnarray}\label{Def:Lim_xito0_Pi_T}
\Pi_{T}^{\rm{QFT}}(\xi)
& \underset{\xi\to0}{\approx} & \Pi_{T,0}^{\rm{QFT}} + \Pi_{T,1}^{\rm{QFT}}\xi,
\end{eqnarray}
\begin{eqnarray}\label{Def:Plasma_Pi_T0}
\Pi_{T,0}^{\rm{QFT}}
& = & \frac{\alpha c\tilde{k}_{\surf}}{4\pi\hbar}\left(\Psi(x) + 4\int_{x}^{\infty}\dd u\,N_{\mu}\left(\tfrac{\tilde{k}_{\surf}}{2}u\right)\Real{\frac{ x^{2} - u^{2} }{\sqrt{ 1 - u^{2} + x^{2} }}} \right).
\end{eqnarray}
\begin{eqnarray}\label{Def:Plasma_Pi_T1}
\Pi_{T,1}^{\rm{QFT}}
& = & \frac{\alpha c}{\pi}\int_{x}^{\infty}\dd u\,N_{\mu}\left(\tfrac{\tilde{k}_{\surf}}{2}u\right)\Imag{ \frac{u\left( 2 - u^{2} + x^{2} \right)}{\left( 1 - u^{2} + x^{2} \right)^{3/2}} }.
\end{eqnarray}
In the zero temperature limit, using $\tilde{k}_{\surf} = \hbar v_{F}k_{\surf}$ we obtain
\begin{eqnarray}
\dlim_{T\to0}\Pi_{T,0}^{\rm{QFT}}
& = & \frac{\alpha c\tilde{k}_{\surf}}{4\pi\hbar}\left[ \Theta(x-y)\Psi(x) + \Theta(y-x)\Theta\left( \sqrt{1 + x^{2}} - y \right) \Psi(x, y) \right].
\end{eqnarray}
\begin{eqnarray}\label{Def:Plasma_Pi_T1}
\dlim_{T\to0}\Pi_{T,1}^{\rm{QFT}}
& = & \frac{\alpha c}{\pi}\frac{ \tilde{K}^{2} }{\tilde{k}_{\surf}\sqrt{ \tilde{K}^{2} - \tilde{k}_{\surf}^{2} } }\Theta\left( \tilde{K} - \tilde{k}_{\surf} \right)\Theta(\abs{\mu} - \abs{\Delta}).
\end{eqnarray}
The direct consequence of this result is that the $n=0$ Matsubara frequency term of the Lifshitz formula applied to graphene is different by using this $\rm{QFT}$ model and the Kubo model showed in App.~\ref{App_Kubo}.
Now, taking into account the effect of losses, the QFT model is modified into (see \Eq{QFT_response_with_dissipation})
\begin{eqnarray}
\mean{J_{i}^{total}(\omega,\bm{k})} = -\Pi_{ij}(\omega + \ii\Gamma,\bm{k})A^{j}(\omega,\bm{k})    
\end{eqnarray}
In this case, in the limit of small $\xi$ limit, we have
\begin{eqnarray}
\Pi_{L}^{\rm{QFT}}(\xi+\Gamma)
& \underset{\xi\to0}{=} & \Pi_{L}^{\rm{QFT}}(\Gamma)\approx \Pi_{L,2}^{\rm{QFT}}\,\Gamma^{2},
\end{eqnarray}
\begin{eqnarray}
\Pi_{T}^{\rm{QFT}}(\xi+\Gamma)
& \underset{\xi\to0}{=} & \Pi_{T}^{\rm{QFT}}(\Gamma) \approx \Pi_{T,0}^{\rm{QFT}} + \Pi_{T,1}^{\rm{QFT}}\,\Gamma.
\end{eqnarray}

\section{Comparison of the experiment with the local Kubo and QFT models}\label{Appendix_Comparison_Local_NR}

In this section we compare the experimental results with calculation of the CLF gradient using the local Kubo conductivity and QFT polarization tensor.
\subsection{Comparison of the experiment using the local Kubo conductivity }\label{Appendix_Comparison_Local}
By using the local Kubo model given in \Eq{local_sigma_realw} \cite{non-local_Graphene_Lilia_Pablo}\cite{PabloMauroComparisonKuboQFT2024} in the CLF gradient $G(d)$ we find the results of figure \Fig{Fig_Comparison_SILocal_Model} and \Fig{Fig_RelError_Local}. We can see the same agreement with the experiments as obtained using the non-local Kubo model shown in \Fig{Fig_Comparison_NOLocal_Model}, \Fig{Fig_RelError_NoLocalPRL} and \Fig{Fig_RelError_NoLocalPRB}.
\begin{figure}[H]
\centering
\begin{subfigure}[b]{0.49\textwidth}
   \centering
   \includegraphics[width=\textwidth]{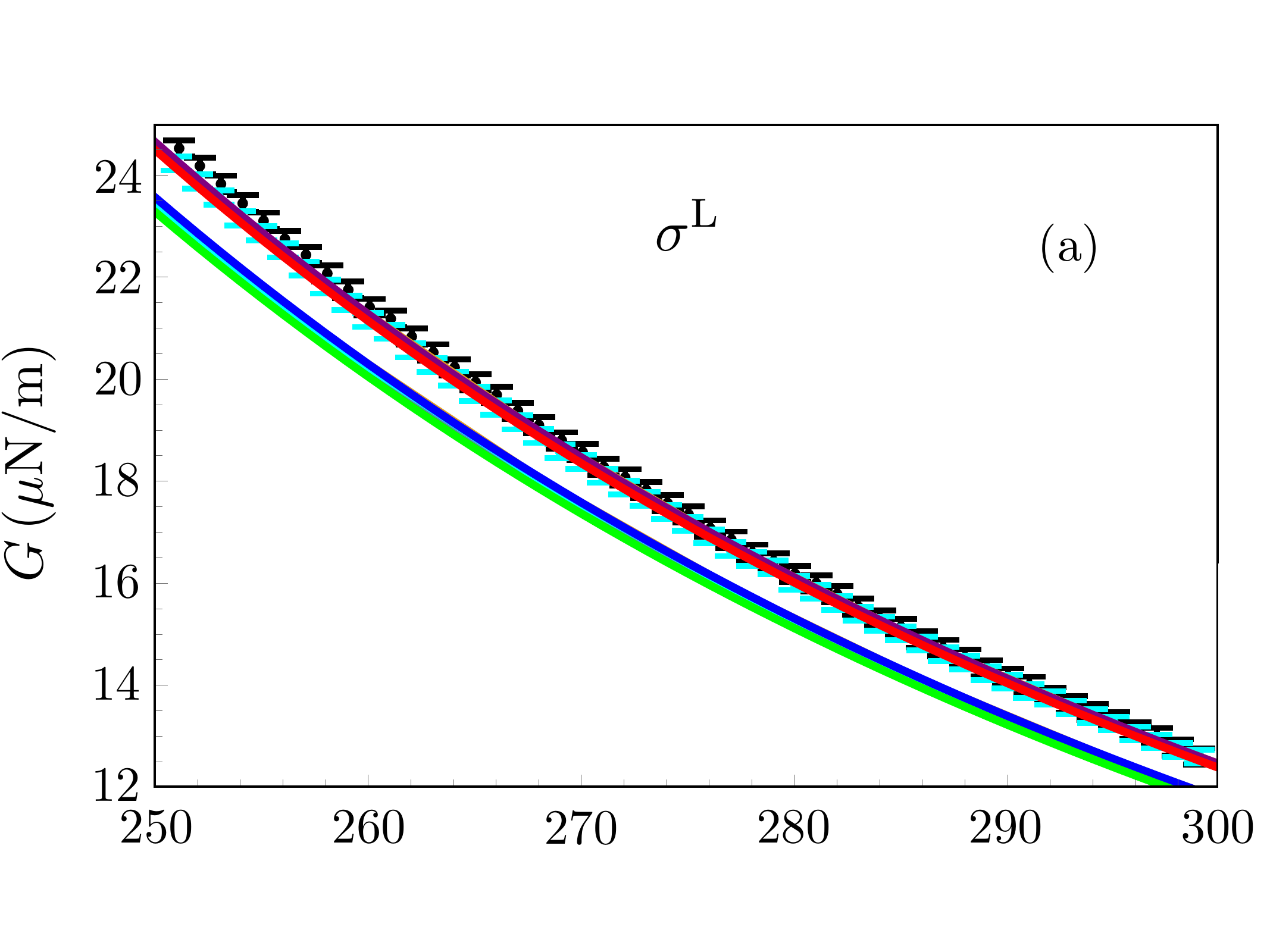}
\end{subfigure}
\hfill
\begin{subfigure}[b]{0.49\textwidth}
   \centering
   \includegraphics[width=\textwidth]{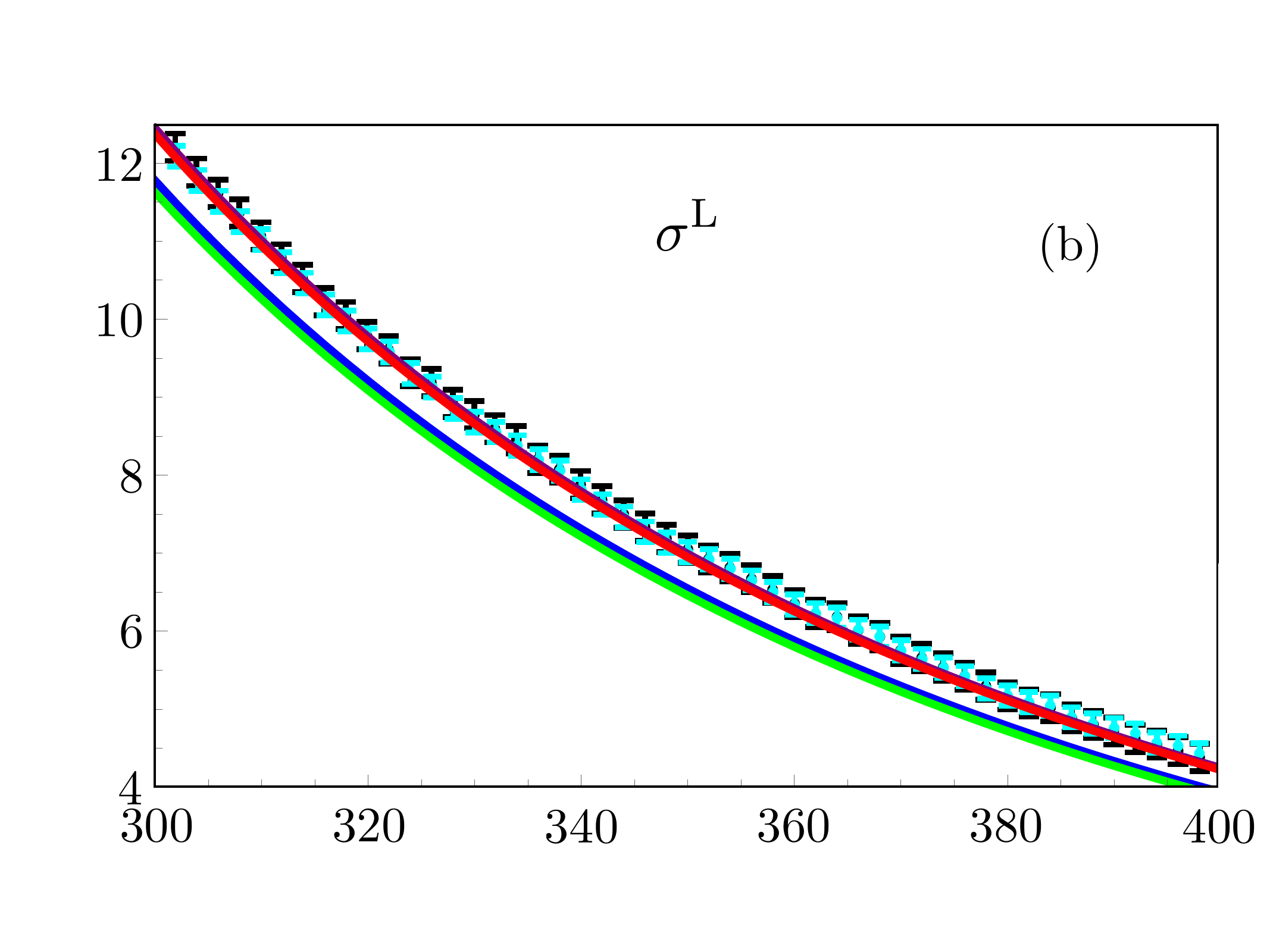}
\end{subfigure}
\hfill
\begin{subfigure}[b]{0.49\textwidth}
   \centering
   \includegraphics[width=\textwidth]{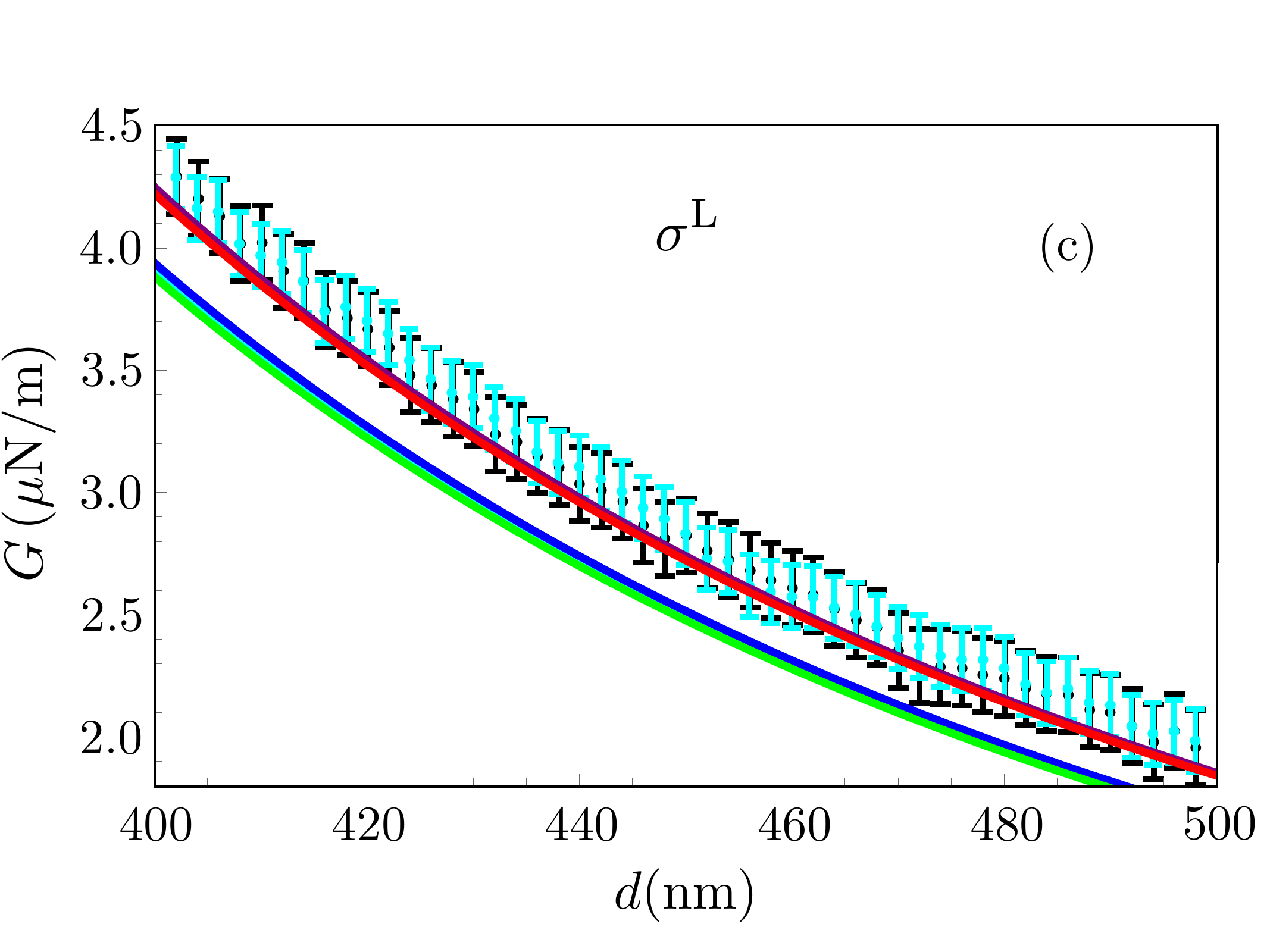}
\end{subfigure}
\hfill
\begin{subfigure}[b]{0.49\textwidth}
   \centering
   \includegraphics[width=\textwidth]{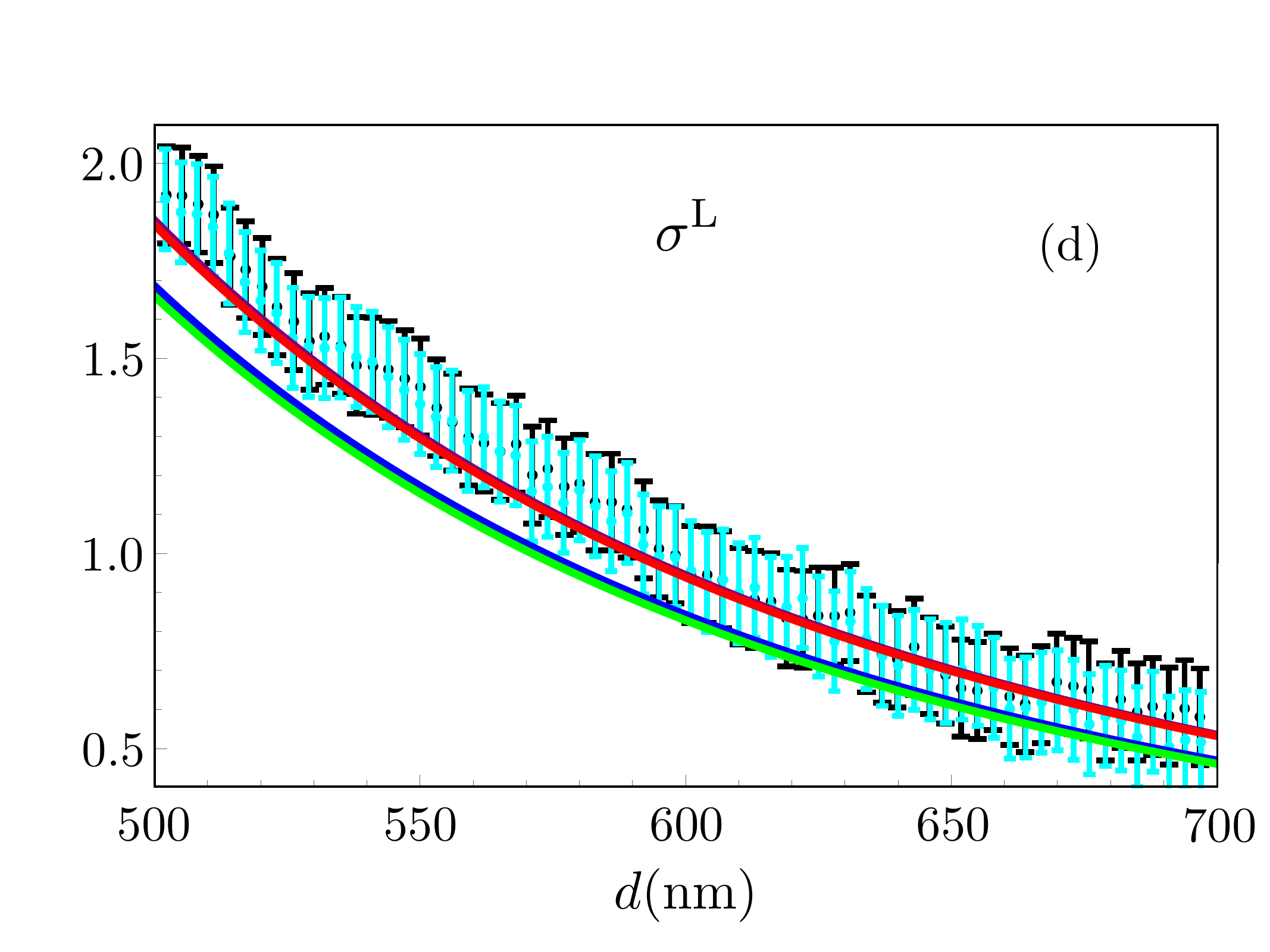}
\end{subfigure}
\caption{  Same as for \Fig{Fig_Comparison_NOLocal_Model}, where instead of $\sigma^{\rm{K}}$ we use here the Local Kubo conductivity model for graphene $\sigma^{\rm{L}}$ using \Eq{local_sigma_realw} and \Eq{Maldague_Formula} to obtain the finite $T$ results.
}
\label{Fig_Comparison_SILocal_Model}
\end{figure}

\begin{figure}[H]
\centering
\includegraphics[width=0.49\linewidth]{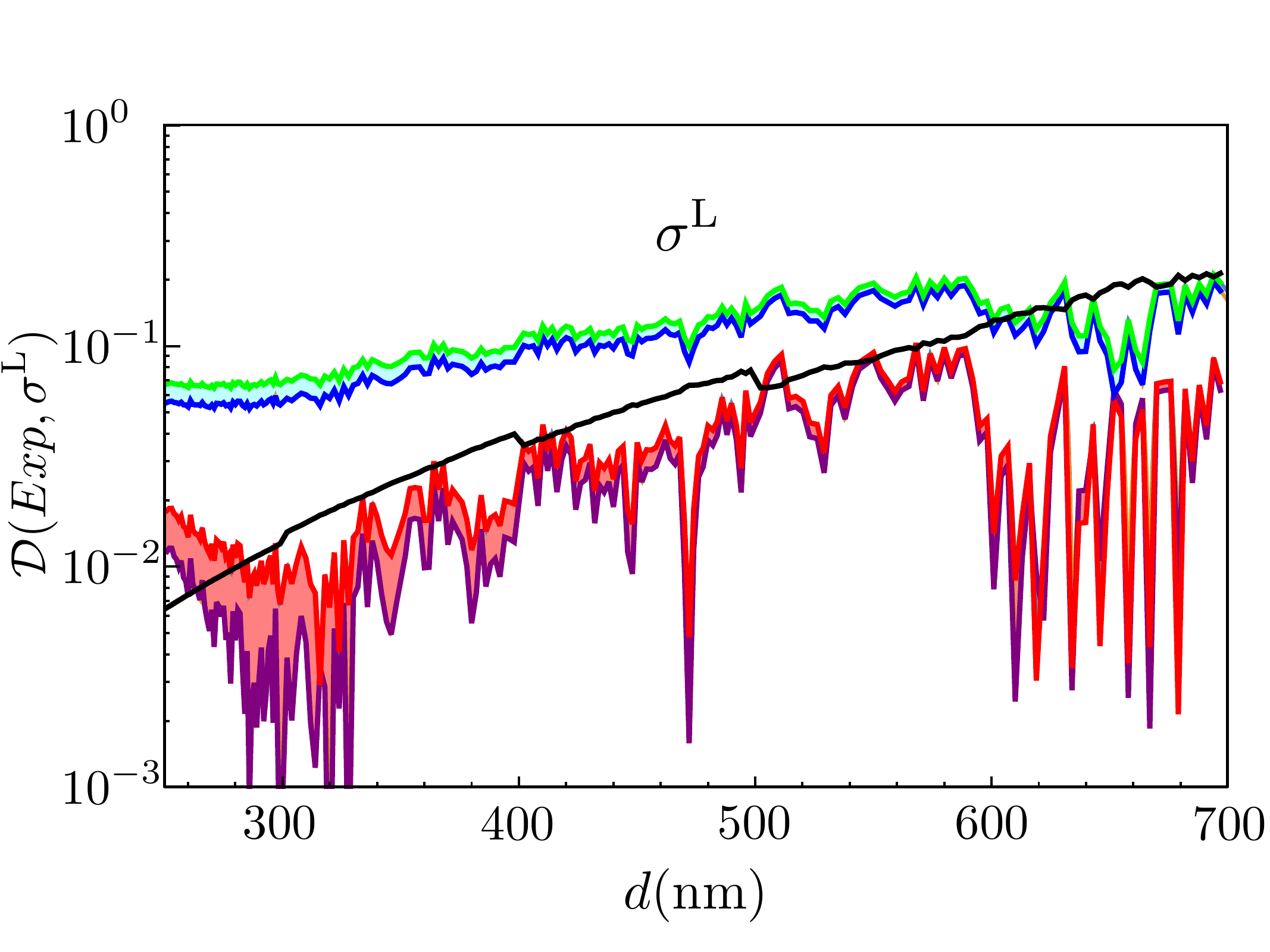}
\includegraphics[width=0.49\linewidth]{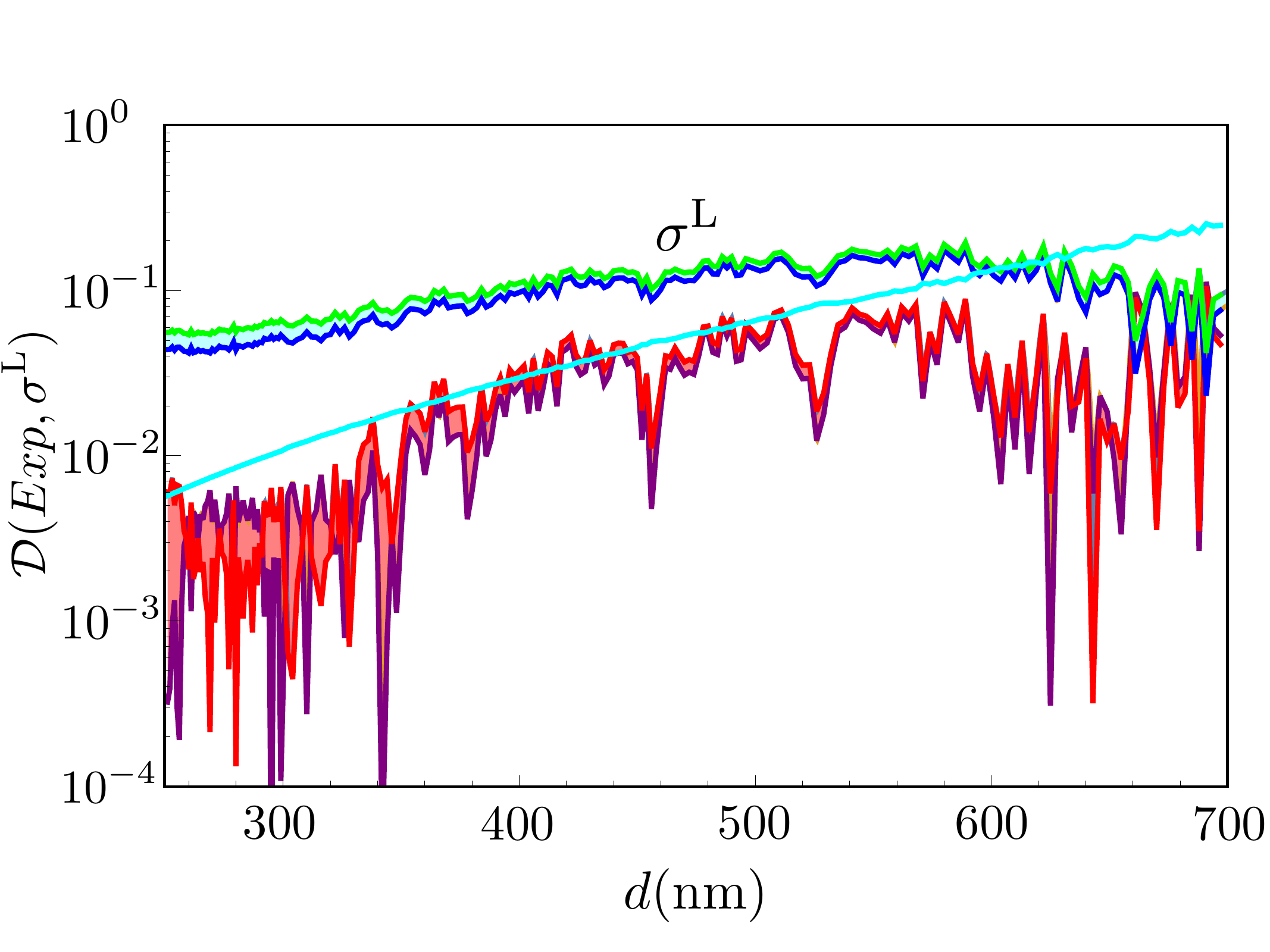}
\caption{ Same as for \Fig{Fig_RelError_NoLocalPRL} and for \Fig{Fig_RelError_NoLocalPRB}, where instead of $\sigma^{\rm{K}}$ we use here the Local Kubo conductivity model for graphene $\sigma^{\rm{L}}$ using \Eq{local_sigma_realw} and \Eq{Maldague_Formula} to obtain the finite $T$ results. Theoretical values below this black (cyan) curve show cases when the theoretical results of \cite{PRL_Mohideen} (\cite{PRB_Mohideen}) are inside the experimental error-bars. We can observe a small discrepancy at short distances for the results of \cite{PRL_Mohideen} that disappears for the results of \cite{PRB_Mohideen}.
}
\label{Fig_RelError_Local}
\end{figure}

\subsection{Comparison of the experiment using the QFT model}\label{Appendix_Comparison_NR}
By using the QFT model \cite{PRL_Mohideen,PRB_Mohideen,PabloMauroComparisonKuboQFT2024,Bordag2015} in the CLF gradient $G(d)$ we find the results of figure \Fig{Fig_Comparison_NRLocal_Model} and \Fig{Fig_RelError_NR}. We can see the same agreement with the experiments as obtained using the non-local Kubo model shown in \Fig{Fig_Comparison_NOLocal_Model}, \Fig{Fig_RelError_NoLocalPRL} and \Fig{Fig_RelError_NoLocalPRB}.
\begin{figure}[H]
\centering
\begin{subfigure}[b]{0.49\textwidth}
   \centering
   \includegraphics[width=\textwidth]{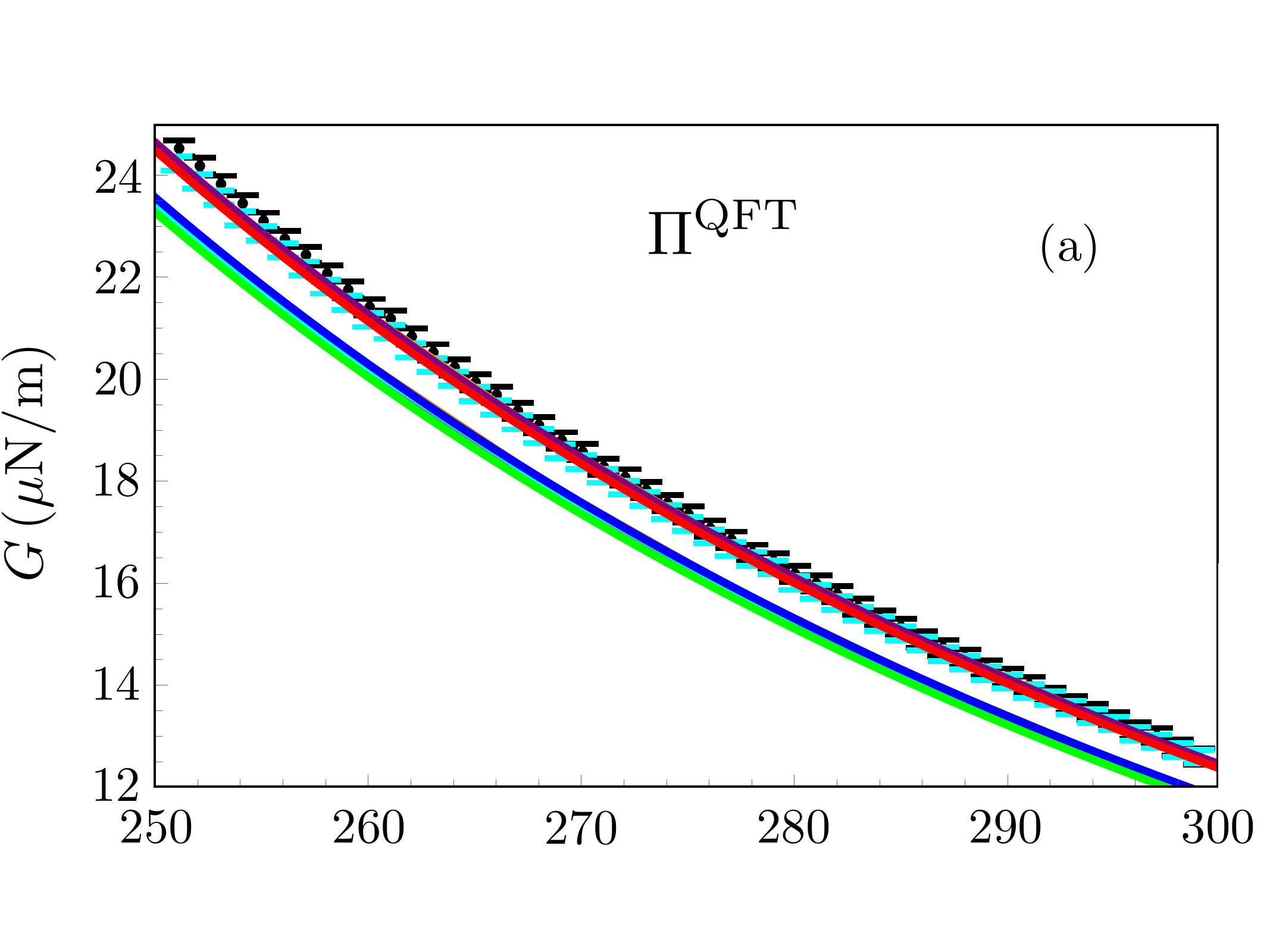}
\end{subfigure}
\hfill
\begin{subfigure}[b]{0.49\textwidth}
   \centering
   \includegraphics[width=\textwidth]{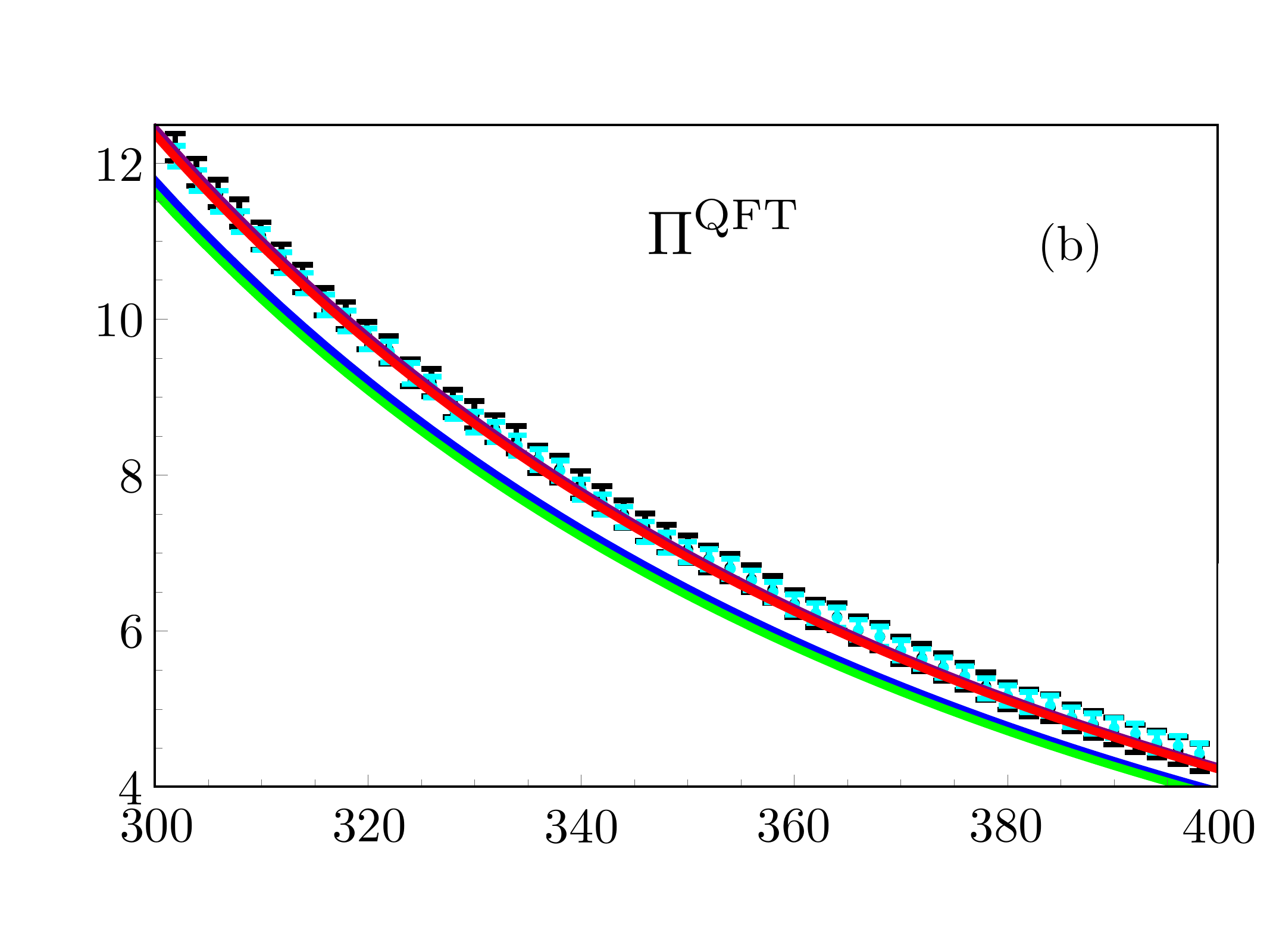}
\end{subfigure}
\hfill
\begin{subfigure}[b]{0.49\textwidth}
   \centering
   \includegraphics[width=\textwidth]{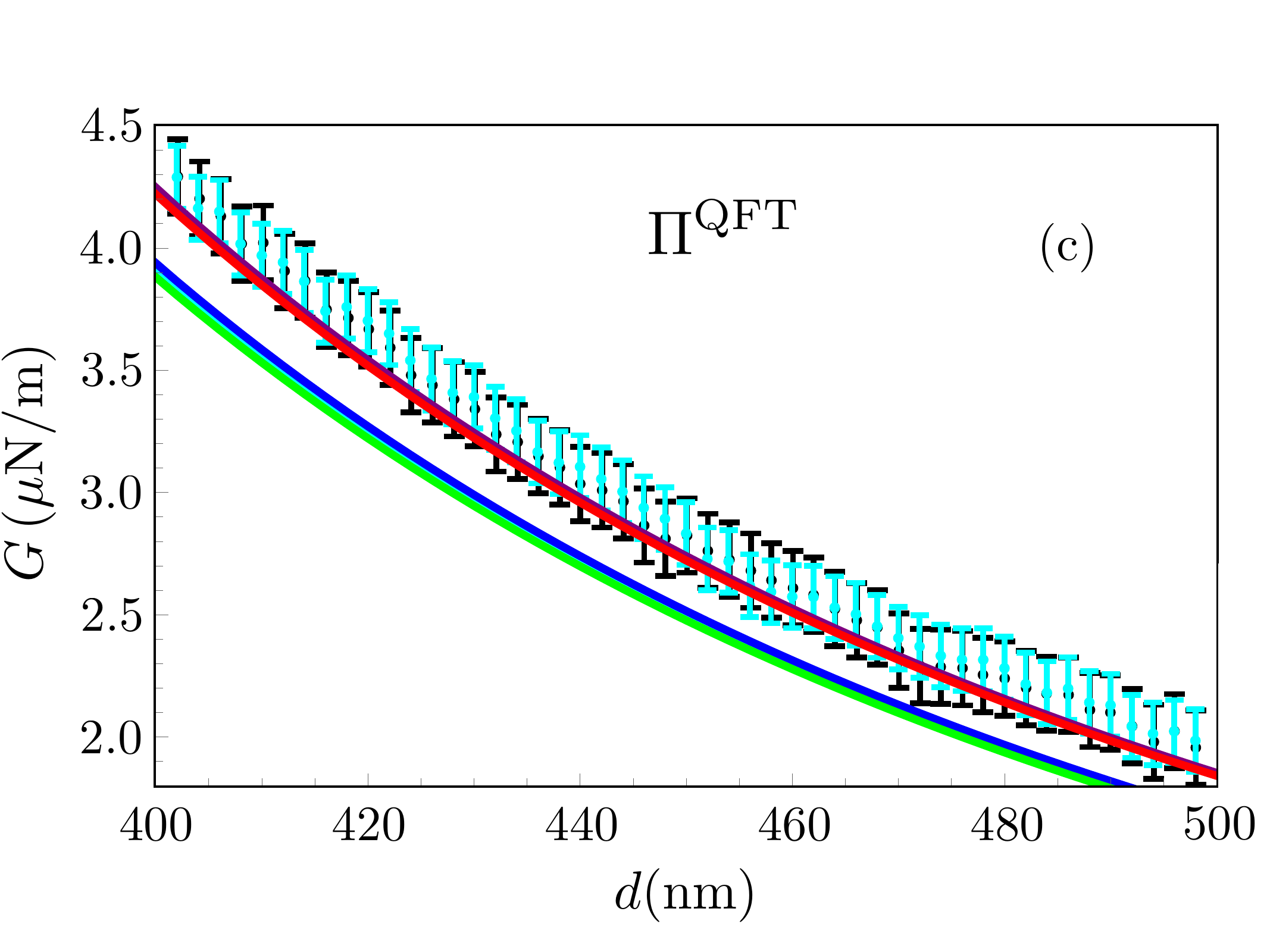}
\end{subfigure}
\hfill
\begin{subfigure}[b]{0.49\textwidth}
   \centering
   \includegraphics[width=\textwidth]{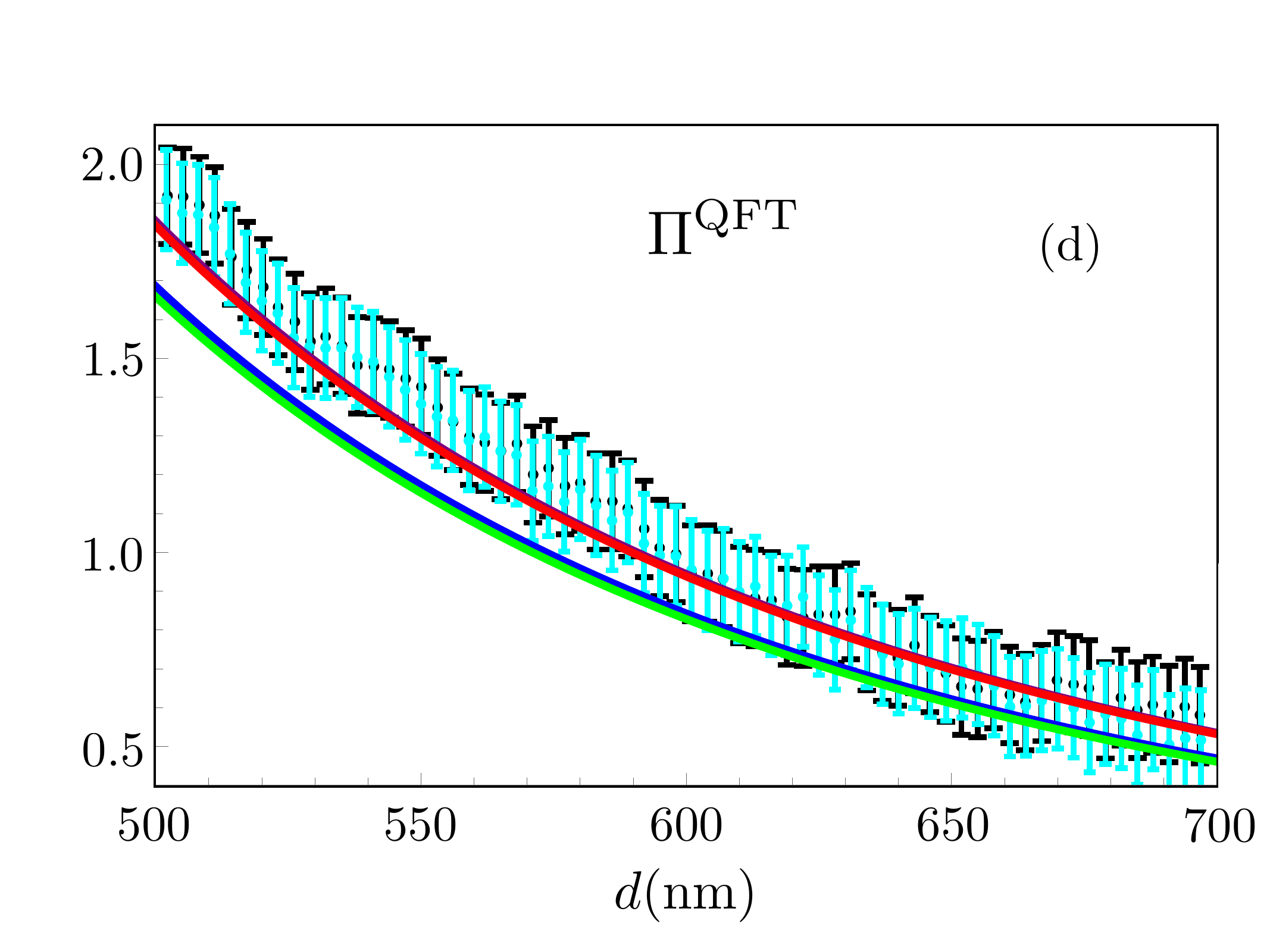}
\end{subfigure}
\caption{  Same as for \Fig{Fig_Comparison_NOLocal_Model}, where instead of $\sigma^{\rm{K}}$ we use here the the QFT polarization operator $\Pi^{\rm{QFT}}$ of Eqs.~(\ref{Def:PiL_QFTb},\ref{Def:Lim_xito0_PiL_QFTb})\cite{Bordag2015}\cite{PRL_Mohideen}\cite{PabloMauroComparisonKuboQFT2024}.
}
\label{Fig_Comparison_NRLocal_Model}
\end{figure}

\begin{figure}[H]
\centering
\includegraphics[width=0.49\linewidth]{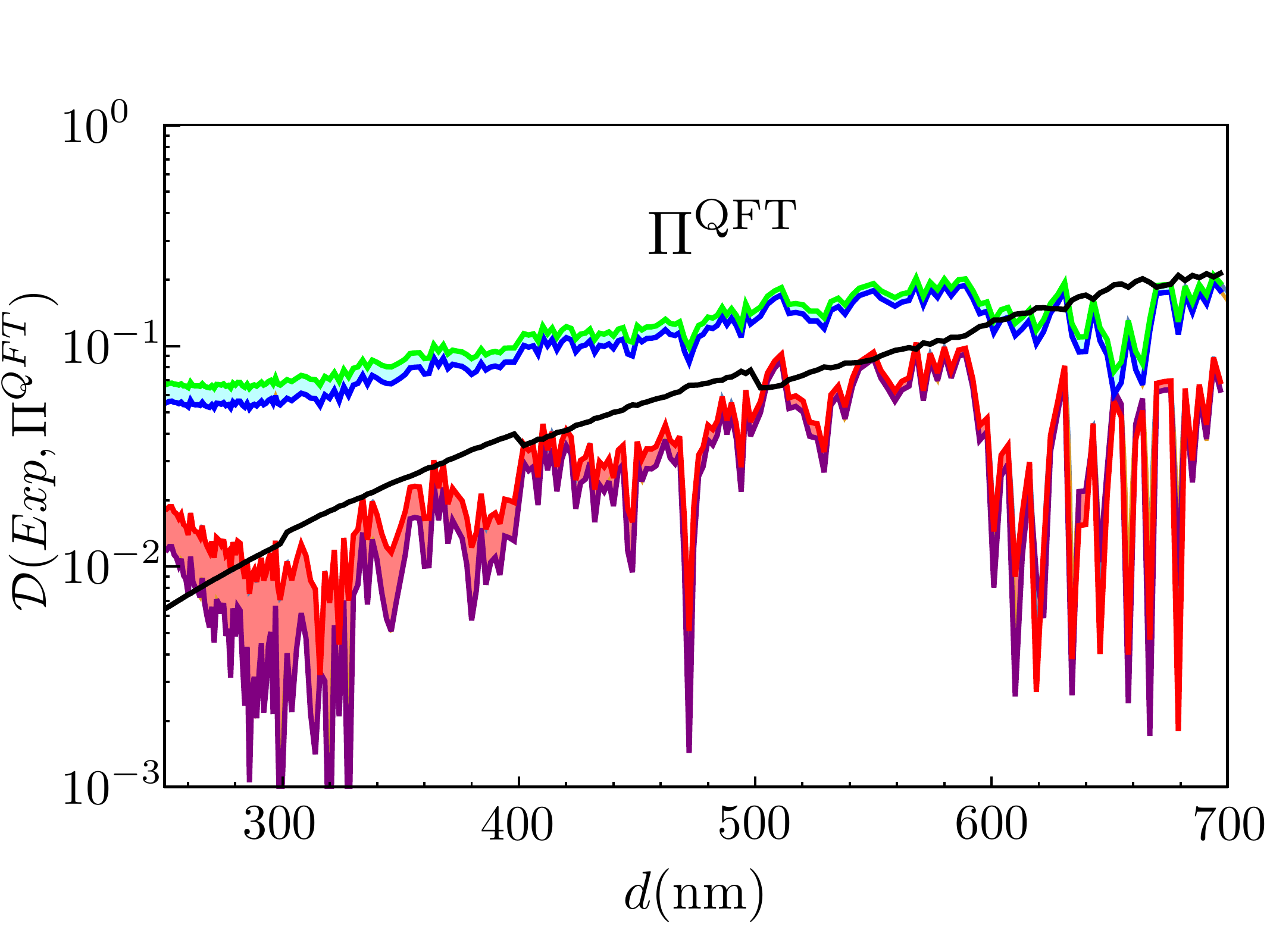}
\includegraphics[width=0.49\linewidth]{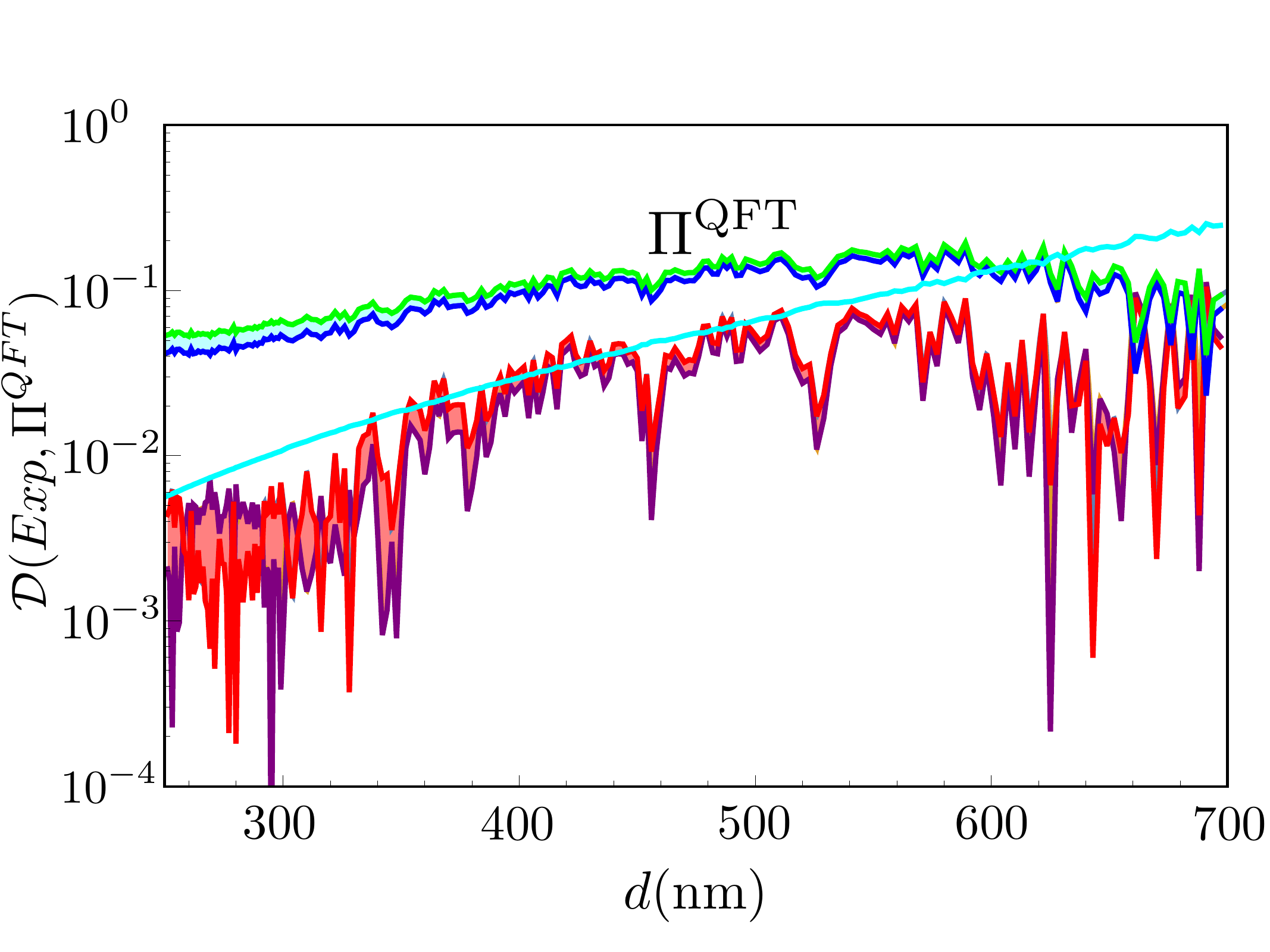}
\caption{ Same as for \Fig{Fig_RelError_NoLocalPRL} and for \Fig{Fig_RelError_NoLocalPRB}, where instead of $\sigma^{\rm{K}}$ we use here the the QFT polarization operator for graphene $\Pi^{\rm{QFT}}$  of Eqs.~(\ref{Def:PiL_QFTb},\ref{Def:Lim_xito0_PiL_QFTb}) \cite{Bordag2015}\cite{PRL_Mohideen}\cite{PRB_Mohideen}\cite{PabloMauroComparisonKuboQFT2024}. Theoretical values below this black (cyan) curve show cases when the theoretical results of \cite{PRL_Mohideen} (\cite{PRB_Mohideen}) are inside the experimental error-bars. We can observe a small discrepancy at short distances for the results of \cite{PRL_Mohideen} that disappears for the results of \cite{PRB_Mohideen}.
}
\label{Fig_RelError_NR}
\end{figure}

\end{widetext}

\end{document}